\documentclass[12pt,preprint]{aastex}

\newcommand{\noprint}[1]{}
\newcommand{\figsetstart}{{\bf Fig. Set} }
\newcommand{\figsetend}{}
\newcommand{\figsetgrpstart}{}
\newcommand{\figsetgrpend}{}
\newcommand{\figsetnum}[1]{{\bf #1.}}
\newcommand{\figsettitle}[1]{ {\bf #1} }
\newcommand{\figsetgrpnum}[1]{\noprint{#1}}
\newcommand{\figsetgrptitle}[1]{\noprint{#1}}
\newcommand{\figsetplot}[2]{\noprint{#1}}
\newcommand{\figsetgrpnote}[1]{\noprint{#1}}

\slugcomment{Accepted to AJ} 

\shorttitle{$\eta$ Carinae 2009.0 Spectroscopic Event}
\shortauthors{Richardson et al.}

\begin{document}
\title{The Optical Wind Line Variability of $\eta$ Carinae During the 2009.0 Event\altaffilmark{1}}

\author{N. D. Richardson\altaffilmark{2}, 
D. R. Gies\altaffilmark{3},
T. R. Gull\altaffilmark{4},
A. F. J. Moffat\altaffilmark{2},
\& L. St-Jean\altaffilmark{2}}
\altaffiltext{1}{Based on observations taken at the Cerro Tololo Inter-American Observatory 1.5m telescope, National Optical Astronomy Observatory, which is operated by the Association of Universities for Research in Astronomy, under contract with the National Science Foundation.}
\altaffiltext{2}{D\'epartement de physique and Centre de Recherche en Astrophysique du Qu\'ebec (CRAQ), Universit\'e de Montr\'eal, C.P. 6128, Succ.~Centre-Ville, Montr\'eal, Qu\'ebec, H3C 3J7, Canada; richardson@astro.umontreal.ca}
\altaffiltext{3}{Center for High Angular Resolution Astronomy, 
Department of Physics and Astronomy, 
Georgia State University, P. O. Box 5060, Atlanta, GA  30302$-$5060, USA} 
\altaffiltext{4}{Astrophysics Science Division, Code 667, NASA Goddard Space Flight Center, Greenbelt, MD 20771, USA}

\setcounter{footnote}{4}

\begin{abstract}
We report on high-resolution spectroscopy of the 2009.0 spectroscopic
event of $\eta$ Carinae collected via SMARTS observations using the CTIO 1.5 m
telescope and echelle spectrograph.  Our observations were made almost every
night over a two-month interval around the photometric minimum of $\eta$ Car
associated with the periastron passage of a hot companion.  The photoionizing
flux of the companion and heating related to colliding winds causes large changes
in the wind properties of the massive primary star.  Here we present an analysis
of temporal variations in a sample of spectral lines that are clearly formed
in the wind of the primary star. These lines are affected by a changing illumination of the
flux of the secondary star during the periastron passage.  We document the sudden onset of blue-shifted
absorption that occurred in most of the lines near or slightly after periastron,
and we argue that these absorption components are seen when we view the relatively
undisturbed wind of the foreground primary star.  We present time series
measurements of the net equivalent width of the wind lines and of the radial
velocities of the absorption trough minima and the emission peak midpoints.
Most lines decrease in emission strength around periastron, and those high
excitation lines formed close to the primary exhibit a red-ward velocity excursion.
We show how these trends can be explained using an illuminated hemisphere model
that is based on the idea that the emission originates primarily from the
side of the primary facing the hot companion.

\end{abstract}


\section{Introduction}
$\eta$ Carinae A (HD 93308) is one of the most massive and luminous stars in the local
region of the Galaxy.  It is surrounded by an hourglass-shaped nebula called the
Homunculus that was ejected during its mid-nineteenth century, luminous blue variable
eruption (Davidson \& Humphreys 1997).  It is also the primary star in a
5.54 year, eccentric orbit with a hot, massive, binary companion ($\eta$ Carinae B).
Many properties of the observed flux are modulated with this 5.54 year periodicity (forbidden and permitted emission lines, e.g., ~Damineli et al.~2000; photometric light curves, e.g., van Genderen et al.\ 2003, Fern\'{a}ndez-Laj\'{u}s et al.\ 2003, 2010, Whitelock et al.\ 2004; and X-ray light curves, e.g., Corcoran 2005) that is attributed to a highly eccentric ($e\sim0.9$) binary orbit and changing illumination of the primary star's wind and extended emission regions (Damineli et al.~1997). The spectroscopic events and other rapid changes in flux occur near the periastron passages. An overview of the observational history is given in our previous paper on $\eta$ Carinae (Richardson et al.~2010, hereafter Paper 1) as well as in Damineli et al.~(2008a,b).

The strongest constraints on the period of the binary system come from the X-ray light curve (e.g.,~Corcoran 2005; Corcoran et al.~2010) and the variability of the \ion{He}{2} $\lambda$4686 line (e.g., Mehner et al.~2011a; Teodoro et al.~2012), caused by the interaction of the winds of the two stars. X-rays from the wind--wind collision region encounter a varying column density of gas along our line of sight with the changing binary orientation. The X-ray maximum occurs shortly before periastron when the rarefied, highly ionized secondary star's wind pushes much of the dense, partially ionized primary wind out of our line of sight (bounded by a Coriolis-deflected bow shock), while the X-ray minimum occurs close to periastron when the collision region is blocked by obscuring, dense gas of the primary's wind. The \ion{He}{2} variability shows some similarities to the X-ray flux variability and indicates formation in a hot plasma.
Numerical models of the wind--wind collision by Okazaki et al.\ (2008) and Parkin et al.\ (2009) can reproduce many of the features of the X-ray light curve, but they also reveal discrepancies from observations made during the intense interaction at closest approach.  Although the details of wind collision need further investigation, the basic geometry of the models helps to explain the spatial variations of the emission spectrum from the resolved, extended wind region surrounding the central binary (Gull et al.\ 2009; Madura et al.~2012). 

$\eta$ Carinae is relatively nearby, situated in the Trumpler 16 cluster at a distance of $2.3\pm0.1$ kpc (Smith 2006), which has made it an excellent target for angularly resolved observations with the {\it Hubble Space Telescope} (HST) and the {\it Space Telescope Imaging Spectrograph} (STIS). With {\it HST/STIS}, we are able to probe structures to a spatial resolution of $0.1\arcsec \approx 230$ AU in the system. Gull et al.~(2009) demonstrate the complicated geometry of the system. Gull et al.~(2011) show that the spatially resolved spectroscopy of $\eta$ Car is best explained through colliding winds and provide 2-D models that explain the geometry of the different emission lines forming in the extended colliding-winds region. Madura et al.~(2012) and Clementel et al.~(2015a, 2015b) have further extended this modeling effort to show that the plane of the binary orbit is closely aligned to the plane of the Homunculus skirt, i.e., the bipolar ejecta of the Homunculus are nearly perpendicular to the binary plane. These models also show that $\eta$ Car B passes behind $\eta$ Car A's thick wind across the observed minimum, consistent with the models of Okazaki et al.~(2008).

The observed spectrum of $\eta$ Car in the optical region is extremely complicated. There are emission lines from the nearby slow-moving ejecta, dominated by the Weigelt knots (Weigelt \& Ebersberger 1986; Zethson et al.~2012). In addition, absorption lines from the Homunculus, Little Homunculus, and interstellar features complicate our understanding of the light emitted from the primary and secondary stars and their winds. Nielsen et al.~(2009) presented a spectral atlas for the ultraviolet through near-infrared wavelength regions. In total, more than 1500 lines are present in the region between 3060 and 10430 \AA, being formed in the ejecta, Weigelt knots, and the primary star's wind. There is no known or measurable spectral contribution from the secondary star in this portion of the spectrum. In the optical, the primary wind has strong emission lines such as the hydrogen Balmer lines. Nielsen et al.~(2007) showed that permitted transitions of \ion{He}{1}, \ion{N}{2}, \ion{Si}{2}, and \ion{Fe}{2}  form in the wind of the primary star and the wind--wind collision zones. The wind lines tend to be extremely broad, and are typically observed to be mostly stable except during the spectroscopic events, when the emission line flux drops, and many lines develop P Cygni type absorptions. In addition, complex, broadened forbidden line emissions most noticeably of [\ion{Fe}{2}], [\ion{Ni}{2}], [\ion{N}{2}] and [\ion{Fe}{3}] are modulated across the 5.54 year period.

The variability of the optical spectrum is dominated by the 5.54 year orbit, with major changes happening near the periastron passage (the spectroscopic events) when the secondary enters the line forming region of the primary (Hillier et al.\ 2001). Nielsen et al.\ (2007) analyzed high spatial resolution spectroscopy obtained with the Hubble Space Telescope and found that the blue-shifted P Cygni absorption troughs of hydrogen, \ion{He}{1}, and \ion{Fe}{2} were dependent on the orbital phase, with large blue-to-red Doppler shifts occurring during the periastron passages. They interpreted the radial velocity and strength variations to be caused by the interaction of the two stellar winds. Damineli et al.\ (2008b) described the orbital variations of the spectrum in two stages.  First, there is a slow variation due to changes in the ionization levels of the wind that are related to the changing binary separation. Then there is a rapid variation near periastron when the stars closely approach each other and cause a global collapse of the wind-wind collision shock zone. Mehner et al. (2011a) presented a detailed record of the spectral variations during the 2009 periastron event.  Their Gemini GMOS observations recorded not only the spectral variations along the line of sight to the central binary, but also at four offset positions that record light scattered at different orientations to the binary. They found (along with Teodoro et al. 2012) that the \ion{He}{2} $\lambda 4686$ emission displays a secondary peak after its collapse due to the geometry of the colliding winds. They also showed that the radial velocity variations of some lines were similar both along our line of sight to the system as well as from scattered light in the Homunculus thought to originate from a polar view of the star, and they suggested that the observed radial velocity variability is not related to binary motion.

In Paper 1, we presented our observations of H$\alpha$ over the 2009.0 event. We showed how the emission strength declined and blue-shifted absorption increased around the time of the X-ray minimum. These echelle spectroscopy observations recorded the spectrum between  H$\beta$ and approximately 7300 \AA. In this paper, we present a more detailed look at the wind lines in order to provide a detailed account of the
variations that occurred in this event and to consider explanations for them.
In Section 2, we outline our observations and reductions. Section 3 details the profile morphologies through the event for H$\beta$, \ion{He}{1}, \ion{N}{2}, \ion{Na}{1} D, \ion{Si}{2}, and \ion{Fe}{2} permitted lines.
Section 4 details the kinematics and variability of the \ion{He}{1} and other lines. In Section 5, we present a discussion focused on a simple model to understand the physics of the variability. We summarize our findings in Section 6.

\section{Observations}

We obtained high signal-to-noise observations with the CTIO 1.5m telescope and fiber-fed echelle spectrograph operated by the SMARTS Consortium ($R\sim 40,000$; Barden \& Ingerson 1998) almost nightly during the 2009 spectroscopic minimum and periastron passage of $\eta$ Car (a total of 40 spectra between 2008 Dec 18 and 2009 Feb 19). The reductions were described in Paper 1. Many of the spectra reported here had longer exposure times that saturated the H$\alpha$ profile but provide a higher signal to noise ratio for the other wind lines. We found that a 120s exposure would allow us to sample many of the wind lines in the spectrum of $\eta$ Car at a good signal to noise (usually at least a S/N $\approx$ 70--100 per pixel in the wavelength range of $\sim$ 4800--7200 \AA), and we began making such exposures regularly after 2009 January 1. For this analysis, we constructed a telluric template spectrum based upon the observed spectrum of the O star $\mu$ Col obtained with the same spectrograph to identify features formed in Earth's atmosphere. Most of the observed line profiles are relatively unaffected by such telluric absorption, but we corrected the line profiles of \ion{He}{1} $\lambda$7065 with the IRAF task {\tt telluric}\footnote{IRAF is distributed by the National Optical Astronomy Observatory, which is operated by the Association of Universities for Research in Astronomy (AURA) under a cooperative agreement with the National Science Foundation.	}. 

We resumed monitoring the star roughly once per week with the same instrumentation in 2009 October, and this continued through 2010 (22 spectra between 2009 Oct 21 and 2010 Jun 12) when this echelle spectrograph was decommissioned in favor of a new echelle spectrograph. During this time, we continued making short exposures for H$\alpha$, and longer exposures (240 s) to obtain higher signal to noise for the other wind lines in the spectrum. The new camera system that was installed at the end of 2009 allowed us to extract an additional 17 orders of data in the blue (for 18 spectra). The new extractions extend the spectral coverage blueward to $\sim$4000 \AA, but the shortest wavelength regions have extremely low signal. 

For comparative purposes, we include here one spectrum obtained with the CTIO 1.5 m telescope and its new echelle spectrograph CHIRON (Tokovinin et al.~2013), which was obtained near apastron on 2012 March 3. This observation had a signal-to-noise of $\sim 150$ per pixel, and a resolving power of 90,000. While the resolving power is much higher, we also note that the recorded spatial distribution of light on the sky is the same, because the new instrument uses the same fiber to input light into the spectrograph. We have continued our monitoring with CHIRON and these results will be shown in future analyses. 

We note that there are several different definitions of phase for this system, which correspond to the different observable minima. We also note that none of these conventions necessarily relates to periastron passage. There is an X-ray minimum (Corcoran 2005), minima from different photometric filters in the ultraviolet through infrared wavelengths, a spectroscopically defined minimum (Damineli et al.~2008a) that corresponds to the time when the \ion{He}{1} narrow emission from the Weigelt knots disappears, as well as a calendar based phase from the $\eta$ Carinae Treasury Project\footnote{http://etacar.umn.edu/}. While there is much debate over which phase is best to use (see the appendix of Mehner et al.~2011a), we adopt the spectroscopic definition of Damineli et al.~(2008a) for this spectroscopic analysis. This corresponds to a zero point of HJD 2,452,819.8 (cycle 11) and a period of 2022.7 d, with the 2009 minimum occurring at HJD 2,454,842.5. The first recorded event where a major change was observed in the spectrum was event \#1 in 1948 (Gaviola 1953), and the 2009.0 event corresponds to event \#12 with this phase convention.

\section{Wind-Line Morphologies During The 2009.0 Event}
Our observations of $\eta$ Carinae through the 2009.0 spectroscopic event recorded several permitted optical wind lines. In this section, we discuss observations of H$\beta$, as well as transitions of \ion{He}{1}, \ion{N}{2}, \ion{Si}{2}, and \ion{Fe}{2}. While our data set is rich in narrow emission lines originating in the Weigelt knots and ejecta, we focus here on the wind line developments during the event for this discussion. We present the rest wavelengths and atomic energy levels of the optical wind lines in Table 1. Table 1 includes the ionization energy of each species from a lower state and the lower and upper energy states of each transition compiled from NIST\footnote{http://nist.gov/pml/data/asd.cfm} in Columns 1--4, the kinematic integration limits for equivalent widths. All the wind line variations are shown in the online Figure Set 1.  Each entry of the figure set presents
an image of the dynamical spectrum through the dense time sampling around
periastron and line plots of representative spectra made before, during, and
after the event.  Arrows on the right hand side of the image indicate
the actual times of observation while white line segments on the right hand
side of the image indicate orbital phases 11.99, 12.00, and 12.01.
The panel below the dynamical image plots the average spectrum.
The panels for the line plots give (from bottom to top) sample spectra
obtained before, during (with offsets related to time of observation),
after periastron, and in 2012 with the CHIRON spectrograph.  The individual
plots are labeled with a date (HJD-2,450,000) and orbital phase $\phi$.
The corresponding figure numbers for each transition are given in column 8 of Table 1. Column 9 of Table 1 lists other features that blend with the line of interest and that complicate measurements.  More complete listings of line blends can be found in the tables of Nielsen et al.\ (2009) and Zethson et al.\ (2012). Brief descriptions of the observed
properties and variations follow in subsections for each line species.

\subsection{H$\beta$ and H$\alpha$}
Our observations of the H$\beta$ profile during the 2009.0 event (Fig.\ Set 1.1) have lower signal to noise because the line is recorded in the low signal wings of the echelle blaze function, and consequently, the profiles are not of high enough quality to make reliable measurements (such as equivalent width or radial velocity) compared to the H$\alpha$ profiles (Paper 1). Nevertheless, these data do qualitatively show the same developments as seen in the H$\alpha$ line presented in Paper 1 (see Fig.\ Set 1.2). The anomalous narrow absorption component at $V_r = -144$ km~s$^{-1}$, attributed to absorption from the Little Homunculus (Ishibashi et al.~2003), as well as a P Cygni type absorption both appear near HJD 2,454,837, just prior to phase 0.0. Mehner et al.\ (2011a) observed the H$\delta$ line during the 2009 event (see their Fig.\ 7), and it showed similar behavior of the P Cygni absorption component, but not in the narrow absorption, as H$\delta$ does not exhibit the narrow absorption component during the spectroscopic events.  Curiously, the H$\delta$ profile observed at offset positions that record scattered light from higher stellar latitudes displays the P~Cyg absorption continuously through the event rather than just appearing near the time of periastron.

\subsection{\ion{He}{1} Profiles}

There are five \ion{He}{1} lines ($\lambda \lambda$4921, 5016, 5876, 6678, and 7065 \AA) present in the parts of our spectra with good signal-to-noise. All the lines are contaminated in some manner by blending with other emission or absorption lines. The 4921 and 5016 \AA~lines are blended with \ion{Fe}{2} wind lines (Fig.\ Set 1.3, 1.4) that actually dominate the appearance and our measurements of the profiles (Zethson et al.~2012). The red wing of the 5876 line (Fig.\ Set 1.5) is blended with the complicated \ion{Na}{1} D doublet (see Section 3.4), which has many absorption components from the intervening gas of the Homunculus, Little Homunculus, and interstellar gas, as well as emission from the wind of the primary (and possibly the secondary) star. The blue wing of this profile is blended with a weak [\ion{Fe}{2}] $\lambda 5870$ emission line. The \ion{He}{1} $\lambda$6678 \AA~line has a [Ni II] line formed in the Weigelt knots on its blue wing, that interferes with the P Cygni absorption during the event (Fig.\ Set~1.6). The 7065 \AA~line (Fig.\ Set. 1.7) is rather clean in comparison, but the spectrum often has a large telluric component from the Earth's atmosphere. 

Fig.\ Set 1.5--1.7  display dynamical spectra and line plots of the \ion{He}{1} $\lambda$5876, 6678, and 7065 profiles across the 2009.0 event. All these profiles show a fading and subsequent brightening in line emission, and the development of a strong P~Cygni absorption component. \ion{He}{1} $\lambda \lambda$5876, 6678, and 7065 \AA~lines all exhibit strong radial velocity shifts during the event, with a range of approximately 200 km s$^{-1}$. The lines all have a narrow emission component (due to the Weigelt knots) that is present in early spectra, but disappears at phase 0.0 (by the definition of Damineli et al.~2008a). The ephemeris of Damineli et al.~(2008a) is confirmed by these data as the narrow emission in the \ion{He}{1} $\lambda$6678 line disappears exactly at our observed phase 12.0. 

Mehner et al.~(2011a) discussed the behavior of \ion{He}{1} $\lambda$4713 during the 2009 spectroscopic event. They show that the equivalent width of the \ion{He}{1} line rises in the months leading up to the event, with a fast decline near phase 0, and a short-lived increase after the minimum. Our results are very similar. Mehner et al. further describe the kinematics of feature, which are fully consistent with our kinematical measurements (Sections 4.2, 4.3).

\subsection{\ion{N}{2} Profiles}

$\eta$ Car is unique among the highest luminosity (and/or highest mass-loss rate) LBVs, in that \ion{N}{2} $\lambda \lambda$5667--5710 is very weak (e.g., Hillier et al.~2001). The comparison of $\eta$ Carinae and HDE 316285 (Hillier et al.~2001) shows how strong these lines are in HDE 316285, but nearly absent in the spectrum of $\eta$ Car. Fig.\ Set 1.8--1.10 present the dynamical spectra and line plots of the region of \ion{N}{2} transitions. We had difficulties normalizing this region due to the large number of absorption and emission features, and a fairly small spectral window, caused by the narrow blaze function of the echelle spectrograph. We performed the normalizations by a localized fit of the data to smooth it, and then we fit a polynomial to regions we deemed to be continuum from the analysis of Mehner et al.~(2011b). Similar difficulties were discussed in the analysis of Mehner et al.~(2011b). 

We observed three transitions in the echelle order best suited to the analysis of \ion{N}{2}. The transitions at 5666, 5676, and 5710 \AA\ always have some P Cygni type absorption present. The onset of deeper absorption is seen for the transitions at 5666, 5676, and 5710 \AA\ just prior to phase zero, with the \ion{N}{2} $\lambda$5676 transition showing the deepest absorption for $\approx 5$ days. All the \ion{N}{2} profiles show a strong radial velocity shift from blue to red with a final reversal at the end of the observation run. From a visual inspection, the P Cygni type absorptions of \ion{N}{2} (Fig.\ Set 1.8--1.10) are seen to show variations similar to those of the \ion{He}{1} transitions (e.g., Fig.\ Set 1.6; \ion{He}{1} $\lambda$6678).

Mehner et al.~(2011a) briefly discussed the \ion{N}{2} multiplet at $\lambda \lambda$ 5666--5710 in relation to the spectroscopic event, and showed how the direct view of the star
and the view from position FOS4 (scattered light from the polar region)
present similar line kinematics. Mehner et al.\ (2011b) extended the analysis and
proposed that the variations were related to the changing illumination
by the secondary star.
They further state that these lines are the only known features originating from the unperturbed primary wind. 

\subsection{\ion{Na}{1} D profiles}

The \ion{Na}{1} D region is one of the most complex portions of the optical spectrum of $\eta$ Car. In addition to the primary wind components, there is a blend of emission and several narrow absorption components from the intervening gas in the Homunculus and Little Homunculus in addition to interstellar absorption. Most of the narrow absorptions are from the great eruption of 1837-1860 ($-$512 km s$^{-1}$) or the lesser eruption of 1885-1895 ($-$146 km s$^{-1}$) and similar line absorption structure for UV resonance lines was described by Gull et al.~(2006). An examination of the profiles (Fig.\ Set 1.11--1.12) shows that there are multiple velocity components between the extreme velocities of $-512$ and $-146$ km s$^{-1}$, as seen in ultraviolet resonance lines (Gull et al.~2005). Other LBVs show this resonance line as a P Cygni profile with both emission and absorption components from the stellar wind as well as the interstellar components. For example, the star HDE 316285 was observed to have such a profile (Hillier et al.~1998; their Fig. 3), and it is nearly a spectroscopic twin to $\eta$ Car (Hillier et al.~2001). 

In total, there are more than a dozen absorption components to the \ion{Na}{1} D complex in our data. The extreme blue portion of the complex blends with \ion{He}{1} $\lambda$5876, and we observe variability in both the stellar and intervening components of the complex. First, in the \ion{Na}{1} D$_1$ line (Fig.\ Set 1.12), we see a narrow absorption component at $\approx -145$ km s$^{-1}$ that strengthens near phase 0.0 ($\sim$HJD 2,454,840). In Paper 1, we saw a similar component in H$\alpha$ at $-144$ km s$^{-1}$ that appeared just prior to phase 0.0, which is attributed to gas in the Little Homunculus. We postulated that when the hot secondary goes behind the wind of the primary in our line of sight, that the intervening gas from the Little Homunculus drops in ionization due to quenching of Lyman continuum from the secondary star by the enveloping primary wind. As the gas cools, more \ion{Na}{1} is in the ground state, allowing for a larger absorption in this resonance line. 
Secondly, the dynamical spectra show that just after phase 0, the deep absorption components at $\sim-500$ km s$^{-1}$ for both \ion{Na}{1} lines seem to become wider and deeper. This is similar to the changes seen with H$\alpha$ and \ion{Fe}{2}, although the ionization energies are very different. 

\subsection{\ion{Si}{2} Profiles} 

The \ion{Si}{2} $\lambda$6371 line is blended with \ion{Fe}{2} lines at 6371 and 6373 \AA\ and is best analyzed as a complex rather than individual lines. However, the \ion{Si}{2} $\lambda$6347 transition is mostly isolated from other lines (Fig.\ Set 1.13). The \ion{Si}{2} lines show a radial velocity movement towards longer wavelengths near phase 0.0, as well as a development of a P Cygni absorption that begins near phase 0.0. Multiple narrow emission lines from the Weigelt knots are present, and these weaken or disappear during the event. The behavior of this profile is reminiscent of that of the \ion{He}{1} lines discussed previously.

\subsection{\ion{Fe}{2} Profiles}

Many \ion{Fe}{2} lines are present in the optical spectrum of $\eta$ Car. We examined the lines at 5169, 5197, 5234, 5316, 6238, 6248, and 6456 \AA. The dynamical spectra and line plots for these transitions are shown in Fig.\ Set 1.15--1.21 respectively. Other \ion{Fe}{2} lines were either blended with nearby transitions or had too low a signal to study their variations. The strongest of these transitions (\ion{Fe}{2} $\lambda$5169, 5234, and 6456) developed P Cygni profiles during the event, and these lines qualitatively show an increase in P Cygni absorption and small increases in emission line strength between the P Cygni absorption trough and Weigelt knot emission. The timing of the P Cygni absorption development happens slightly later for these transitions (between HJD 2,454,842--2,454,852) than for the Balmer lines ($\sim$2,454,837--2,454,840) and the \ion{He}{1} lines (HJD 2,454,838) discussed previously.

\section{Wind Line Measurements} 
\subsection{Equivalent Widths}

We measured the equivalent widths of ten optical wind lines during the time of the photometric
minimum, and these are listed in column 4 of Table 2. These lines were selected based upon
transitions that were relatively isolated with relatively little contamination from blends.
Because the profiles have complicated shapes in general, we simply performed
a numerical integration across the entire blue absorption trough and red
emission peak using the integration limits listed in columns 6 and 7 of Table~1.
The net equivalent widths have typical uncertainties of $\pm 0.5$~\AA\ or larger
in strong lines.  The somewhat arbitrary method of rectifying the spectra
means that there may be systematic differences with measurements made by
other investigators, but relative variations in our measurements are reliable.
As none of these transitions have the remarkable strength of H$\alpha$
and often have P Cygni absorption at nearly the same strength as the emission,
we did not correct these equivalent widths for a changing continuum flux as we
did for H$\alpha$ measurements in Paper~1.

We show in Figure 2 the time evolution of the equivalent widths for these optical
lines and H$\alpha$ (Paper 1). The largest relative changes occurred in the
\ion{He}{1} and \ion{Si}{2} lines (and probably also in the \ion{N}{2} lines
that were too weak to measure securely).  All these lines experienced a drop in
emission strength near phase 0.0 due the emergence of an absorption component
and a decrease in the emission peak.  A similar decrease occurred in the
H$\alpha$ emission strength but with a lower fractional change.
The variations in the \ion{Fe}{2} emission strength were all small, due in
part to a near balance between increases in both the absorption trough and
emission peak strength.  Note that we do not include in Figure~2 our results
for \ion{He}{1} $\lambda\lambda 4922, 5016$ because these features are
significantly blended with \ion{Fe}{2} lines (Zethson et al. 2012).

\subsection{P Cygni Absorption Velocities}

We measured a radial velocity for the P~Cygni absorption components by
determining the minimum flux position as we did for the H$\alpha$ transition of P Cygni in Richardson et al.~(2011).
This velocity $V_{\rm min}$ was set by finding the zero crossing in the
numerical derivative of a smoothed version of the spectral lines.
The S/N ratio was generally sufficient to determine $V_{\rm min}$ reliably
in those cases where the P~Cygni absorption was present.
We estimate the uncertainty as approximately $\pm 5$ km~s$^{-1}$ based
upon the night-to-night scatter after removal of the velocity trends.

We measured $V_{\rm min}$ for 14 transitions, and these measurements are listed
in column 5 of Table~2.  The temporal variations of $V_{\rm min}$ are shown
in Figure~3 in panels for each line species.
We caution again that the results for \ion{He}{1} $\lambda\lambda 4922, 5015$
are significantly affected by blends with \ion{Fe}{2} lines.
The largest variations occurred
in the \ion{He}{1}, \ion{N}{2}, and \ion{Si}{2} lines, all of which showed
a systematic red-ward shift of $\approx 200$ km~s$^{-1}$ after phase 0.0.
Inspection of the actual profiles in Figure Set 1 suggests that this
velocity shift is caused by a systematic reduction in absorption strength
with time that starts at the most extreme negative velocities and progresses
towards zero velocity.  Thus, these changes might be better regarded as
representative of an absorption component that becomes narrower over time,
rather than a wholesale shift of a fixed-width absorption feature.
The measured changes in the other lines with absorption components are
generally small, consistent with their constant appearance in Figure Set 1.

\subsection{Emission Bisector Velocities}

In Paper 1, we presented emission line bisector velocities for H$\alpha$
across the 2009 spectroscopic event in order to search for evidence of spatial asymmetries
in the emitting volume. Here we present emission line bisector velocities for
\ion{He}{1} and several other lines with well-defined emission peaks.
These were measured by estimating a bisector velocity $V_b$ at $\sim 20\%$ of the
peak height above the continuum.  The results appear in column 6 of Table~2.
We estimate that the uncertainties are generally $\pm 5$ km~s$^{-1}$ for
the most reliable cases based upon the scatter around the de-trended curves.
However, many of these lines are fairly weak, so the measurement is complicated
by the low contrast between the emission line and continuum and by blending
with components from the Weigelt knots or other emission lines.
Therefore, we only present in Figure~4 measurements of $V_b$ for those
cases with clearly defined emission peaks.  The largest changes occurred in
the \ion{He}{1} $\lambda\lambda 6678, 7065$ and \ion{Si}{2} $\lambda 6347$ lines.
All these lines show an increase in velocity of $\approx 200$ km~s$^{-1}$
that begins around phase 0.0.  Curiously there was also a local flattening
in the velocity curves of these lines around phase 0.005 that is suggestive
of some structure in the wind geometry.   The emission bisector velocities
of H$\alpha$ and \ion{Fe}{2} $\lambda 5316$ show similar trends but with
much smaller amplitude.

\section{Illuminated Hemisphere Model for \ion{He}{1}}

The wind line variations documented above reflect how the hot secondary
alters the wind properties as the companion plunges deep into the envelope of
the primary.  We can begin to understand the wind and line variations by
considering the large amplitude changes observed in the \ion{He}{1} lines.
The \ion{He}{1} lines probe the inner part of the stellar wind.
The ionization energy needed for a recombination line such as
\ion{He}{1} $\lambda 6678$ is 24 eV, and the models of
Hillier et al. (2001) show that these lines should form at a
radius of $\sim 11 R_\star$ (3 AU) from the primary.
Madura et al. (2012) estimate that the binary semi-major axis
is 15.4 AU and the eccentricity is 0.9, so that the periastron
separation is only 1.5 AU, similar to or smaller than the
\ion{He}{1} line forming region.  This implies that the \ion{He}{1}
lines should form near the region of a wind-wind collision zone at periastron, and this expectation is borne out in
detailed models of the wind ionization zones (Clementel et al.\ 2015b).

The main orbital-phase related variations of the \ion{He}{1} lines
are probably linked to changing illumination of the primary star
by the flux from the companion and the wind-wind collision region
between the stars.  We show in Figure 5 the orientation of
the orbit of the secondary relative to the primary as projected onto
the sky.  The orbital elements are from Madura et al.\ (2012):
$a=15.4$ AU, $d=2.3$ kpc, $e=0.9$, $i=138^\circ$,
$\omega_p = 263^\circ$ (longitude of periastron for the primary), and
$\Omega = {\rm PA}_z - 90^\circ = 227^\circ$ (longitude of the ascending
node).
The top four panels show the position of the secondary (marked by a plus
sign)
at four orbital phases (arranged in clockwise order like the sense of the
orbit
in the sky). The open circle at the bottom left represents the primary
star, and
the brighter portion of the disk shows the hemisphere facing the secondary
(mostly hidden behind the primary at phase 0.0).

If emission lines like \ion{He}{1} $\lambda 6678$ form in the illuminated
hemisphere of the primary wind,
then we expect that the emission would be stronger (weaker) when that
hemisphere is
orientated towards (away) from us.  Let us assume for simplicity that the
emission
flux varies with the projected area of the illuminated region.  Then the
fraction of
the projected disk illuminated by the secondary is given by
$$F={1\over 2} (1 + \cos a)$$
where $a$ is the angle from the observer through the primary to the
secondary.
This angle depends on the orbital geometry as
$$\cos a = -\sin(\nu + \omega_s) \sin i$$
where $\nu$ is the time dependent true anomaly, $\omega_s$ is the
longitude of periastron for the secondary star ($\omega_s = \omega_p
+180^\circ)$,
and $i$ is the inclination.

We compare the orbital variation of the illuminated fraction with the
the equivalent width measurements of \ion{He}{1} $\lambda 6678$
in Figure 6 (left).  A single linear scaling parameter was selected
to match the amplitudes of both quantities.  We assumed that periastron occurs
at phase 0.0, and we adopted an eccentricity $e=0.9$ and three trial
values of longitude of periastron, $\omega_p = 200^\circ$ (solid line),
$\omega_p = 240^\circ$ (dashed line), and $\omega_p = 263^\circ$ (dotted line).
The latter estimate of $\omega_p = 263^\circ$ was derived from the analysis of 
Madura et al.\ (2012) from their analysis of the forbidden line
emission resolved by HST/STIS, with $\omega_p = 240^{\circ}$ derived from an analysis of the \ion{He}{2}
$\lambda 4686$ variability by M. Teodoro et al.\ (2015, in preparation).
This model does fit the time and approximate duration of the emission
weakening observed around periastron which occurs when the
companion in almost behind the primary (assuming a longitude of periastron
for the primary of $\omega_p = 263^\circ$, which is close to the value for
periastron at conjunction, $\omega_p = 270^\circ$). However, that model
suggests
that the drop is more or less symmetrical around periastron, while the
observations
suggest a slower recovery. A better match is made with a smaller longitude of
periastron, $\omega_p = 240^\circ$ (dashed line). The overall good match of this simple model
with the observed emission variations around periastron suggests that the
emission is localized in the hemisphere facing the companion. We note that this illumination would arise from a combination of both the incident flux of the secondary star and the apex of the wind-wind collision zone. The deviation between the model and
the observations at later time (near phase 12.2) may be due to
the presence of strong P~Cygni absorption at these epochs that causes
a decrease in emission line flux.

The illuminated hemisphere model can also address the issue of the radial
velocity
changes observed around periastron.  If the emission lines form predominantly
in the wind of the primary that is facing the companion, then
we would expect to see outflow directed away from the illuminated
hemisphere and towards the companion, and at periastron the wind flow would
be directed away from us.  We calculated the mean radial velocity by adopting 
a constant radial wind outflow of $v_\infty = 500$ km~s$^{-1}$
(assuming the wind has reached terminal speed in the line forming region).
The observed mean Doppler shift over the illuminated hemisphere is
then given by
$$v_r = -v_\infty {\int \cos \theta dA} / {\int dA}$$
where $\theta$ is the angle (in radians) from the stellar surface normal to the
observer and the area integral is taken over the illuminated region
of the disk.  The analytical solution of the integral is
$$v_r = v_\infty {{4}\over{3\pi}} {{\pi + \cos a \sin a - a}\over{1+\cos
a}}$$
where again $a$ is the angle from the observer through the
primary to the secondary (with a time dependence as given above).

The predicted wind velocity as a function of orbital phase is shown in
Figure 6 (right) and is compared to the measured line
bisector velocities for \ion{He}{1} $\lambda 6678$.
Here we applied a single constant offset velocity to match
the predictions with the observed radial velocities.
This constant was taken as the difference between the mid-range
velocity of the bisector measurements (for the red-shifted position
of the emission peak) and the mid-range velocity of the model.
We also included the relatively
small orbital motion of $\eta$~Car~A with the mean wind velocity.
The nominal model with $\omega_p = 263^\circ$ (dotted line) shows
the right amplitude, but it appears more symmetrical than the
observations that show a more extended decline.
A model with a smaller longitude of periastron, $\omega_p = 240^\circ$
(dashed line), does a better job of matching the observations.  
The agreement is suggestive that the simple model is on
the right track.  The differences at later time may again be
related to the increased P~Cygni absorption that causes a
a red-ward (positive) shift in the bisector velocity.

We emphasize that this model is purely geometric and does
not account for physical processes that must be important.
For example, we have neglected the change in separation with
orbital phase that is clearly related to the flux from the
secondary incident on the primary star, and we ignored
radiative transfer issues related to heating of the primary's
wind and photosphere.  More fundamentally, we have treated the
secondary as a point source, whereas in fact it is a hot star
with a fast wind that comes so close to the primary at periastron
that the wind interaction zone reaches close to where the \ion{He}{1}
line formation zone occurs.  The location and structure of the
zones of singly-ionized He in the wind of $\eta$~Car~A have
recently been investigated in numerical models by Clementel et al.\
(2015a,b).  In particular, Clementel et al.\ (2015b) have
calculated the He ionization structure near periastron, and
they find potential emission zones in the inner wind of A,
the pre-shock wind of A near the apex of the wind-wind collision zone,
and in post-shock wind gas that is also photoionized by B.
The combination of emission from all these zones might indeed
cause an asymmetry in the wind emission in the sense of enhanced
emission in the direction of companion.  Consequently, our illuminated
hemisphere model might be regarded as a first order approximation
of a highly structured wind and interaction region.

\section{Conclusions}

The data shown in Figure Set 1 present a detailed account of the optical spectral variability
of the system during the 2009 event, with the best timing cadence of any previous campaign.
These data portray how the wind lines of the primary are influenced by the
close proximity of the hot companion at periastron.  The line variations
are different in the transitions of ions that form preferentially at
different heights in the expanding atmosphere of the primary star.
We found that all the lines display the emergence of a blue-shifted,
absorption trough near or shortly after phase 0.0 (near periastron).
The absorption component lasts only 10 days or so in some ions (\ion{He}{1}, \ion{N}{2})
but remains throughout the event in others (H$\alpha$, H$\beta$).
We measured the net combined equivalent width of the blue absorption trough and
red emission peak (plus other blends in some cases) for the selected lines,
and in most cases we find that the equivalent width dropped around periastron
due to the combined effects of increased blue absorption and decreased emission.
The largest relative changes occurred in the lines of \ion{He}{1} and \ion{Si}{2}
(and probably \ion{N}{2} but these lines are difficult to measure).
Some of the lines displayed Doppler shift changes that we measured in two ways.
First, we determined the minimum flux position $V_{\rm min}$ in the absorption trough
whenever possible.  This absorption velocity is relatively constant in the
H and \ion{Fe}{2} lines, but it increases with time in the lines of
\ion{He}{1}, \ion{N}{2}, and \ion{Si}{2} as the absorption progressively
disappears at the lower, more extreme Doppler shifts.
Second, we derived a bisector velocity $V_b$ that measures the Doppler shift of
the red emission peak, and these show a large change of $\approx +200$ km~s$^{-1}$
during the event for the lines of \ion{He}{1} and \ion{Si}{2}.  On the other hand,
changes in the H and \ion{Fe}{2} lines were quite small.

All these variations are probably related to changes in the
ionization state of the wind gas caused by the close proximity of
the hot companion star around periastron.  The orbit of the companion
has yet to be directly measured, but the changes in the illumination
of the hot circumstellar gas by the companion can set important
constraints on the orbital elements. 
The documented illumination changes for the spectral lines in this data set can successfully rule out the family of orbital elements that have the companion star between the observer and the primary at periastron.
Madura et al.\ (2012) developed
a three-dimensional model for the circumstellar gas that they
fit to angularly resolved spectroscopic observations from HST/STIS,
and their work indicates that periastron occurs near companion
superior conjunction.  Thus, we expect that the portion of the wind
illuminated by the hot companion will change quickly around periastron,
and the hemisphere facing the companion at periastron will be
directed away from us.  The idea that the hemisphere facing the
companion will be most influenced by its ionizing flux was the basis
for our simple model (\S5) for the variations in the \ion{He}{1}
$\lambda 6678$ line.  The model shows that for reasonable values
of the longitude of periastron of the primary $\omega_p$ we can
approximately explain the decline and recovery of the emission
equivalent width as the result of the changing orientation of
the emitting hemisphere that faces the companion.  Likewise,
the model accounts for the red-ward Doppler shift of the emission
as the result of the changing orientation away from us of the radial
outflow in the illuminated hemisphere of the primary star.
The reality of \ion{He}{1} line formation is complicated by heating in
the wind-wind collision zone (Clementel et al.\ 2015b), but we suspect
that any large scale asymmetry related to an emitting gas distribution
in the vicinity of the hemisphere of the primary facing the secondary
will create line strength and Doppler shift variations similar to
those predicted by the simple model.

The success of our illuminated hemisphere model and the three-dimensional
colliding winds model of Madura et al.\ (2012) in explaining the spectral
variations that occur around periastron results from our adoption of a
longitude of periastron that places the companion beyond the primary at
the time of periastron.  The observed equivalent width and radial velocity
variations of the wind lines can only be matched if the illuminated
hemisphere of the primary is partially directed away from us at periastron
towards a more distant companion.  Thus, earlier suggestions that companion
is in the foreground at periastron (Kashi et al.\ 2011) are no longer tenable.

The differences in the amplitudes of the emission strength and Doppler shift variations
among the various line species probably result from differences in the
radius of line formation.  Hillier et al.\ (2001) present
calculations of the distribution of the line emission contributions
as a function of distance from the center of the primary
(see their Figs.\ 15 and 16), and they find that lines like
H$\alpha$ and \ion{Fe}{2} $\lambda 4923$ form relatively far out
in the wind (at $\log (r/R_\star) = 1.7$ and 2.2, respectively),
while other, higher excitation lines like \ion{He}{1} $\lambda 5876$
form closer to the star (at $\log (r/R_\star) = 0.6$).
Based upon the ionization and excitation energies (Table~1),
we expect that the \ion{N}{2} lines form in a similar location
to the \ion{He}{1} lines, while the \ion{Si}{2} lines originate
at a larger radius intermediate between those of \ion{He}{1}
and H$\alpha$.   Thus, the similarities in the spectral variations
of the \ion{He}{1} and \ion{N}{2} lines probably result from
their formation in a similar location (and likewise for the
H$\alpha$ and \ion{Fe}{2} lines).  The fact that the wind line
variations are much smaller for lines formed far out in the wind
(H$\alpha$ and \ion{Fe}{2}) probably indicates that the
photoionization of the wind by the companion is quite localized
around periastron.  The companion attains such a small periastron
separation ($\log (d/R_\star) = 0.6$) that it must be deeply
immersed in the wind of the primary then.  We suspect that the
ionizing photons from the companion only penetrate a small
distance in the envelope of the primary, so that conditions
in the outer parts of the wind (where the lower excitation lines
form) see relatively little change in ionization state,
leading to relatively small variations in the H and \ion{Fe}{2} lines.

The primary of $\eta$~Car acts as an occulting disk to the
flux of the hot companion, so gas in the shadow cone on the side
of primary facing away from the companion may experience
conditions of a stellar wind undisturbed by the companion
(Richardson et al.\ 2010; Groh et al.\ 2012).  This part of
the undisturbed wind will appear projected against the
photosphere of the primary around periastron when the
companion is beyond the plane of the sky.   We suggest that
the appearance of the P~Cygni absorption components near
periastron is the result of our line of sight crossing
through the undisturbed wind then.  With the decline in
photoionizing flux in this region, the atomic populations
may favor absorption or scattering of photons from the
primary that are directed toward us.  The shorter duration of the
absorption component phase in the \ion{He}{1} and \ion{N}{2} lines
is probably due to the close proximity of the hot companion
and line forming region as the companion reappears
above the optically thick horizon of the primary
as seen by the wind gas along our line of sight to the primary.
On the other hand, the longer duration of the absorption phase
in H$\alpha$ may be due to the longer path length of the ray
from the primary along our line of sight and to the longer
timescale for the wind opacity decline that would allow the
photoionizing flux of the companion to reach large distances
out into the wind of the primary.

The suggestion that the presence of an absorption trough in H$\alpha$
represents our temporary sampling of the undisturbed wind of
$\eta$~Car~A is supported by angularly resolved spectroscopy of
light scattered by dust in the Homunculus nebula.
For example, Mehner et al. (2011a) show how the H$\delta$ line
presents a P~Cygni-type absorption at most orbital phases at
an offset position that scatters light from the polar direction
of the primary.  If this polar wind is relatively unaffected
by the changing photoionization of the companion, then perhaps
primary's wind only suffers significant ionization changes in
sectors centered on the orbital plane.  The similarity in the
appearance of H$\delta$ in the star's spectrum just after periastron
with that for the polar wind (see Fig.\ 7 of Mehner et al.\ 2011a)
suggests that our line of sight to the primary traverses comparable
wind properties at that time.  Thus, the particular orientation
of the companion lying beyond the primary at periastron probably
allows us to view the primary's wind for a short while then without the
ionization changes imposed by the secondary's flux.

Our program to monitor $\eta$~Car with the CTIO 1.5m telescope and fiber fed echelle
spectrograph was very successful in documenting the spectral changes through the 2009 event.
We have continued this monitoring program through to the 2014 periastron passage
and beyond.  In future work we plan to make a detailed comparison of the
wind line variations between these two events in order to investigate the
issue of long-term variations in the mass loss rate (Mehner et al.\ 2012)
and to confront the predictions of hydrodynamical models and radiative
transfer calculations for the ionization zones created by colliding winds
(Clementel et al.\ 2015b).

\acknowledgments 
These spectra were collected with the CTIO 1.5~m telescope, which is operated by the SMARTS Consortium. We thank Todd Henry for his assistance in obtaining the spectra during the spectroscopic event. We are extremely grateful to Fred Walter (Stony Brook University) for his careful scheduling of this program and to the CTIO SMARTS staff for queue observing support. The spectra analyzed in the post-event state were obtained with NOAO programs 09b-153, 12a-0216, and 12b-0194. This research has made use of the data archive for the HST Treasury Program on Eta Carinae (GO 9973) which is available on-line at http://etacar.umn.edu. The archive is supported by the University of Minnesota and the Space Telescope Science Institute under contract with NASA. NDR~gratefully acknowledges his CRAQ fellowship. This work was supported by the National Science Foundation under grants AST-1009080 and AST-1411654. AFJM is grateful for support from NSERC (Canada) and FQRNT (Quebec). Institutional support has been provided from the GSU College of Arts and Sciences and from the Research Program Enhancement Fund of the Board of Regents of the University System of Georgia, administered through the GSU Office of the Vice President for Research and Economic
Development.

\clearpage

\figsetstart
\figsetnum{1}
\figsettitle{Observed line profiles during $\eta$ Car's 2009 event}

\figsetgrpstart
\figsetgrpnum{1.1}
\figsetgrptitle{H$\beta$ dynamical representation (logarithmic, left) and line plots (right). The white horizontal bars on the dynamical representation indicate phases 11.99, 12.00, and 12.01.}
\figsetplot{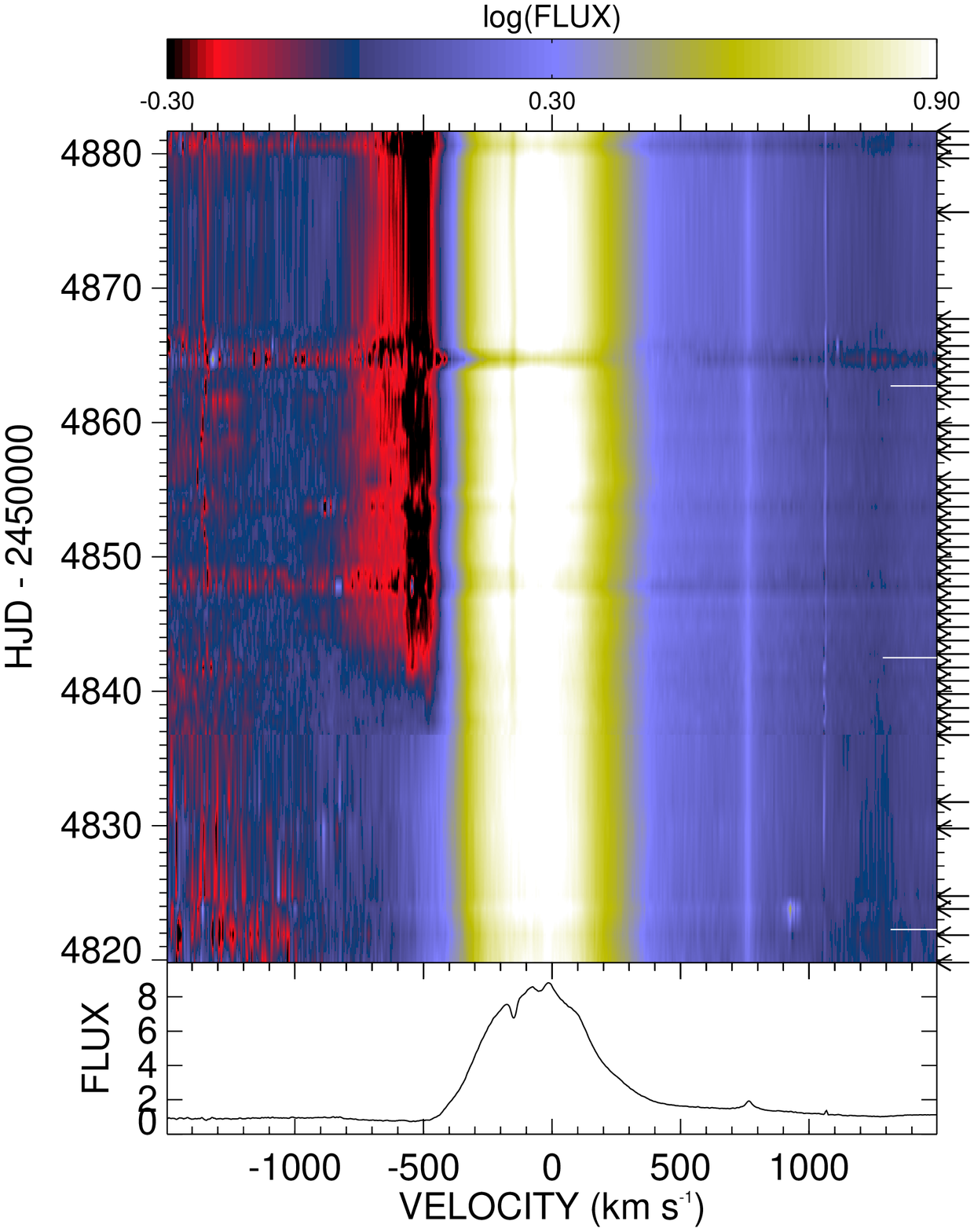}{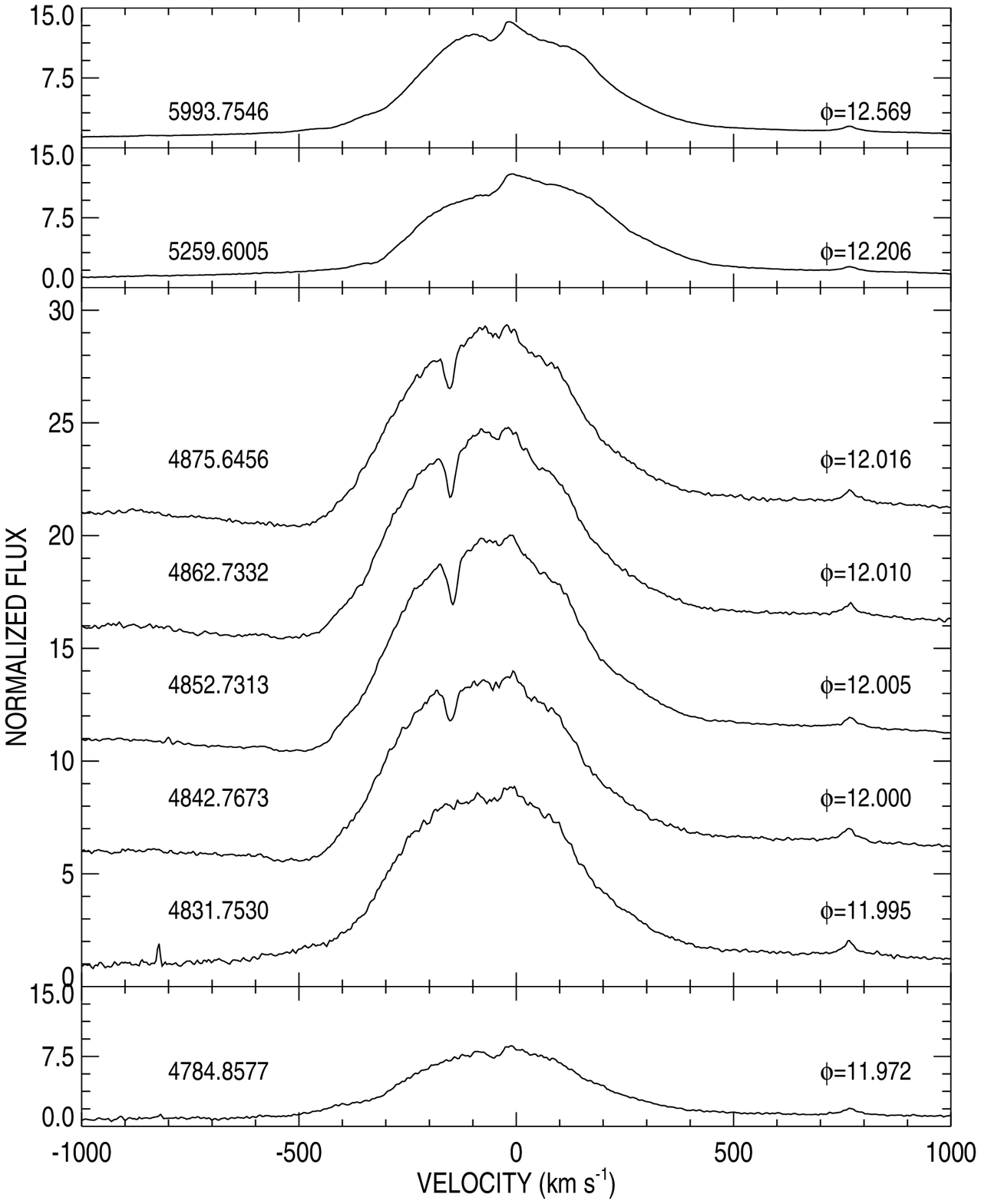}
\figsetgrpnote{Dynamical (left) and line plots (right) of H$\beta$ during the 2009 event. The white horizontal bars on the dynamical representation indicate phases 11.99, 12.00, and 12.01.}
\figsetgrpend

\figsetgrpstart
\figsetgrpnum{1.2}
\figsetgrptitle{H$\alpha$ dynamical representation (logarithmic, left) and line plots (right). The white horizontal bars on the dynamical representation indicate phases 11.99, 12.00, and 12.01.}
\figsetplot{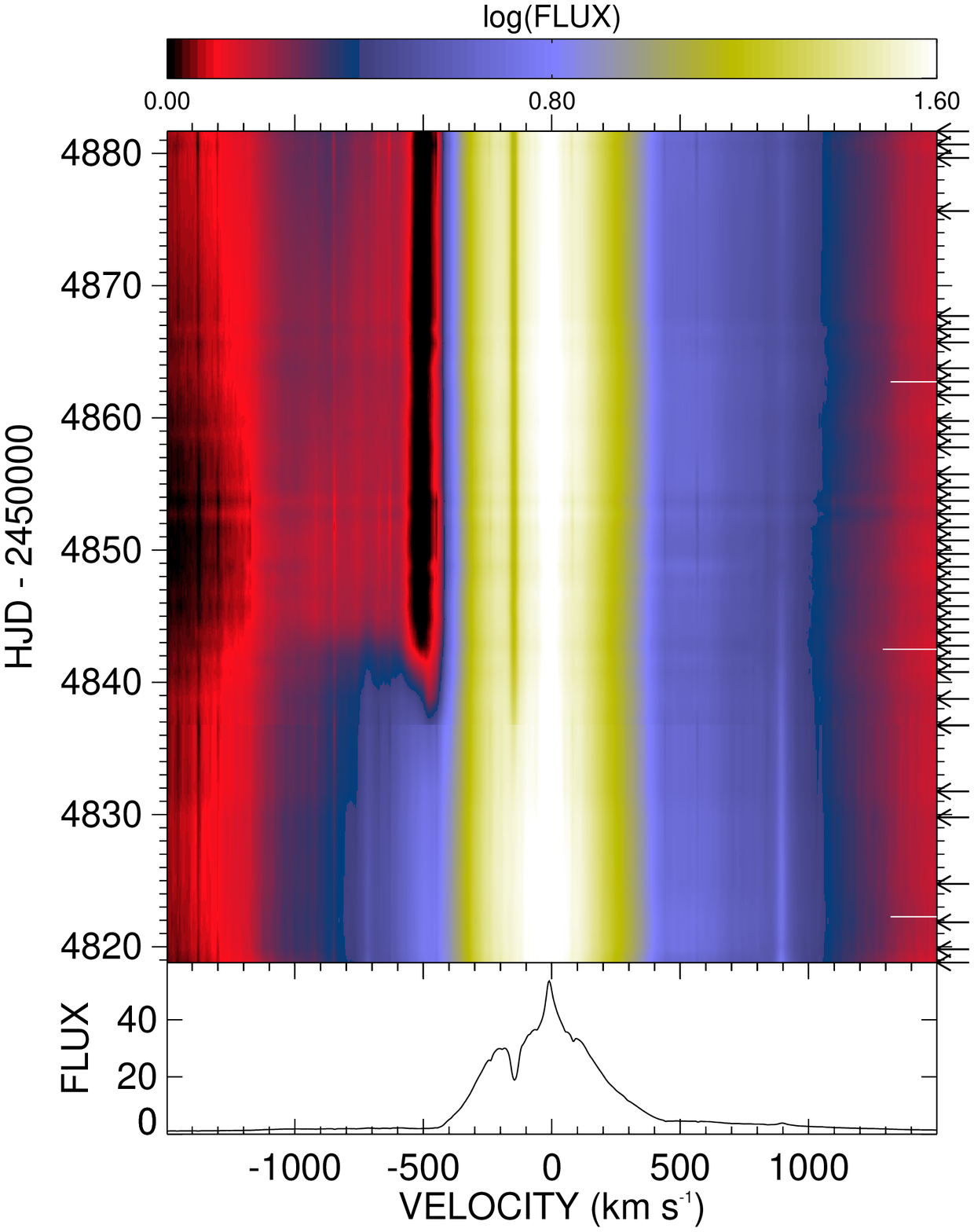}{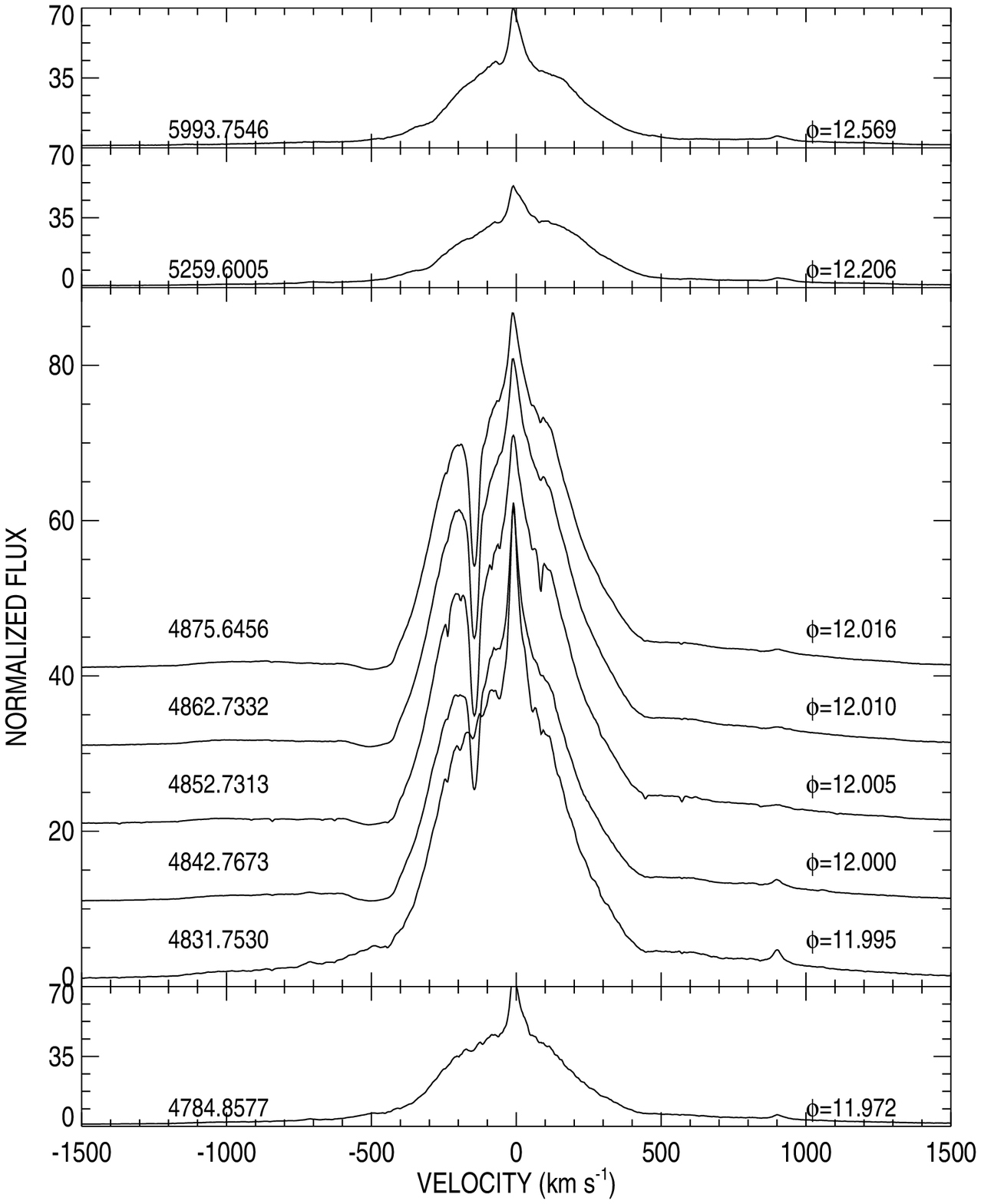}
\figsetgrpnote{Dynamical (left) and line plots (right) of H$\alpha$ during the 2009 event. The white horizontal bars on the dynamical representation indicate phases 11.99, 12.00, and 12.01.}
\figsetgrpend

\figsetgrpstart
\figsetgrpnum{1.3}
\figsetgrptitle{\ion{He}{1} $\lambda$4922 dynamical representation (left) and line plots (right). The white horizontal bars on the dynamical representation indicate phases 11.99, 12.00, and 12.01.}
\figsetplot{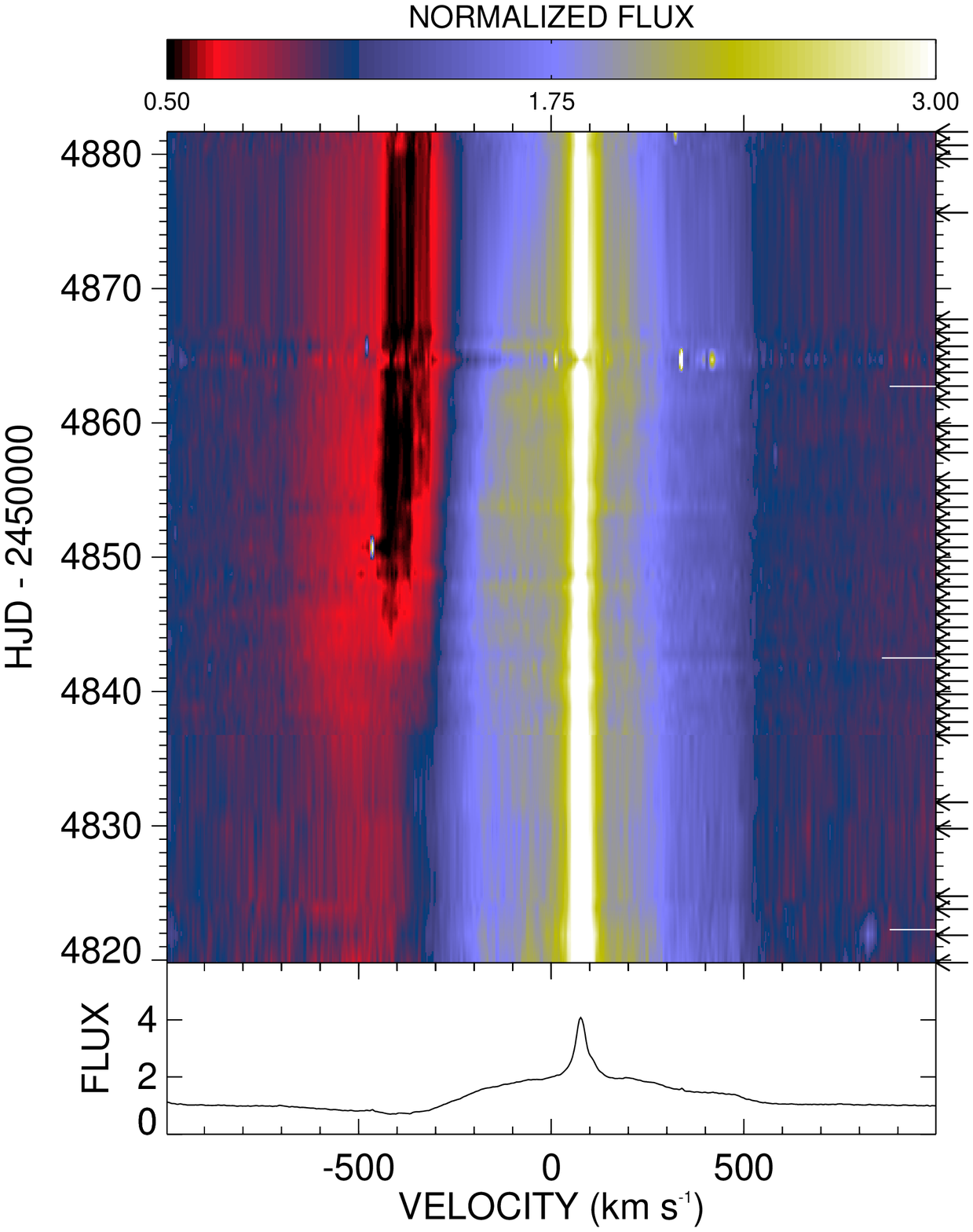}{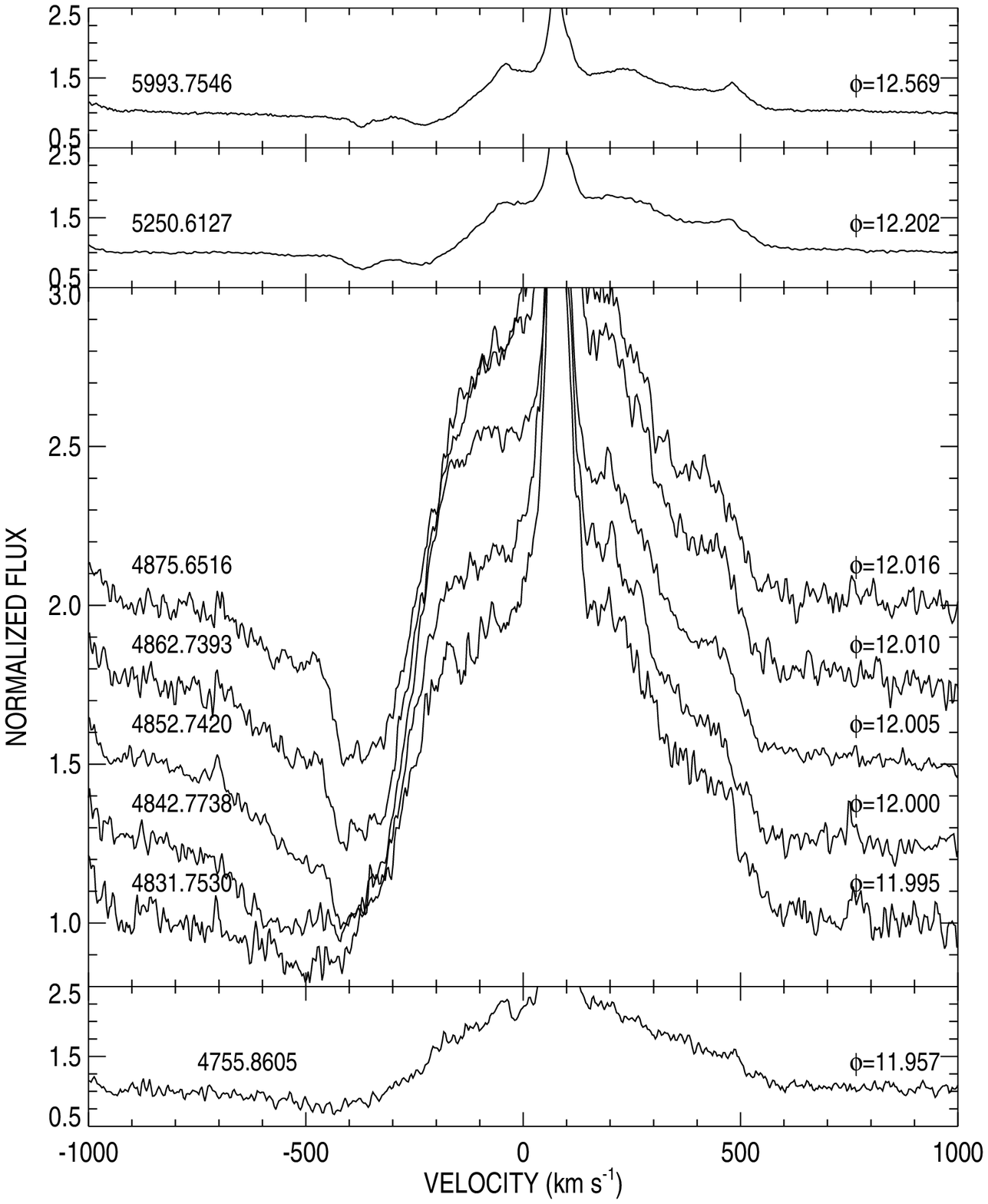}
\figsetgrpnote{Dynamical (left) and line plots (right) of \ion{He}{1} $\lambda$4922 during the 2009 event. The white horizontal bars on the dynamical representation indicate phases 11.99, 12.00, and 12.01.}
\figsetgrpend

\figsetgrpstart
\figsetgrpnum{1.4}
\figsetgrptitle{\ion{He}{1} $\lambda$5015 dynamical representation (left) and line plots (right). The white horizontal bars on the dynamical representation indicate phases 11.99, 12.00, and 12.01.}
\figsetplot{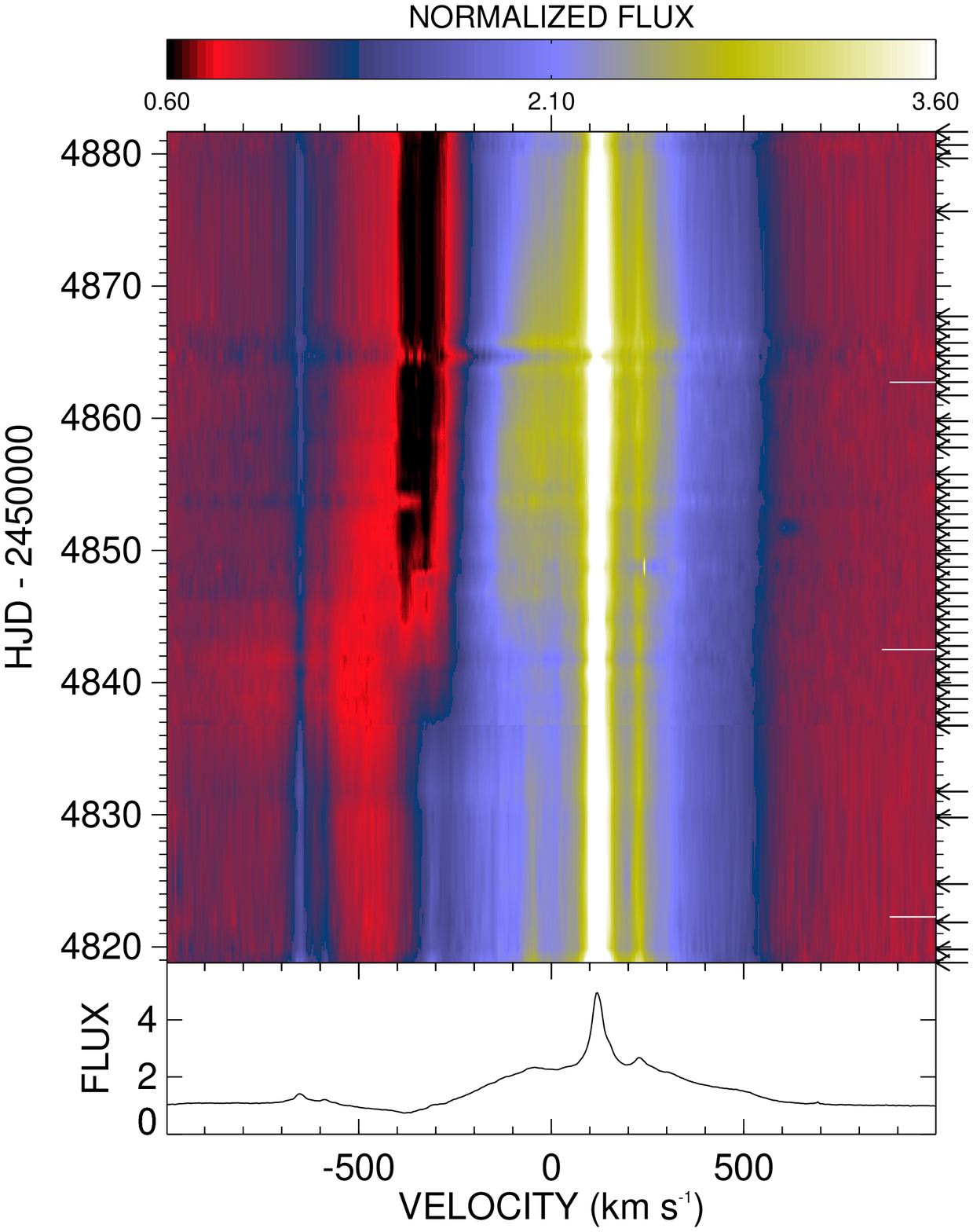}{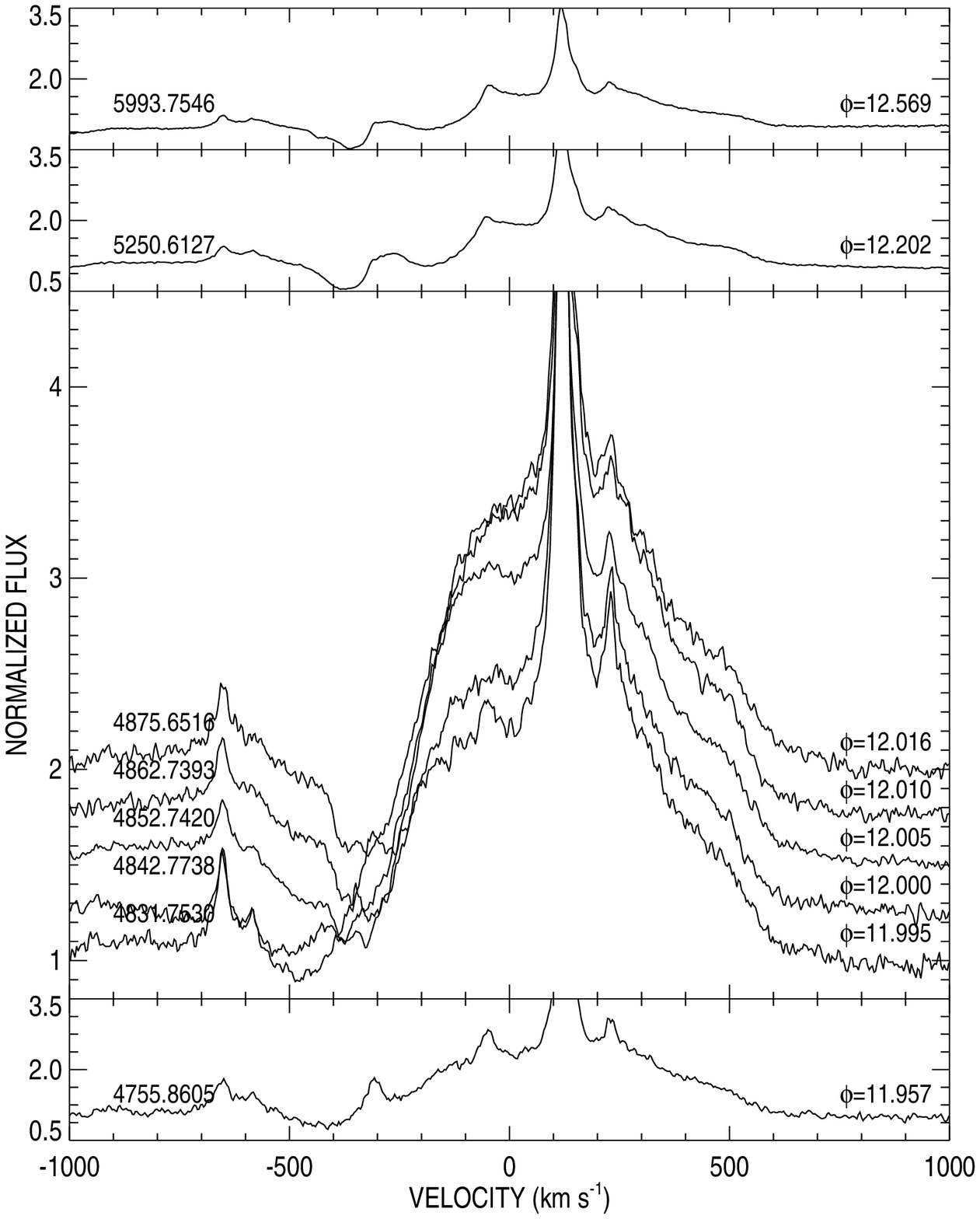}
\figsetgrpnote{Dynamical (left) and line plots (right) of \ion{He}{1} $\lambda$5015 during the 2009 event. The white horizontal bars on the dynamical representation indicate phases 11.99, 12.00, and 12.01.}
\figsetgrpend

\figsetgrpstart
\figsetgrpnum{1.5}
\figsetgrptitle{\ion{He}{1} $\lambda$5876 dynamical representation (left) and line plots (right). The white horizontal bars on the dynamical representation indicate phases 11.99, 12.00, and 12.01.}
\figsetplot{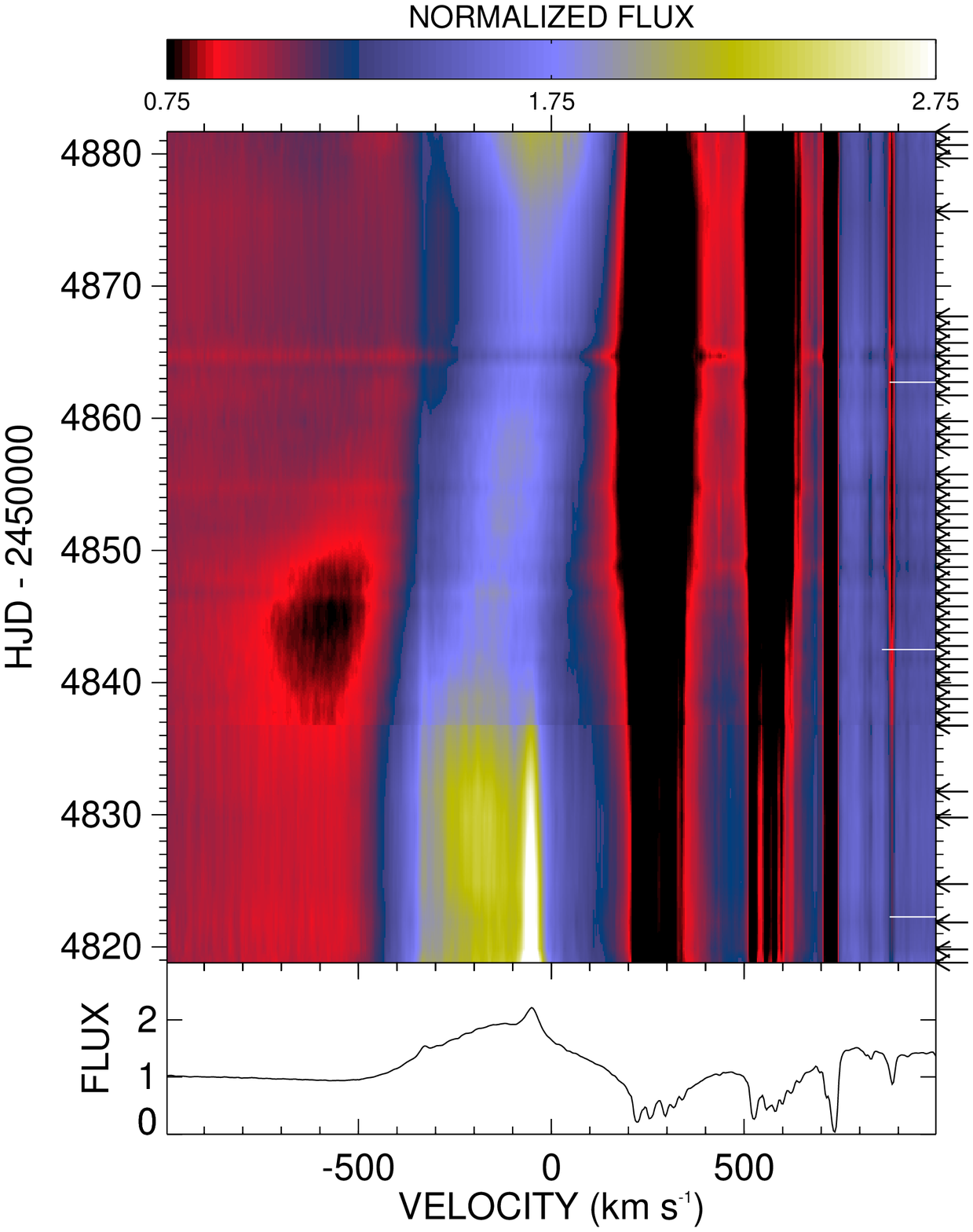}{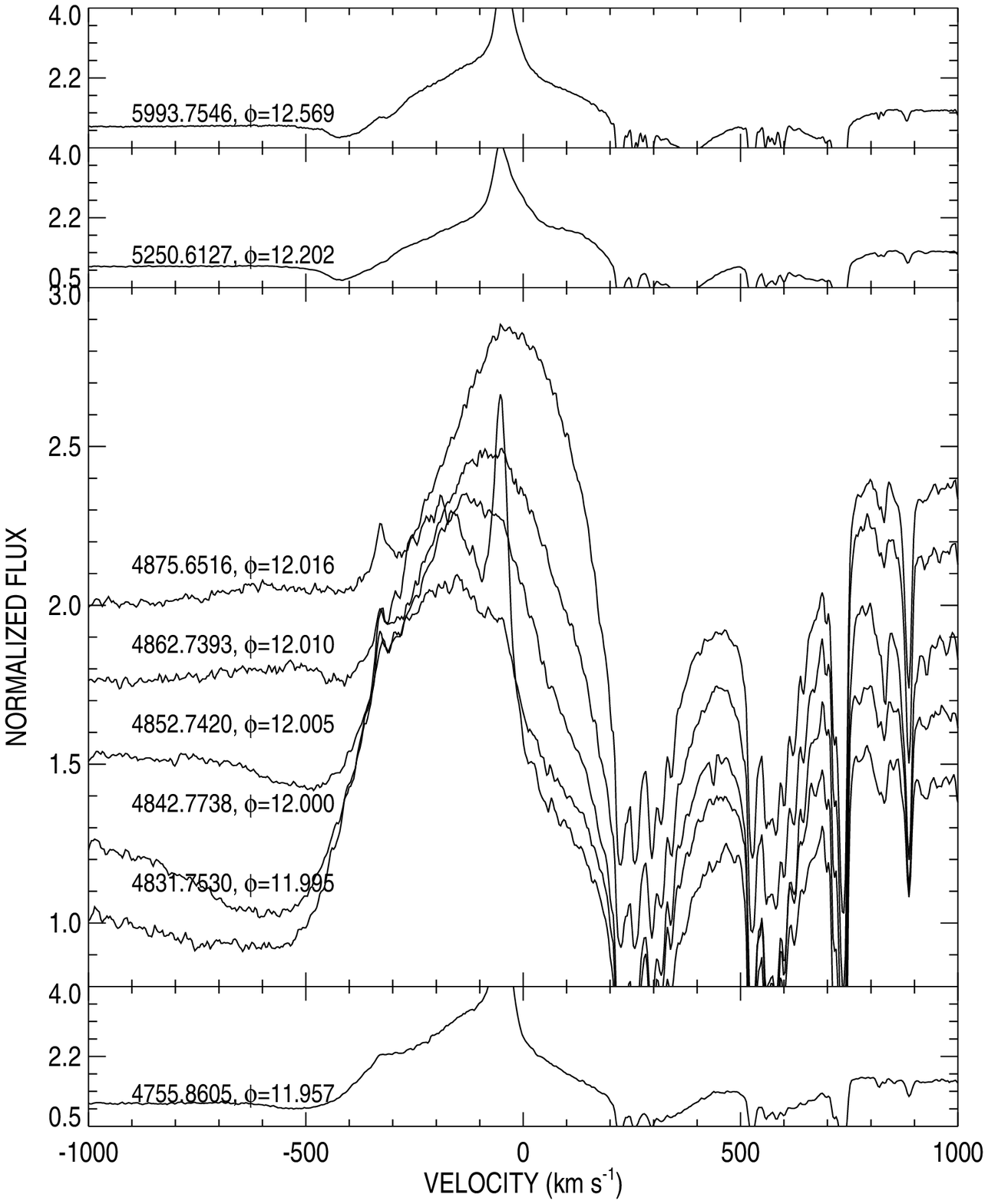}
\figsetgrpnote{Dynamical (left) and line plots (right) of \ion{He}{1} $\lambda$5876 during the 2009 event. The white horizontal bars on the dynamical representation indicate phases 11.99, 12.00, and 12.01.}
\figsetgrpend

\figsetgrpstart
\figsetgrpnum{1.6}
\figsetgrptitle{\ion{He}{1} $\lambda$6678 dynamical representation (left) and line plots (right). The white horizontal bars on the dynamical representation indicate phases 11.99, 12.00, and 12.01.}
\figsetplot{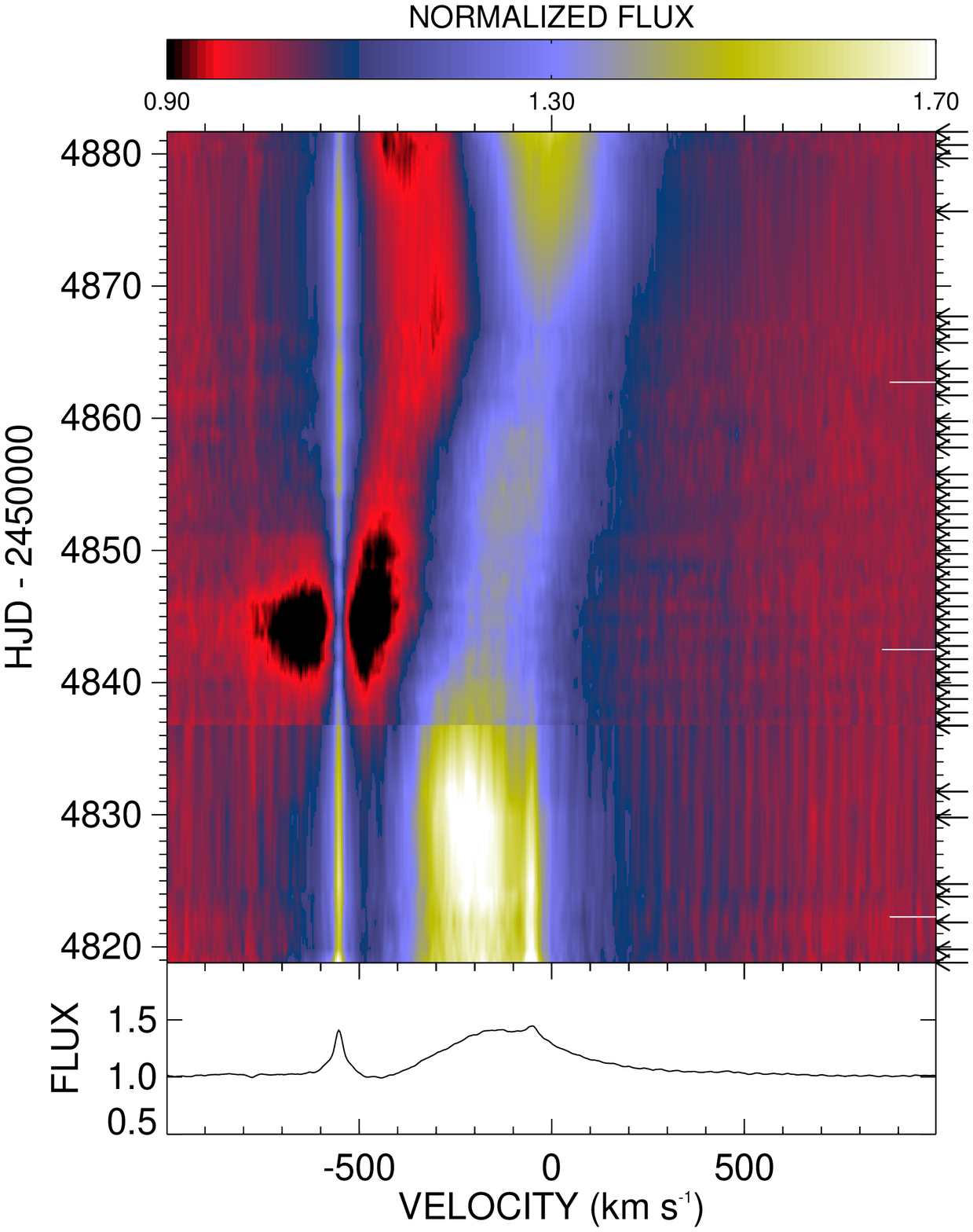}{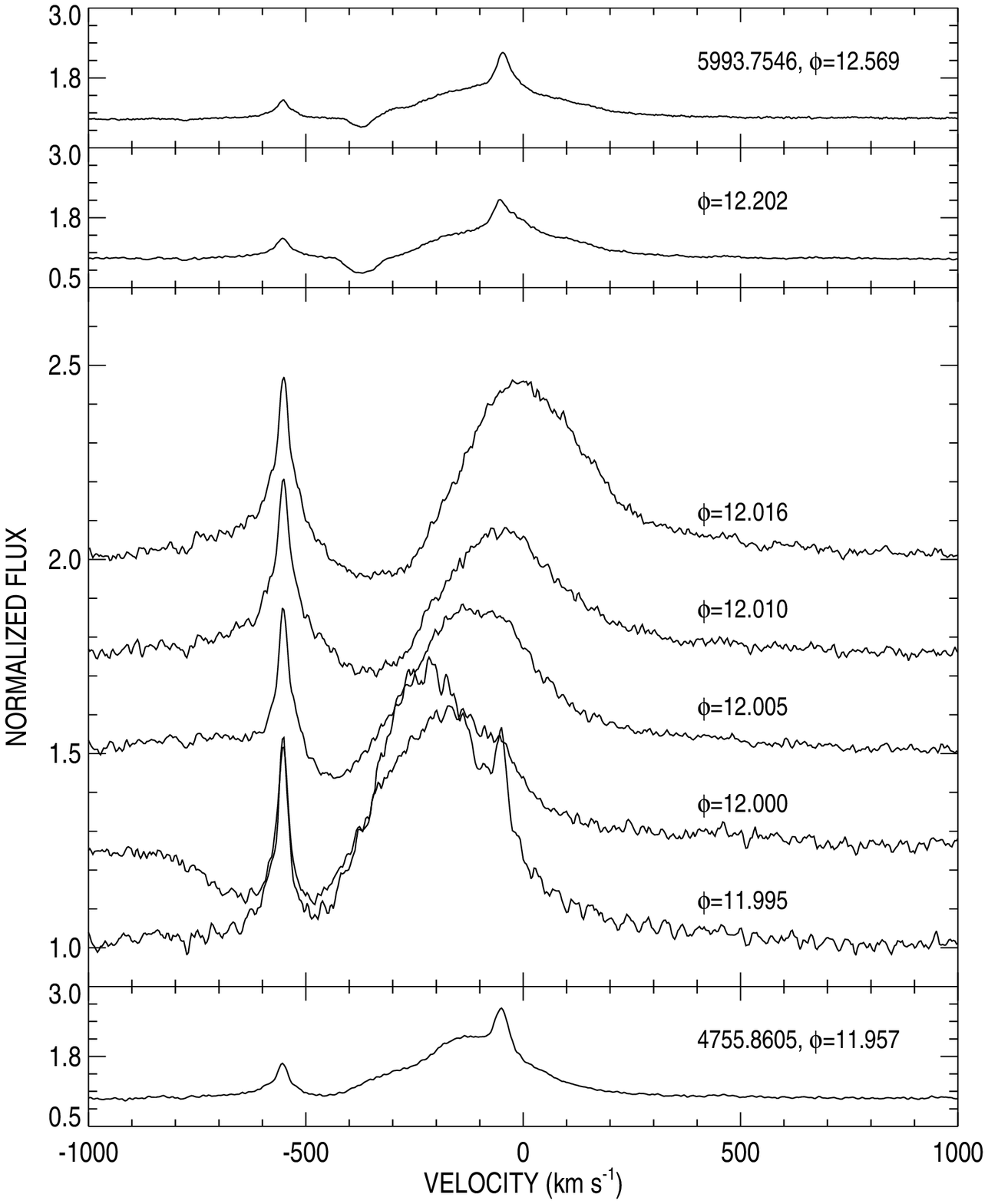}
\figsetgrpnote{Dynamical (left) and line plots (right) of \ion{He}{1}$\lambda$ 6678 during the 2009 event. The white horizontal bars on the dynamical representation indicate phases 11.99, 12.00, and 12.01.}
\figsetgrpend

\figsetgrpstart
\figsetgrpnum{1.7}
\figsetgrptitle{\ion{He}{1} $\lambda$7065 dynamical representation (left) and line plots (right). The white horizontal bars on the dynamical representation indicate phases 11.99, 12.00, and 12.01.}
\figsetplot{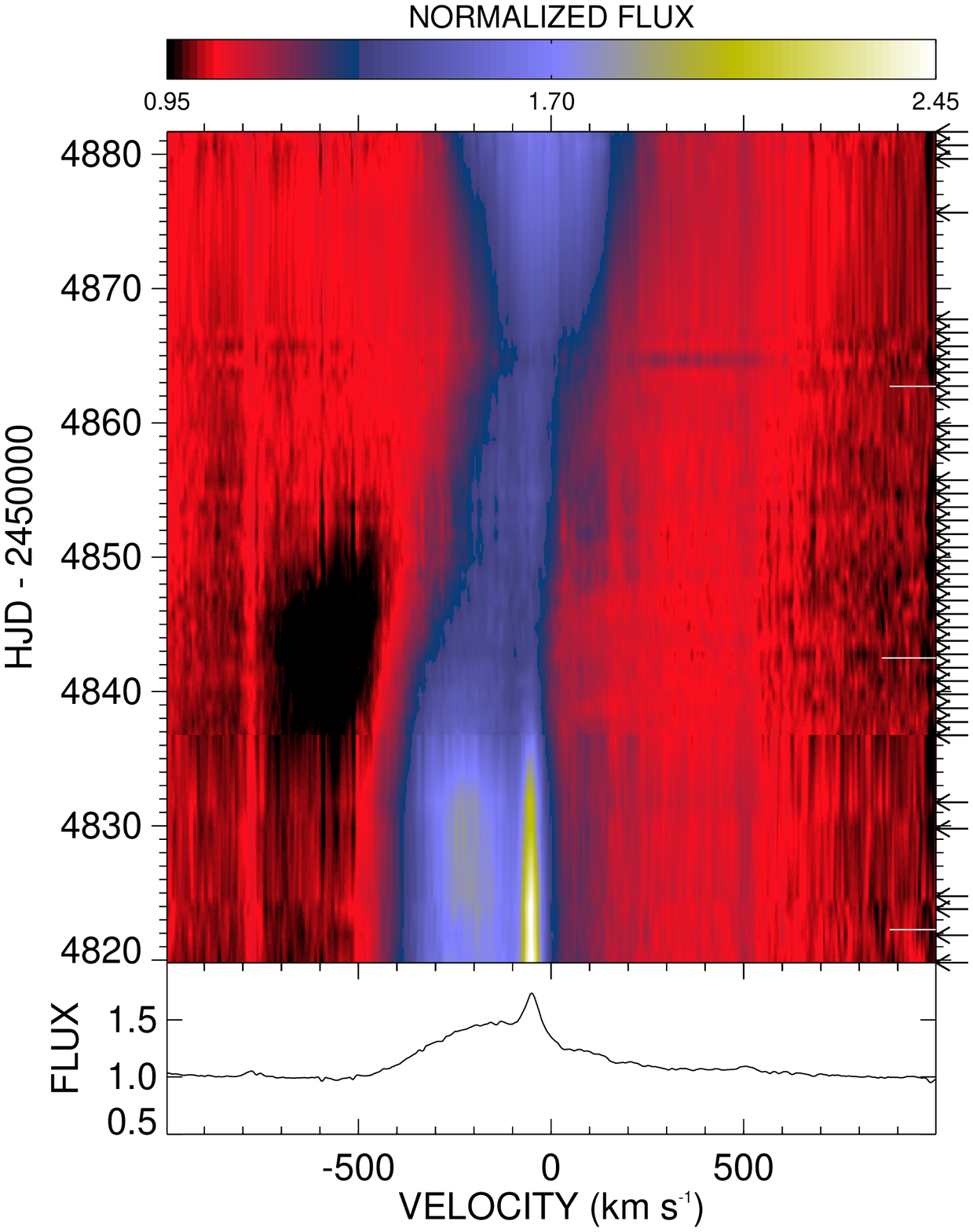}{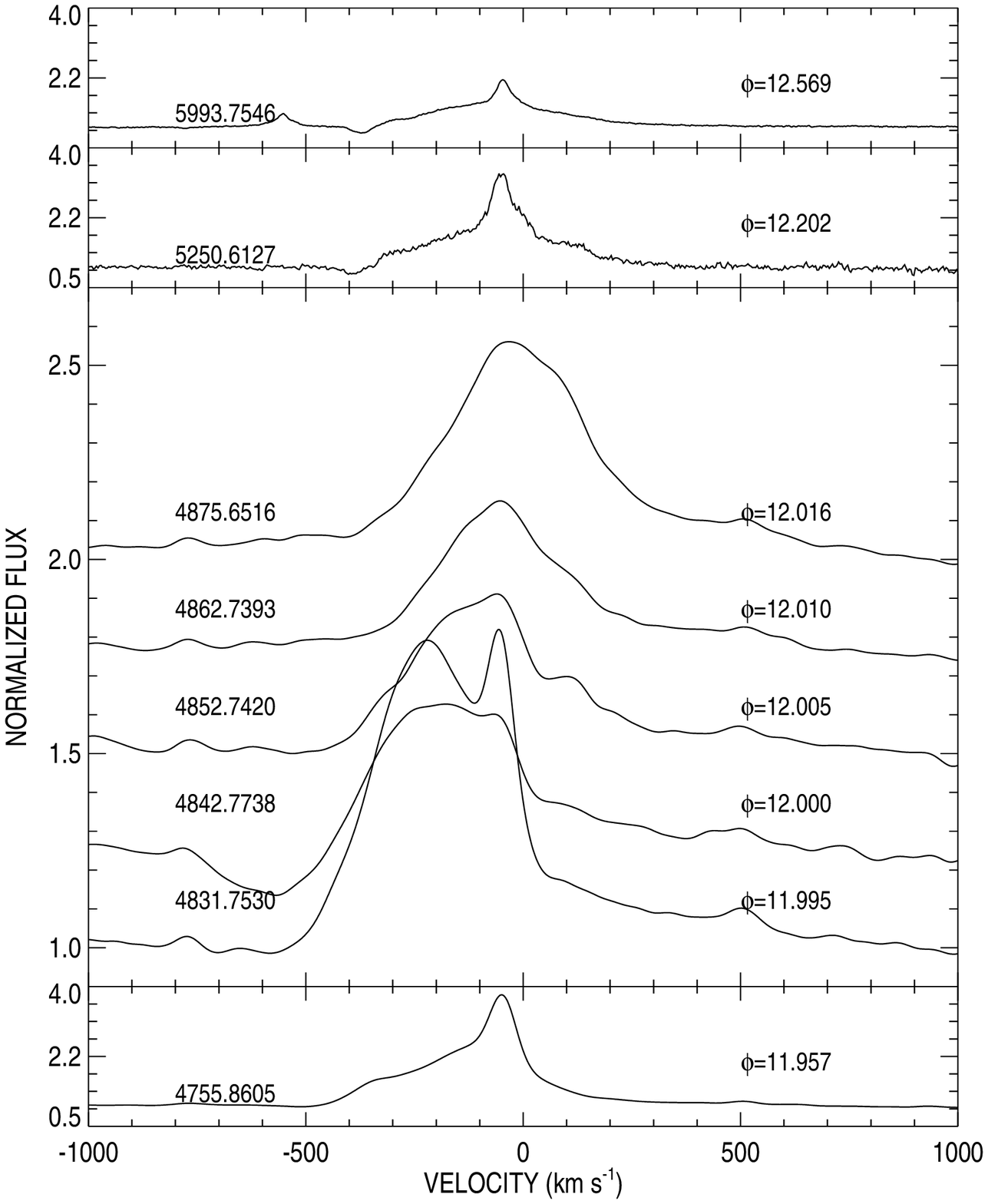}
\figsetgrpnote{Dynamical (left) and line plots (right) of \ion{He}{1} $\lambda$7065 during the 2009 event. The white horizontal bars on the dynamical representation indicate phases 11.99, 12.00, and 12.01.}
\figsetgrpend

\figsetgrpstart
\figsetgrpnum{1.8}
\figsetgrptitle{\ion{N}{2} $\lambda$5667 dynamical representation (left) and line plots (right). The white horizontal bars on the dynamical representation indicate phases 11.99, 12.00, and 12.01.}
\figsetplot{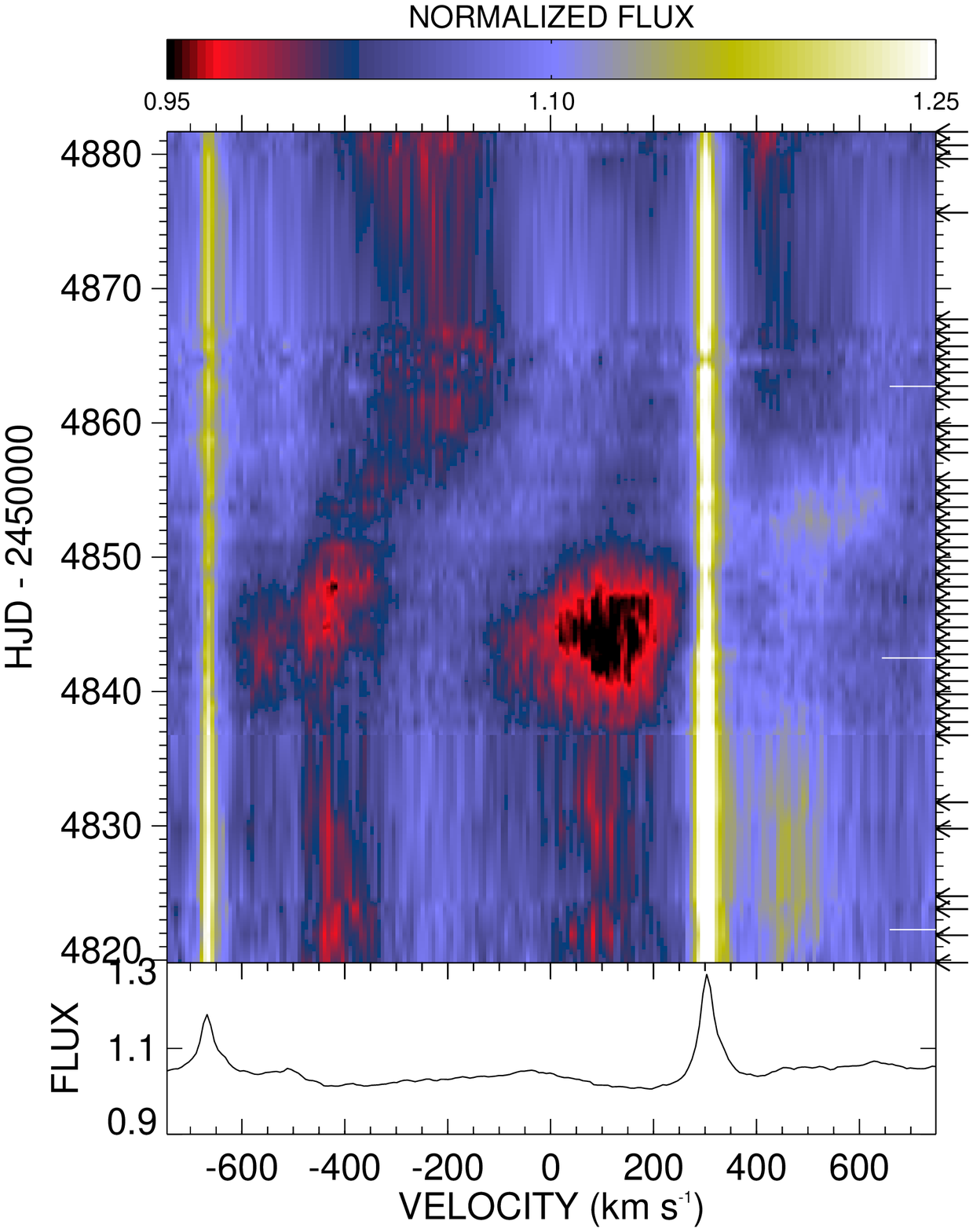}{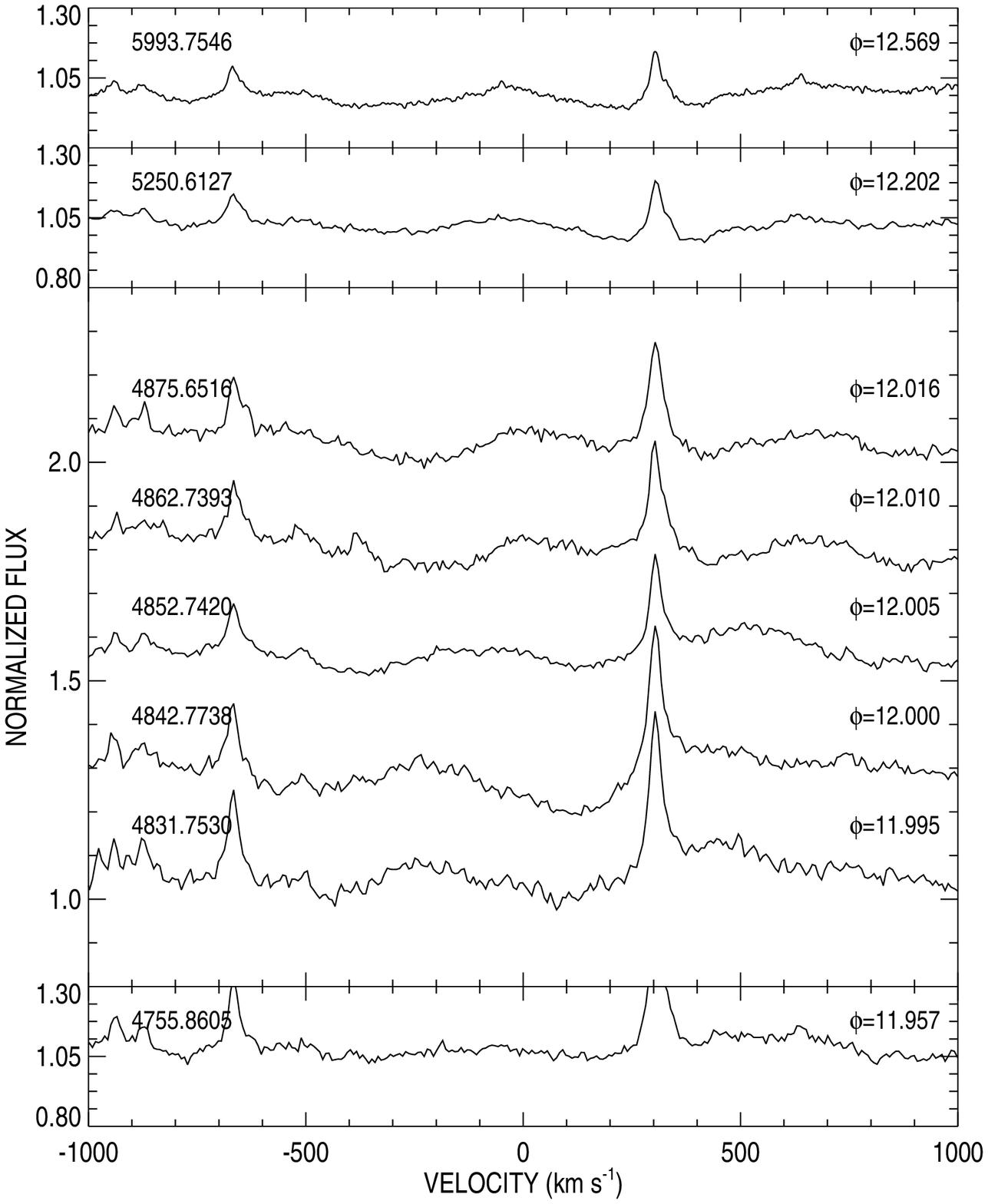}
\figsetgrpnote{Dynamical (left) and line plots (right) of \ion{N}{2}$\lambda$ 5667 during the 2009 event. The white horizontal bars on the dynamical representation indicate phases 11.99, 12.00, and 12.01. Note that \ion{N}{2} $\lambda5676 is visible on the right.}
\figsetgrpend

\figsetgrpstart
\figsetgrpnum{1.9}
\figsetgrptitle{\ion{N}{2} $\lambda$5676 dynamical representation (left) and line plots (right). The white horizontal bars on the dynamical representation indicate phases 11.99, 12.00, and 12.01.}
\figsetplot{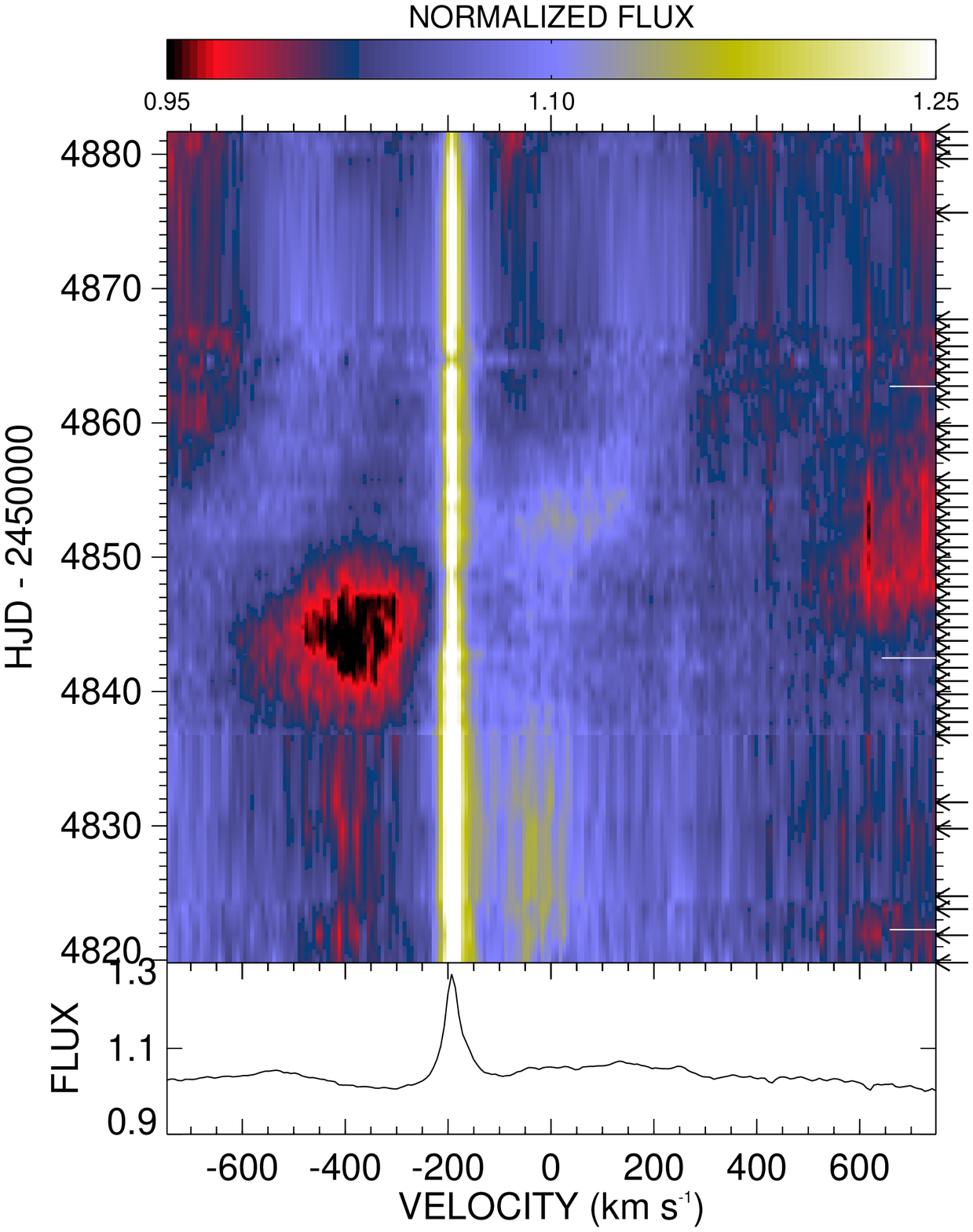}{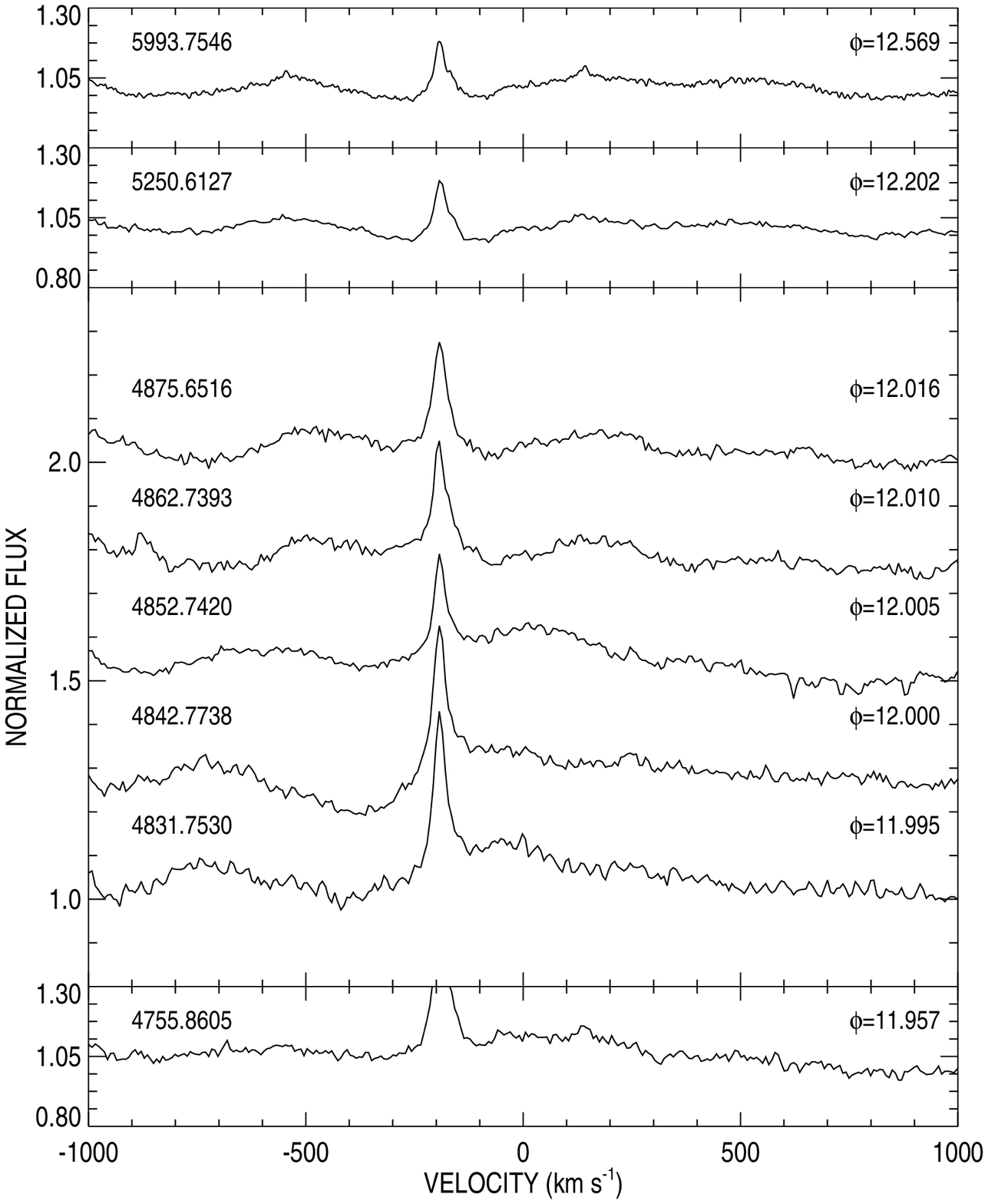}
\figsetgrpnote{Dynamical (left) and line plots (right) of \ion{N}{2}$\lambda$ 5676 during the 2009 event. The white horizontal bars on the dynamical representation indicate phases 11.99, 12.00, and 12.01.}
\figsetgrpend

\figsetgrpstart
\figsetgrpnum{1.10}
\figsetgrptitle{\ion{N}{2} $\lambda$5711 dynamical representation (left) and line plots (right). The white horizontal bars on the dynamical representation indicate phases 11.99, 12.00, and 12.01.}
\figsetplot{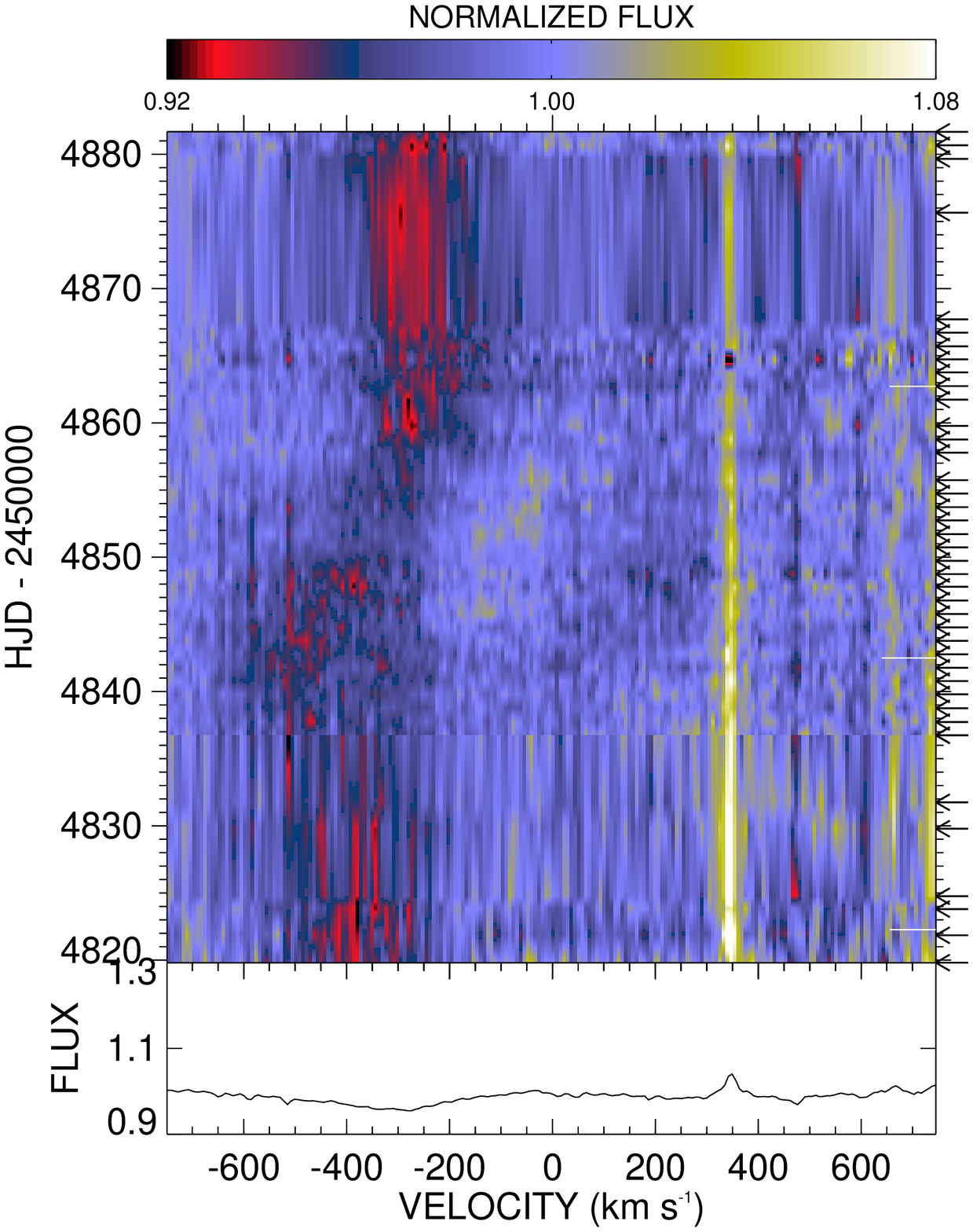}{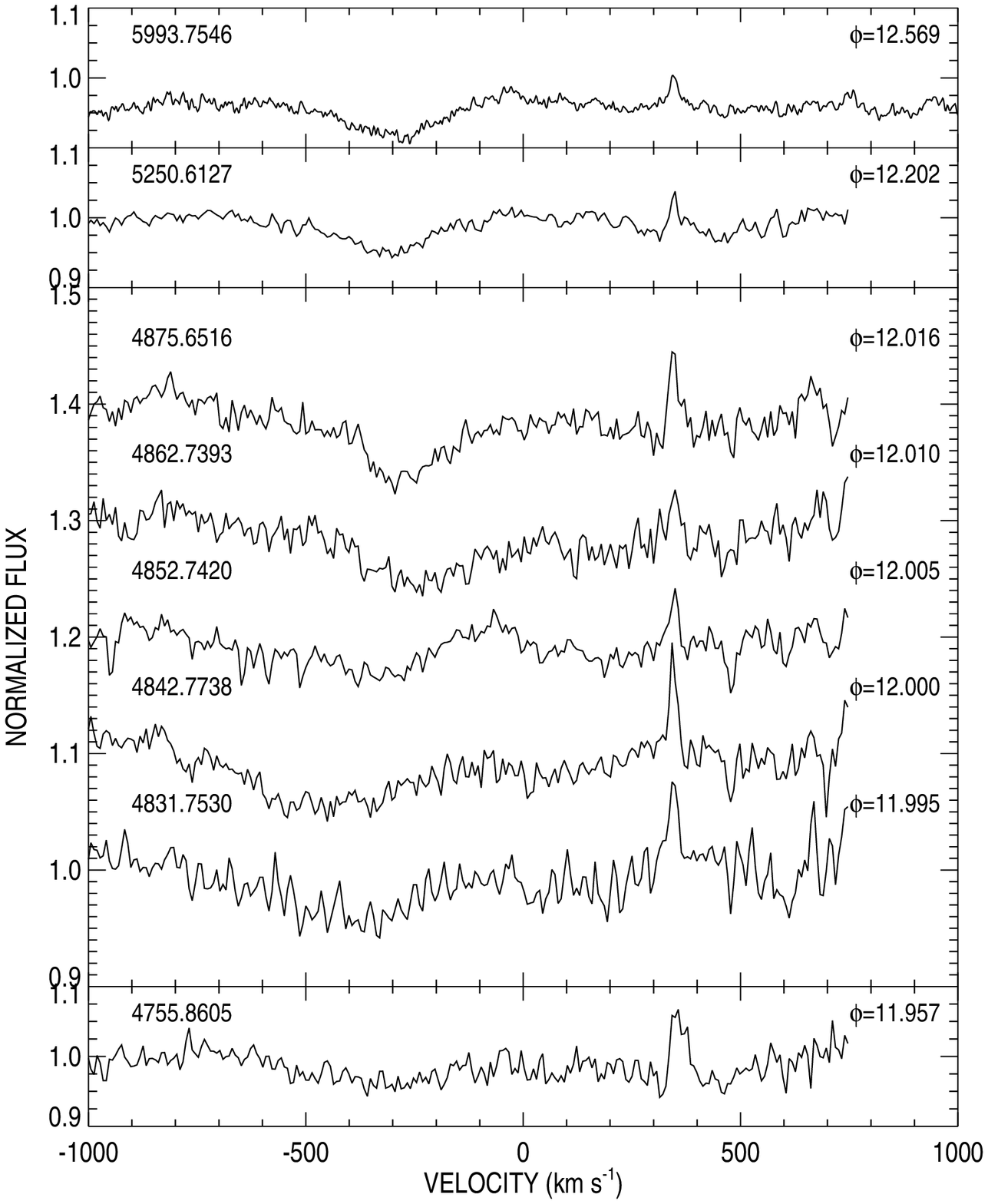}
\figsetgrpnote{Dynamical (left) and line plots (right) of \ion{N}{2} $\lambda$5711 during the 2009 event. The white horizontal bars on the dynamical representation indicate phases 11.99, 12.00, and 12.01.}
\figsetgrpend

\figsetgrpstart
\figsetgrpnum{1.11}
\figsetgrptitle{\ion{Na}{1} D$_2 \lambda 5890$ dynamical representation (left) and line plots (right). The white horizontal bars on the dynamical representation indicate phases 11.99, 12.00, and 12.01.}
\figsetplot{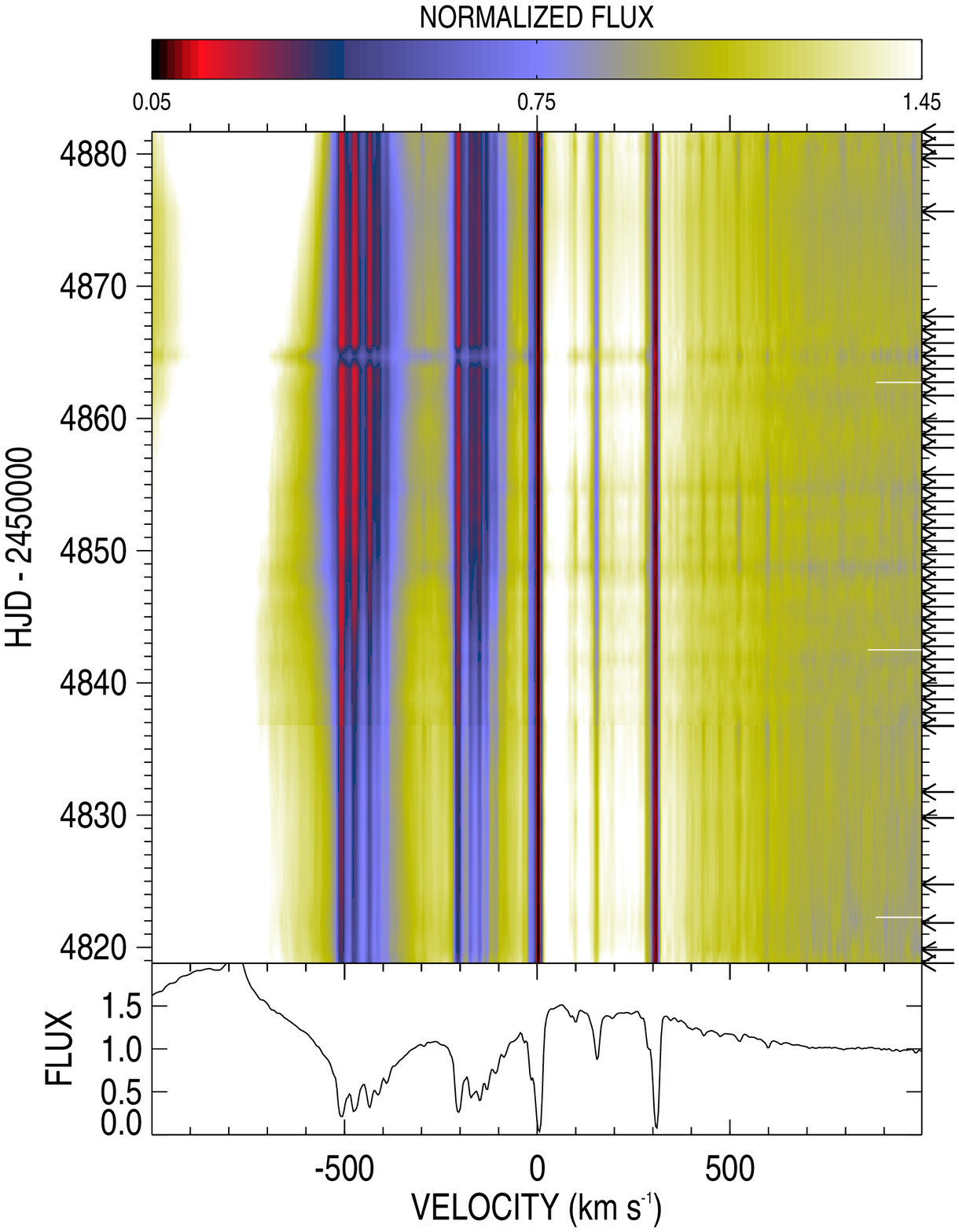}{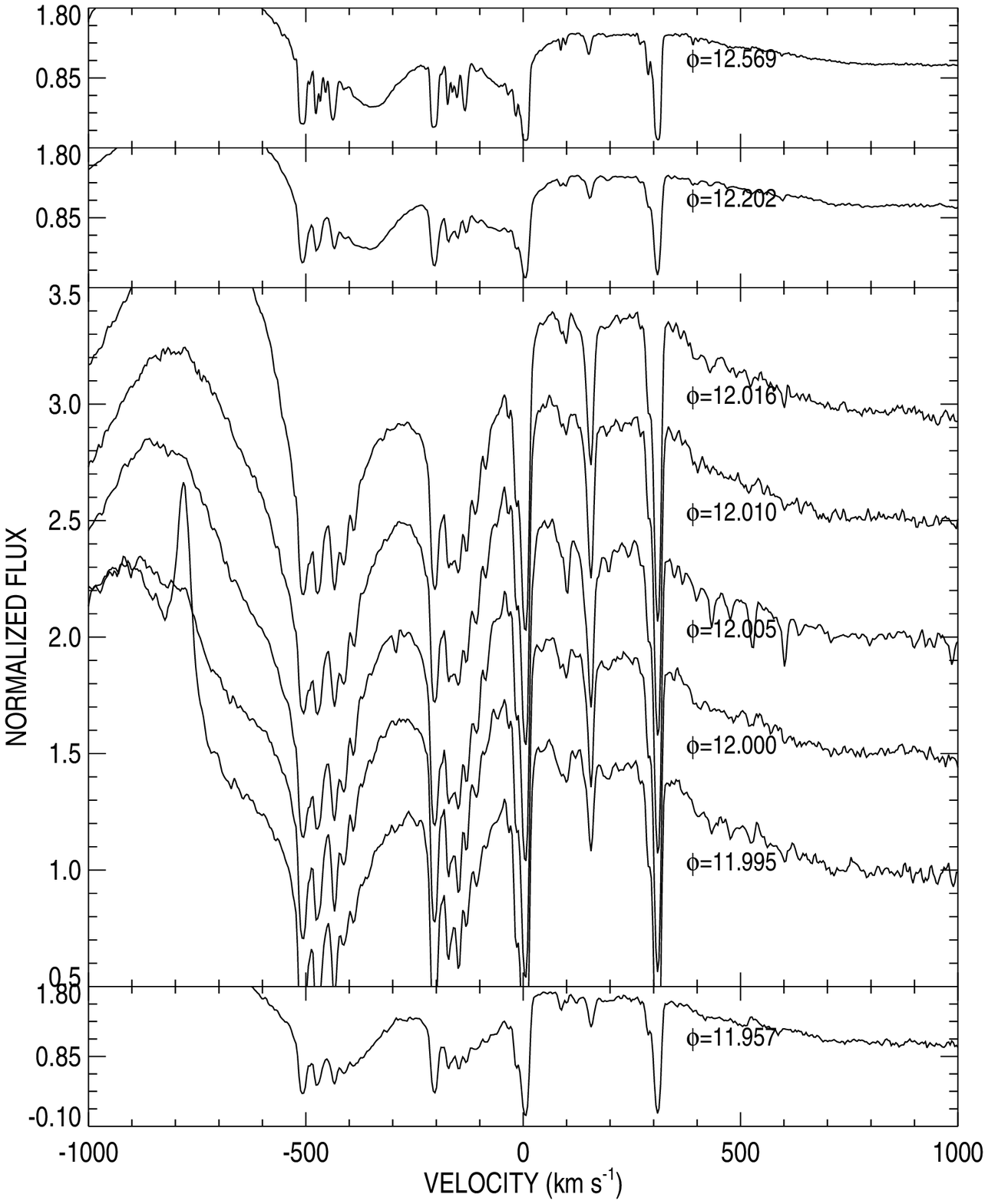}
\figsetgrpnote{Dynamical (left) and line plots (right) of \ion{Na}{1} D$_2$ during the 2009 event. The white horizontal bars on the dynamical representation indicate phases 11.99, 12.00, and 12.01.}
\figsetgrpend

\figsetgrpstart
\figsetgrpnum{1.12}
\figsetgrptitle{\ion{Na}{1} D$_1 \lambda 5896$ dynamical representation (left) and line plots (right). The white horizontal bars on the dynamical representation indicate phases 11.99, 12.00, and 12.01.}
\figsetplot{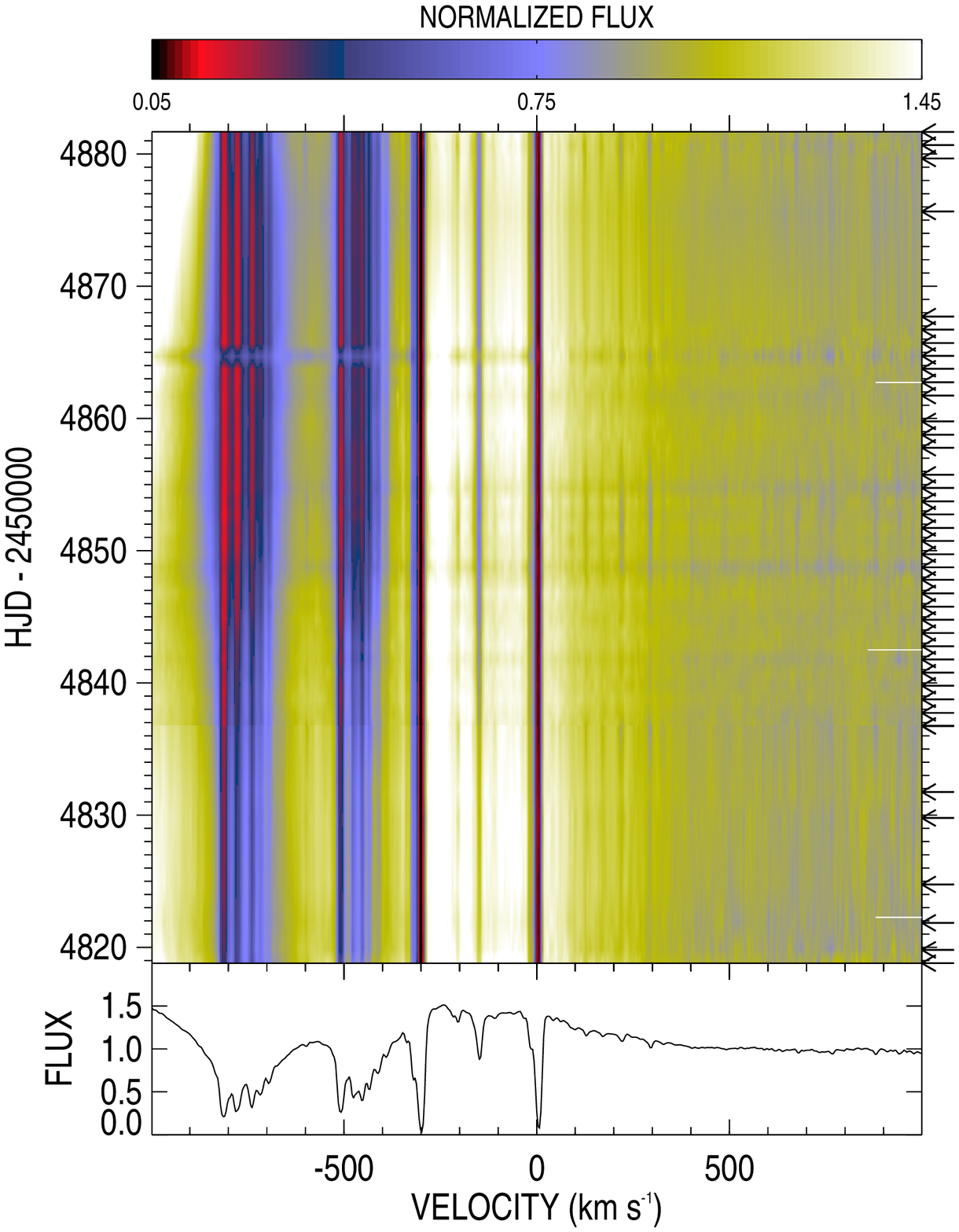}{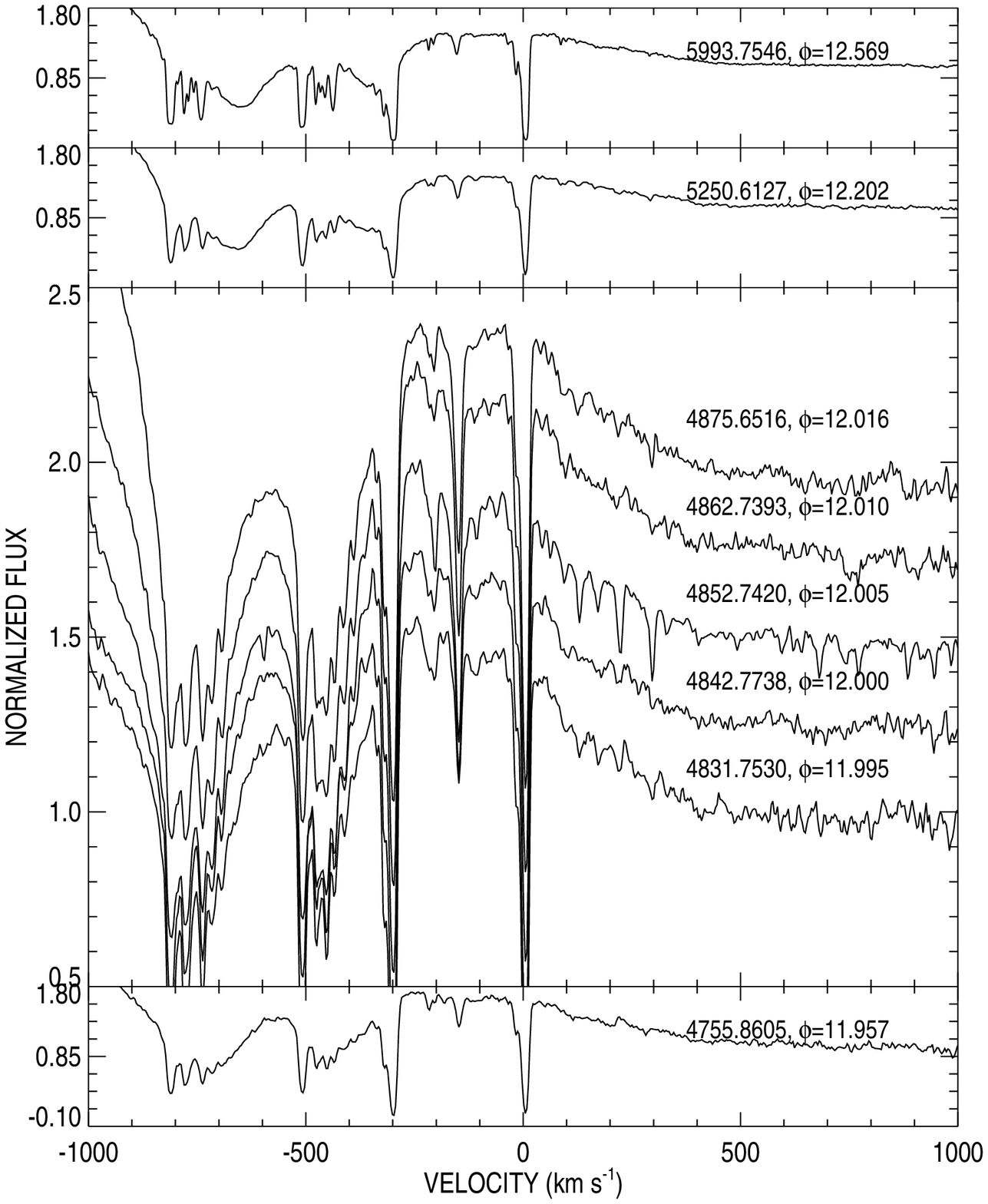}
\figsetgrpnote{Dynamical (left) and line plots (right) of \ion{Na}{1} D$_1$ during the 2009 event. The white horizontal bars on the dynamical representation indicate phases 11.99, 12.00, and 12.01.}
\figsetgrpend

\figsetgrpstart
\figsetgrpnum{1.13}
\figsetgrptitle{\ion{Si}{2} $\lambda$6347 dynamical representation (left) and line plots (right). The white horizontal bars on the dynamical representation indicate phases 11.99, 12.00, and 12.01.}
\figsetplot{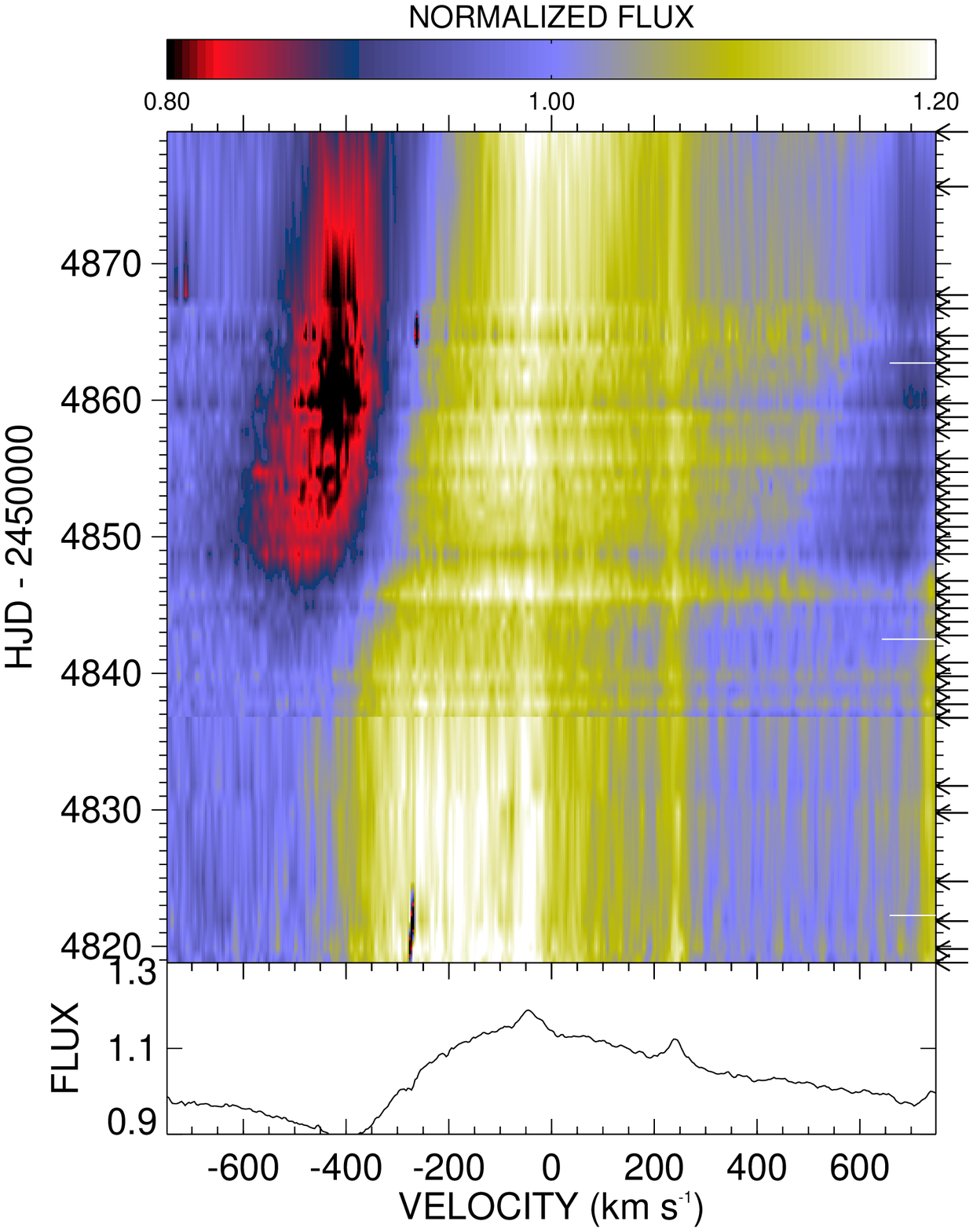}{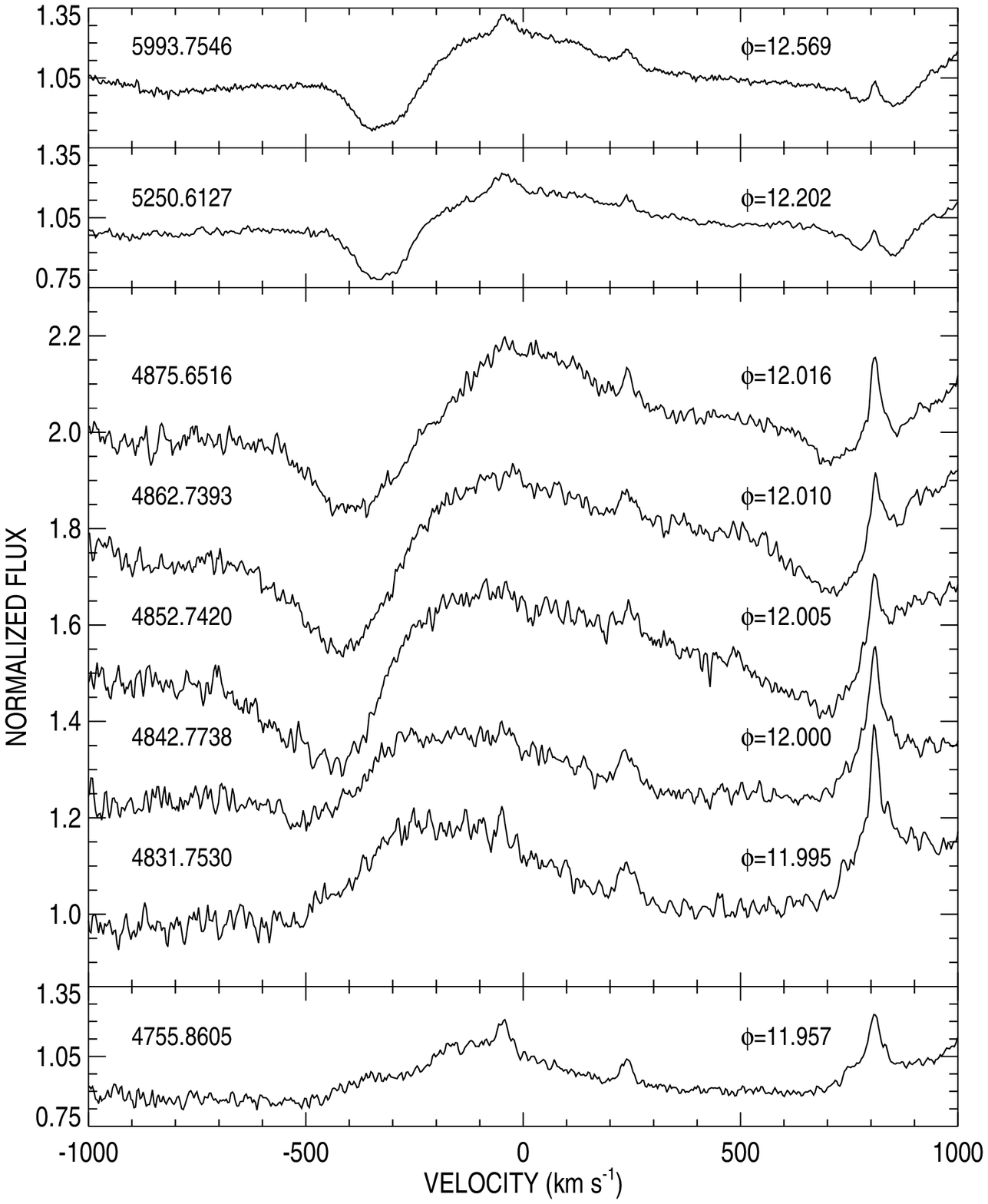}
\figsetgrpnote{Dynamical (left) and line plots (right) of \ion{Si}{2} $\lambda$6347 during the 2009 event. The white horizontal bars on the dynamical representation indicate phases 11.99, 12.00, and 12.01.}
\figsetgrpend

\figsetgrpstart
\figsetgrpnum{1.14}
\figsetgrptitle{\ion{Si}{2} $\lambda$6371 dynamical representation (left) and line plots (right). The white horizontal bars on the dynamical representation indicate phases 11.99, 12.00, and 12.01.}
\figsetplot{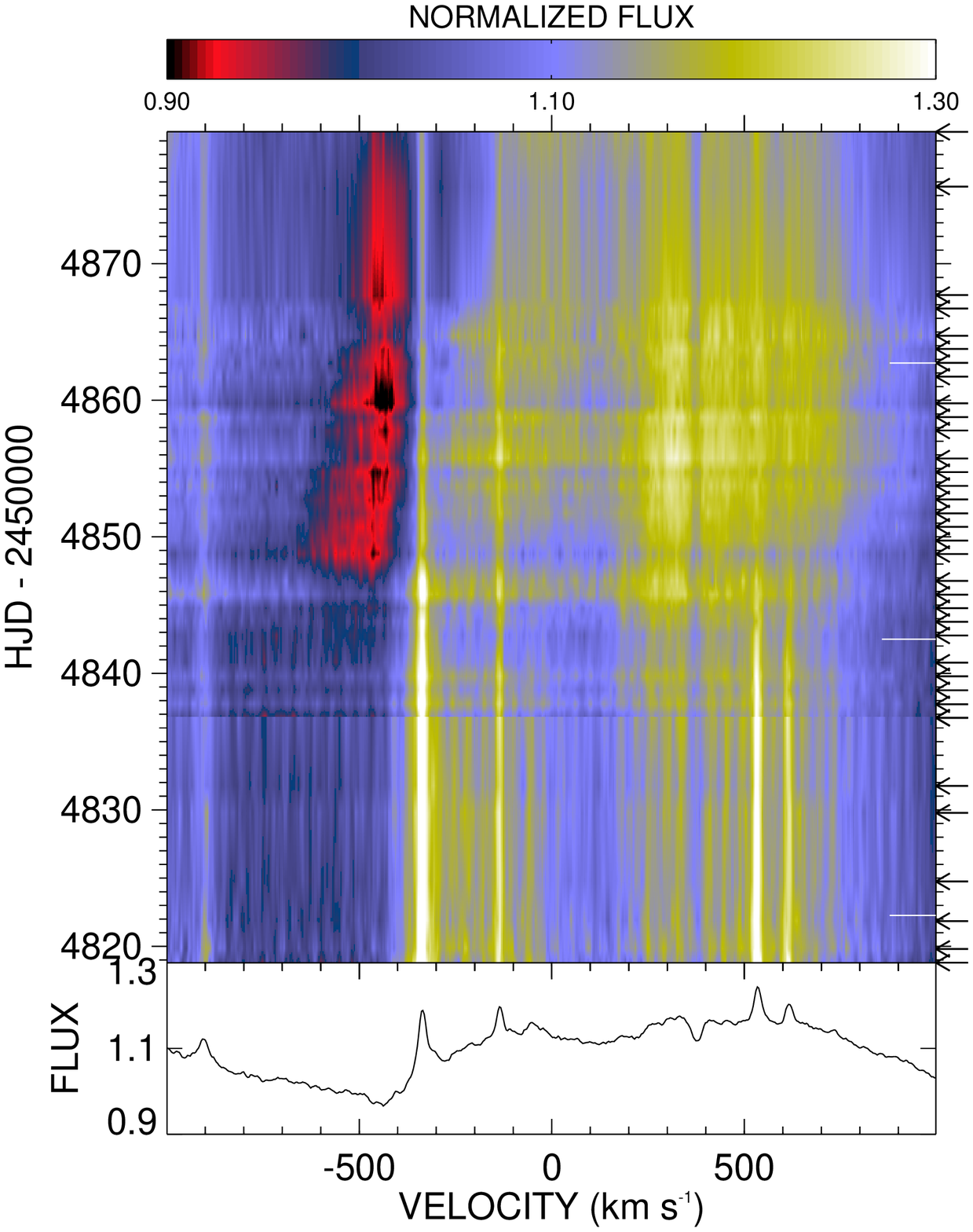}{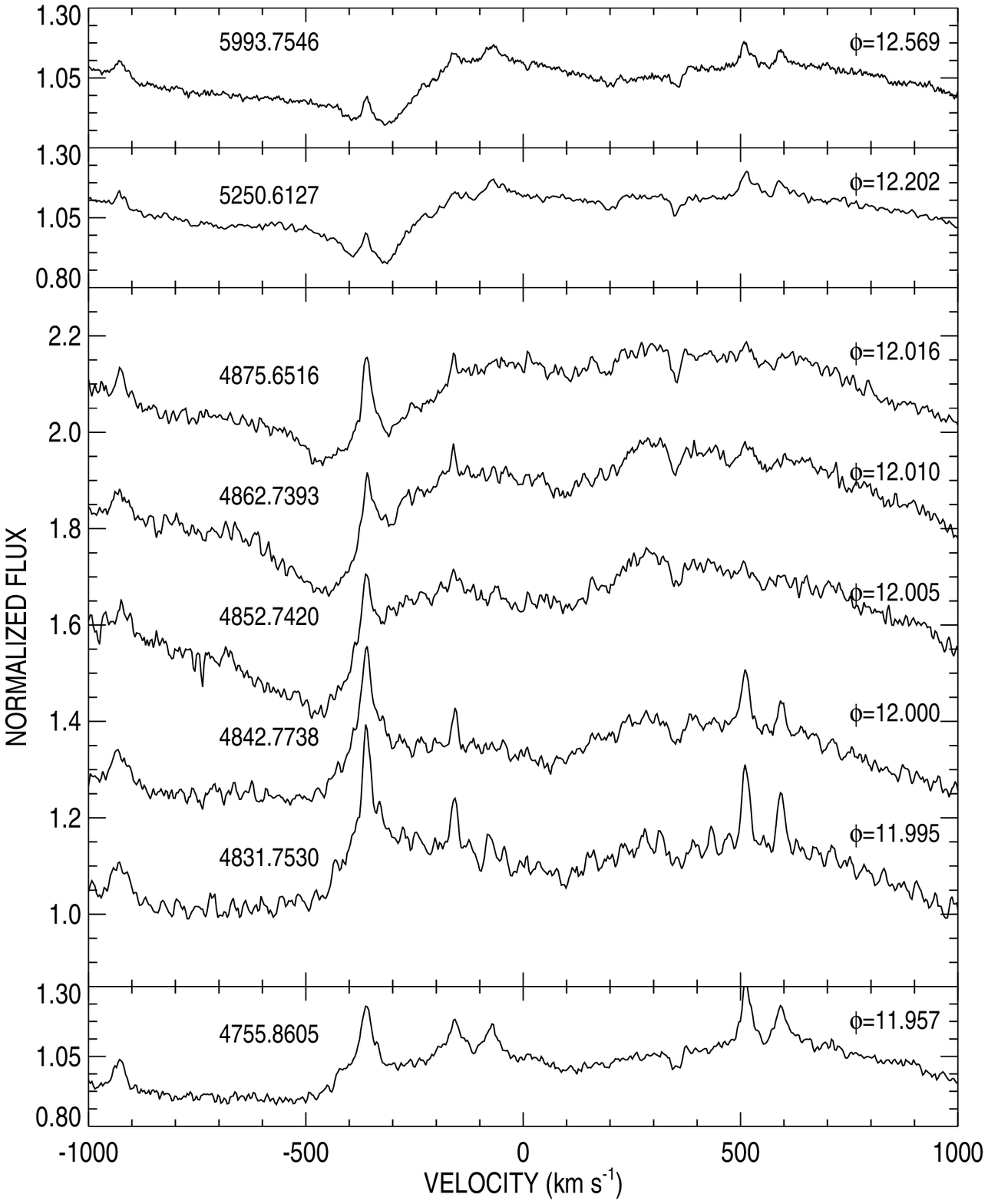}
\figsetgrpnote{Dynamical (left) and line plots (right) of \ion{Si}{2} $\lambda$6371 during the 2009 event. The white horizontal bars on the dynamical representation indicate phases 11.99, 12.00, and 12.01.}
\figsetgrpend

\figsetgrpstart
\figsetgrpnum{1.15}
\figsetgrptitle{\ion{Fe}{2} $\lambda$ 5169 dynamical representation (left) and line plots (right). The white horizontal bars on the dynamical representation indicate phases 11.99, 12.00, and 12.01.}
\figsetplot{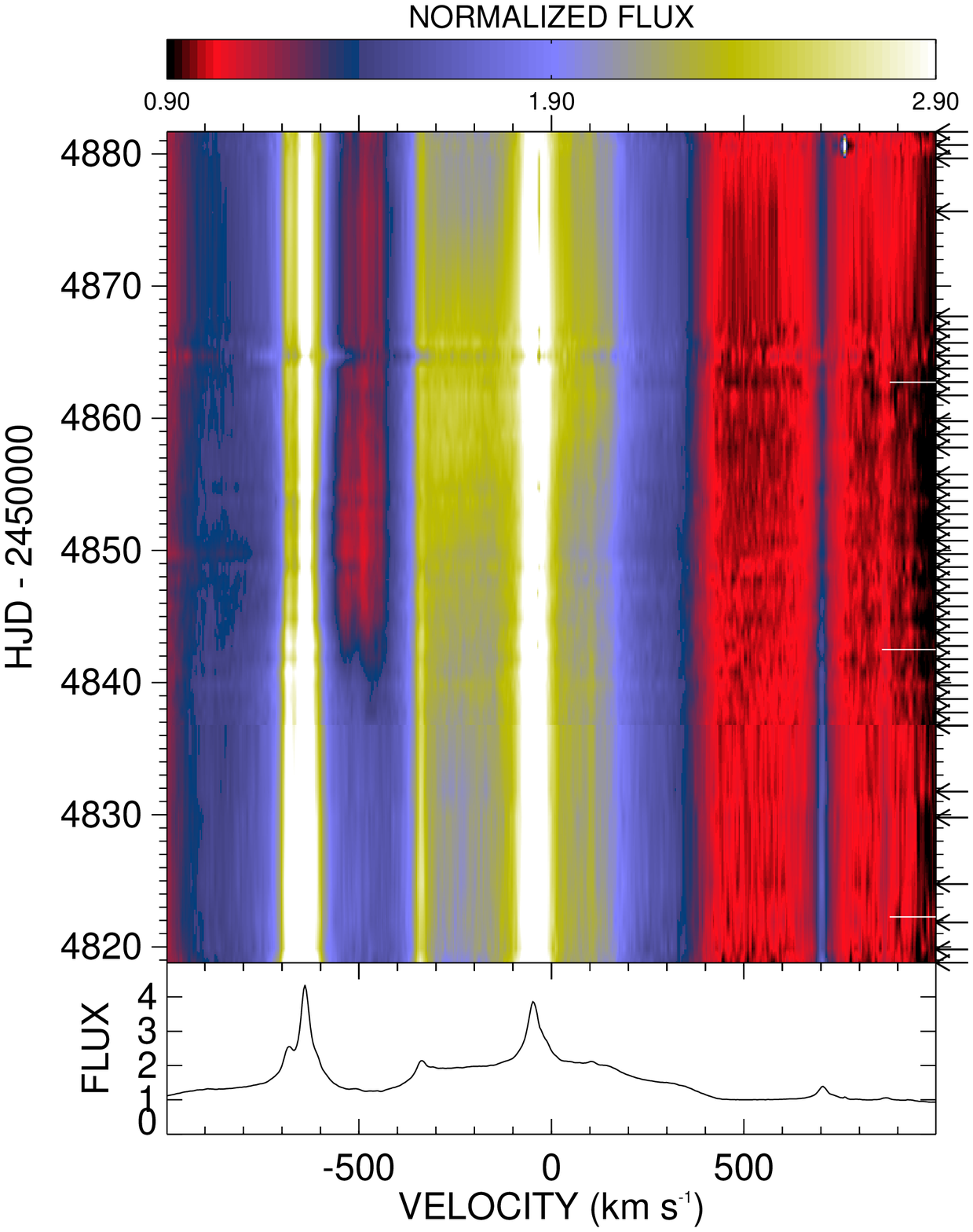}{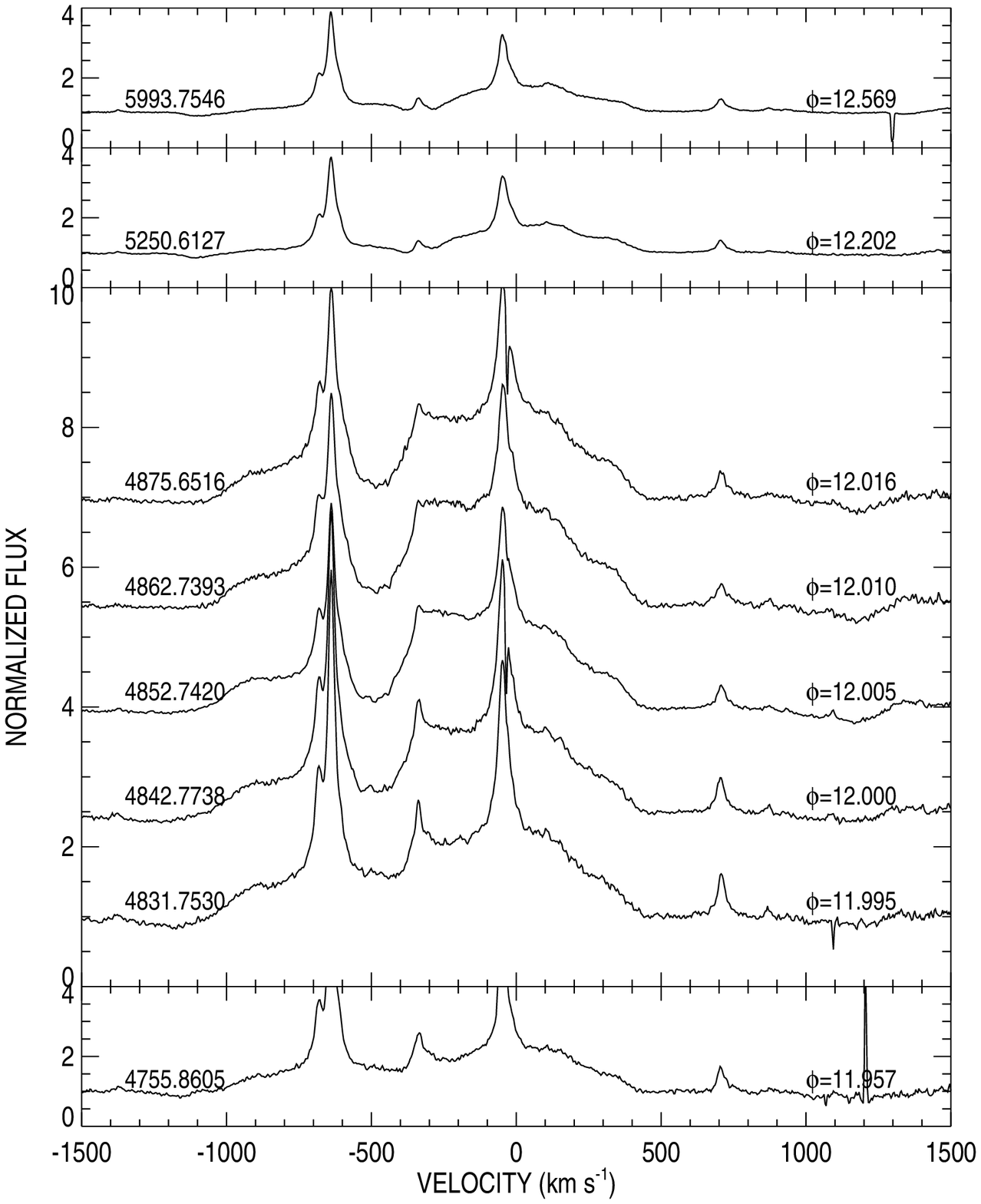}
\figsetgrpnote{Dynamical (left) and line plots (right) of \ion{Fe}{2} $\lambda$ 5169 during the 2009 event. The white horizontal bars on the dynamical representation indicate phases 11.99, 12.00, and 12.01.}
\figsetgrpend

\figsetgrpstart
\figsetgrpnum{1.16}
\figsetgrptitle{\ion{Fe}{2} $\lambda$5197 dynamical representation (left) and line plots (right). The white horizontal bars on the dynamical representation indicate phases 11.99, 12.00, and 12.01.}
\figsetplot{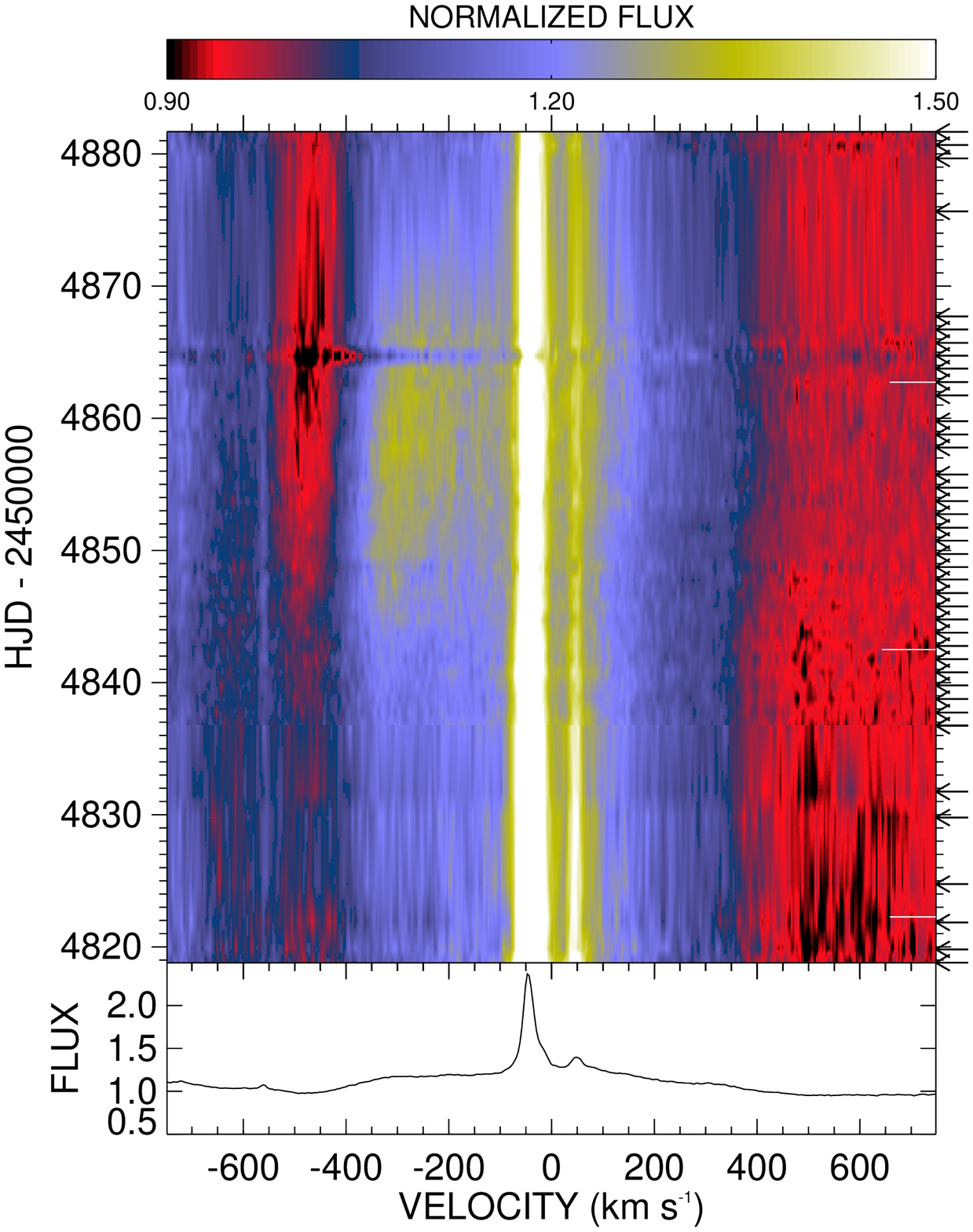}{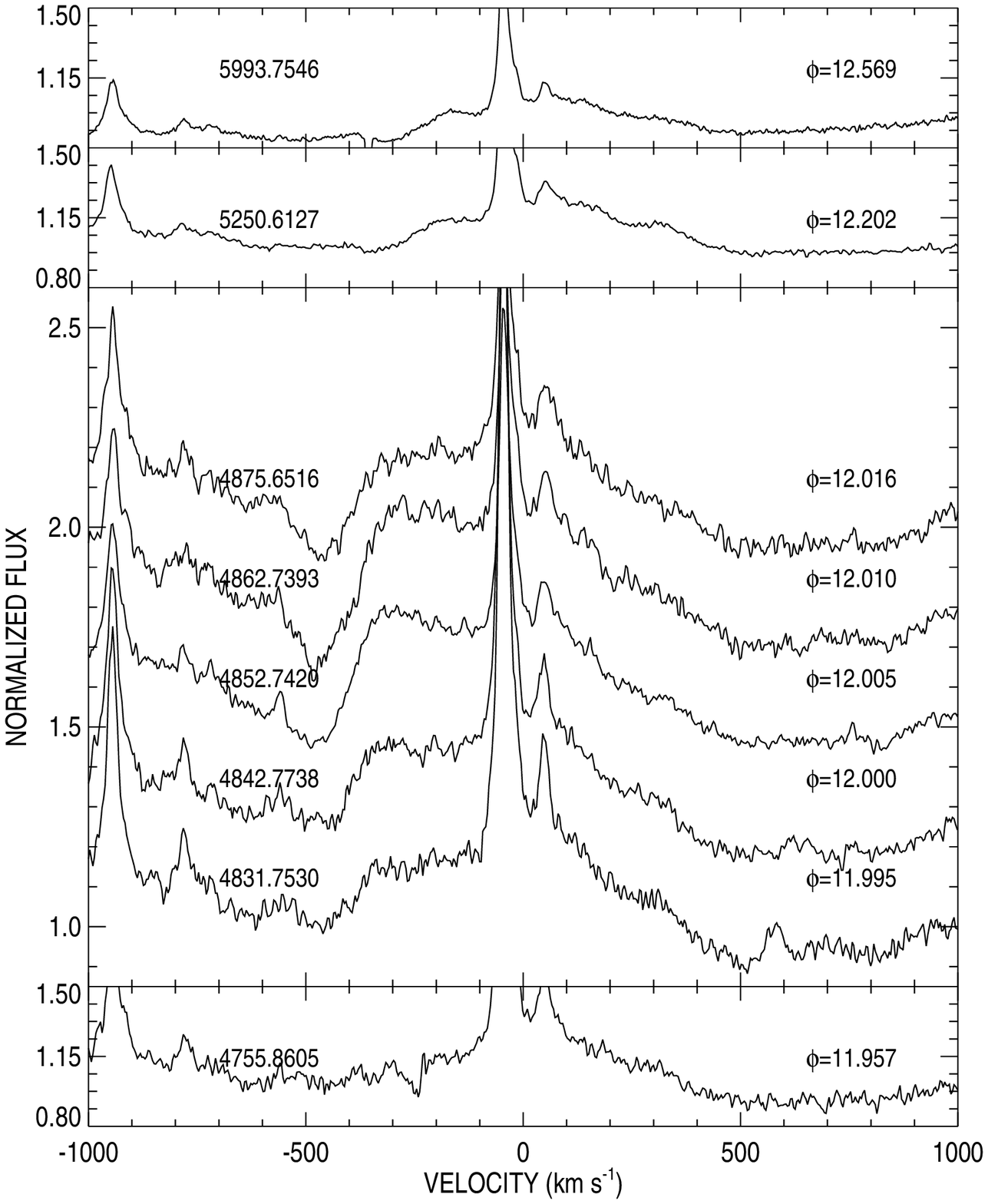}
\figsetgrpnote{Dynamical (left) and line plots (right) of \ion{Fe}{2} $\lambda$5197 during the 2009 event. The white horizontal bars on the dynamical representation indicate phases 11.99, 12.00, and 12.01.}
\figsetgrpend

\figsetgrpstart
\figsetgrpnum{1.17}
\figsetgrptitle{\ion{Fe}{2} $\lambda$5235 dynamical representation (left) and line plots (right). The white horizontal bars on the dynamical representation indicate phases 11.99, 12.00, and 12.01.}
\figsetplot{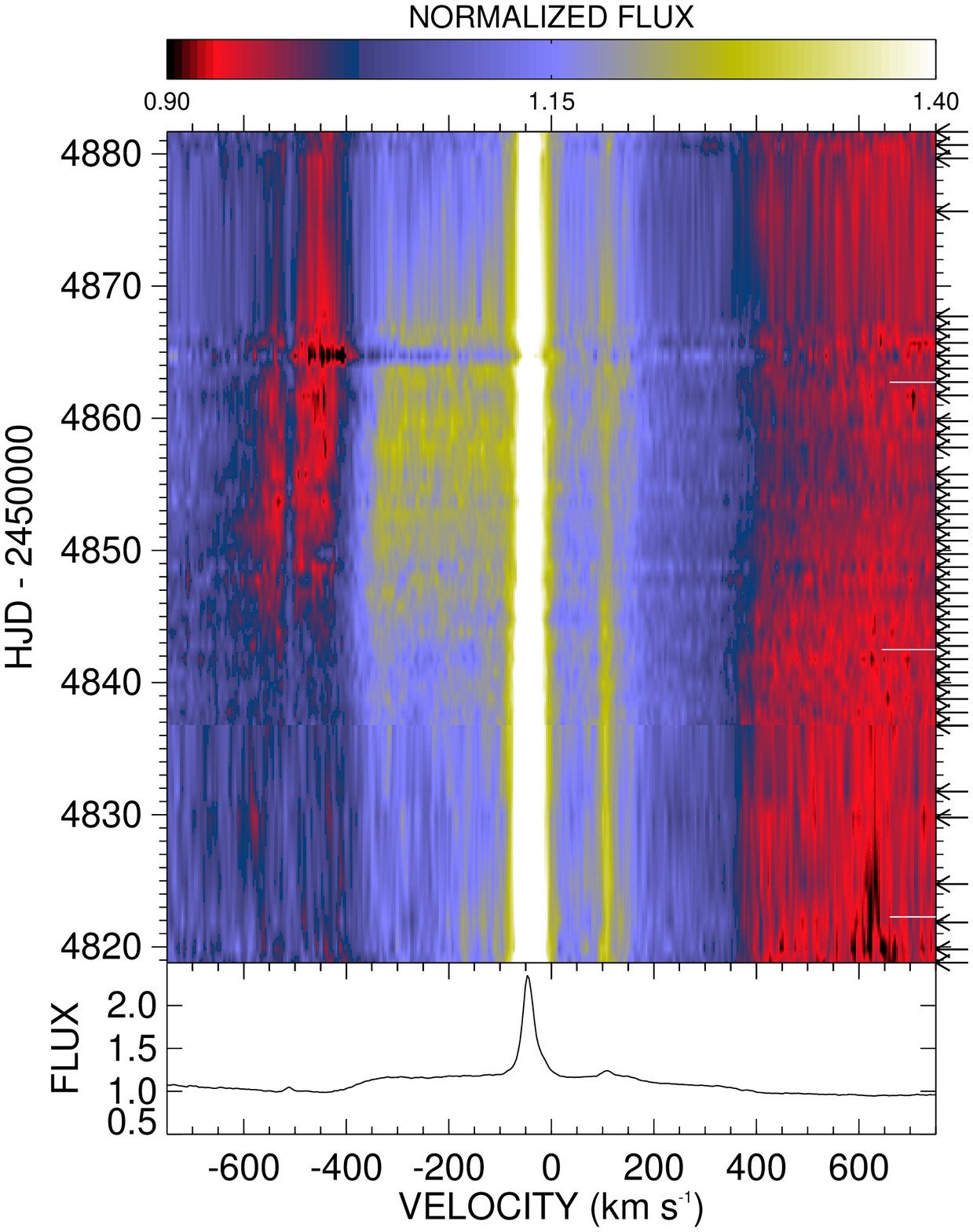}{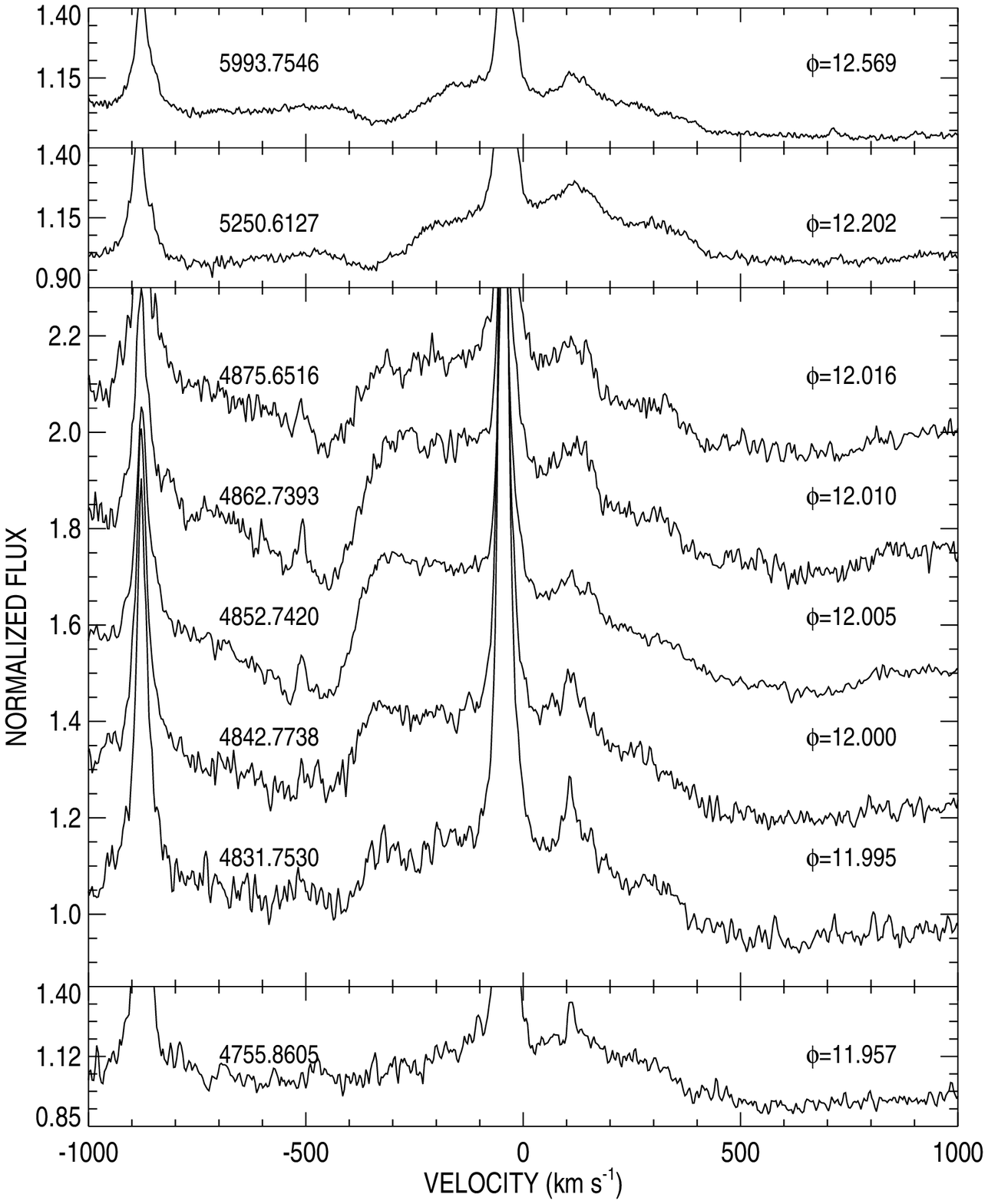}
\figsetgrpnote{Dynamical (left) and line plots (right) of \ion{Fe}{2} $\lambda$5235 during the 2009 event. The white horizontal bars on the dynamical representation indicate phases 11.99, 12.00, and 12.01.}
\figsetgrpend

\figsetgrpstart
\figsetgrpnum{1.18}
\figsetgrptitle{\ion{Fe}{2} $\lambda$5316 dynamical representation (left) and line plots (right). The white horizontal bars on the dynamical representation indicate phases 11.99, 12.00, and 12.01.}
\figsetplot{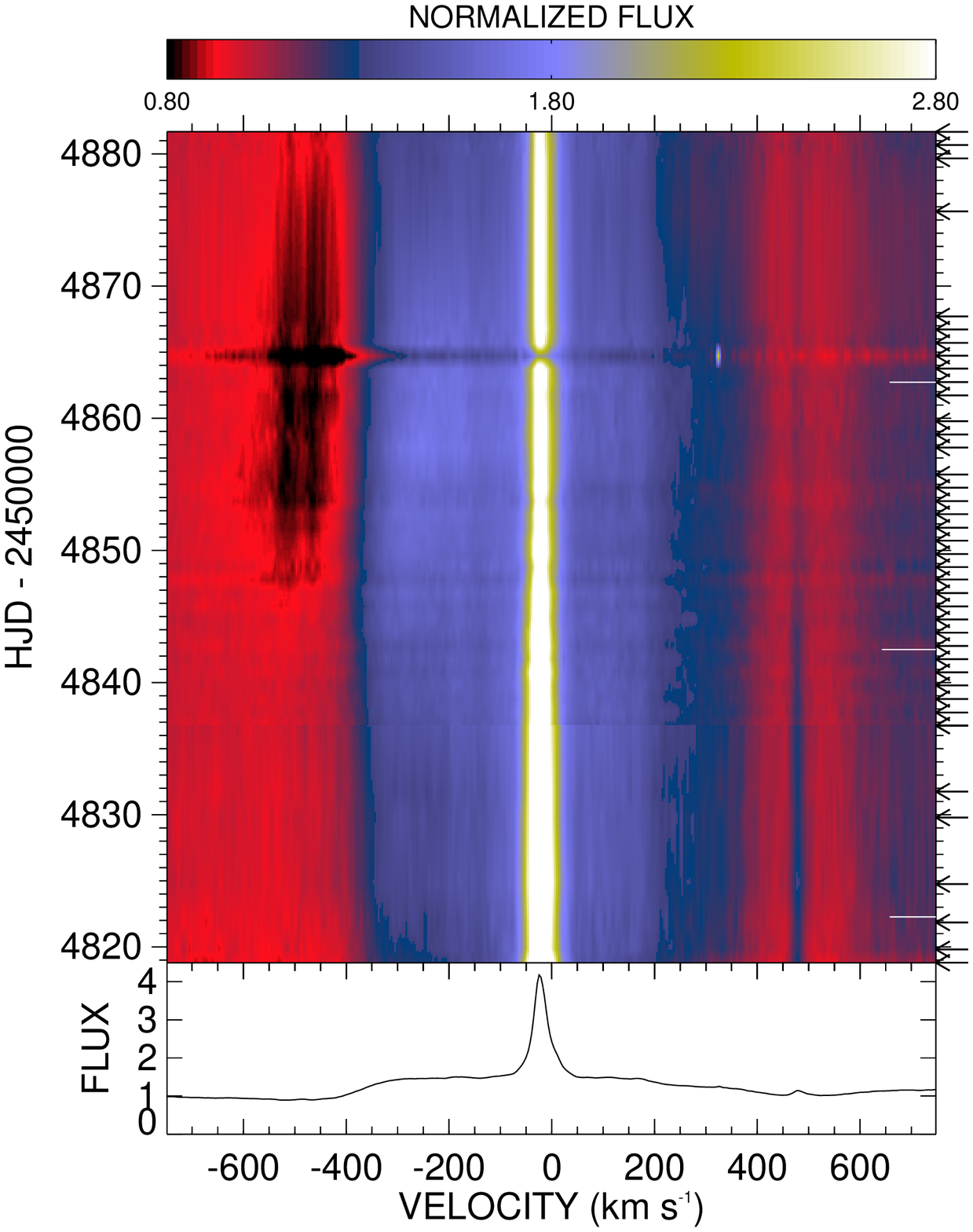}{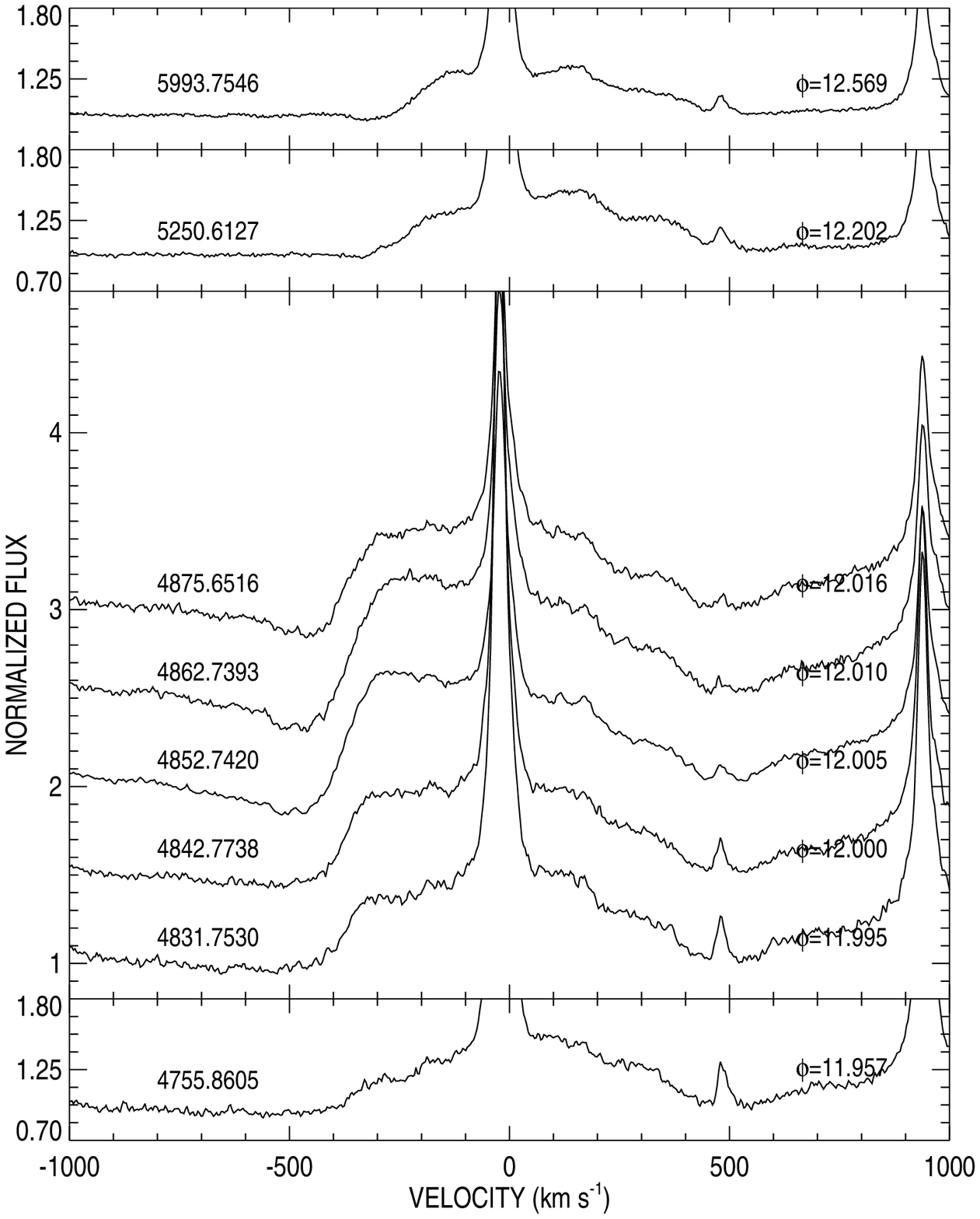}
\figsetgrpnote{Dynamical (left) and line plots (right) of \ion{Fe}{2} $\lambda$5316 during the 2009 event. The white horizontal bars on the dynamical representation indicate phases 11.99, 12.00, and 12.01.}
\figsetgrpend

\figsetgrpstart
\figsetgrpnum{1.19}
\figsetgrptitle{\ion{Fe}{2} $\lambda$6238 dynamical representation (left) and line plots (right). The white horizontal bars on the dynamical representation indicate phases 11.99, 12.00, and 12.01.}
\figsetplot{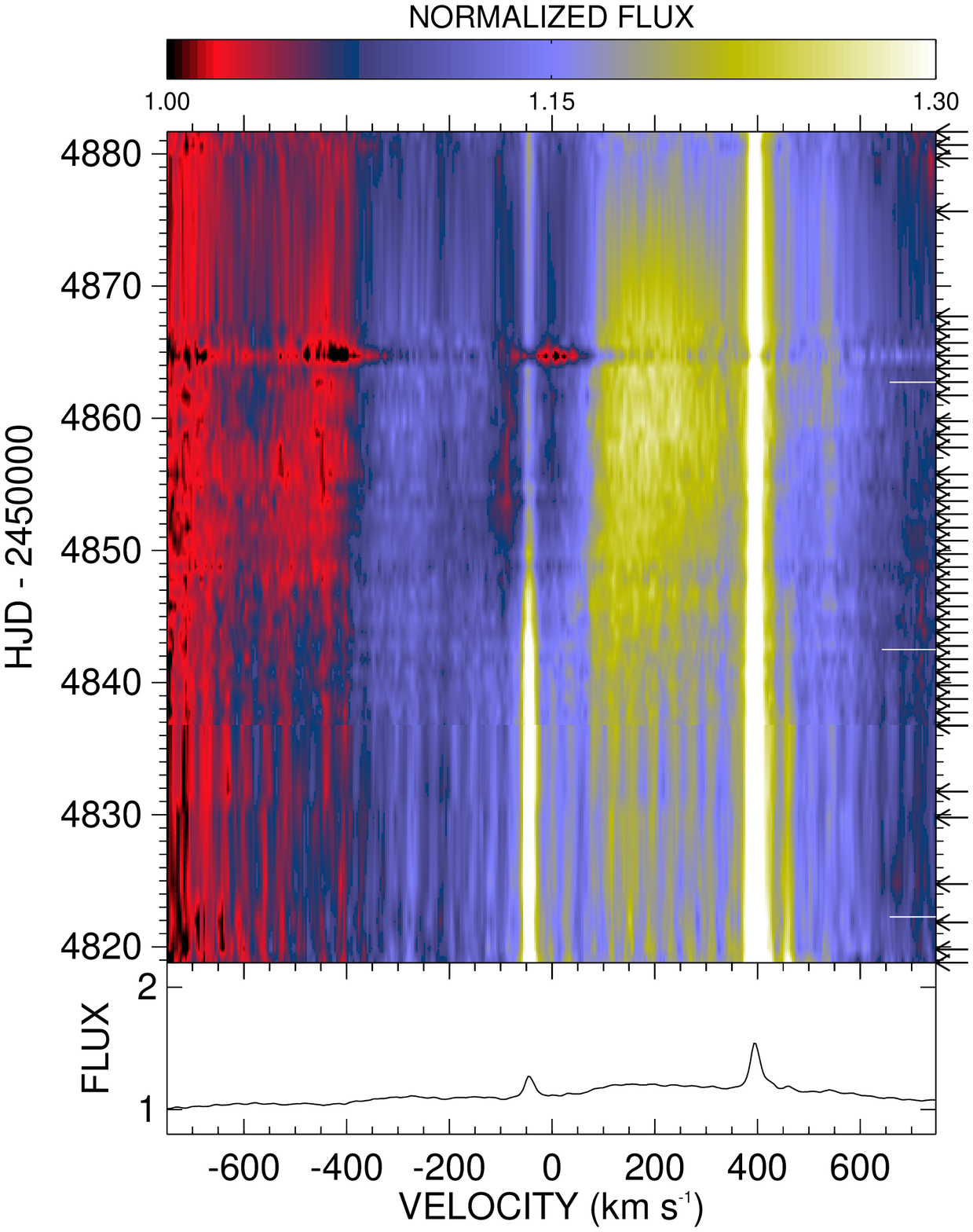}{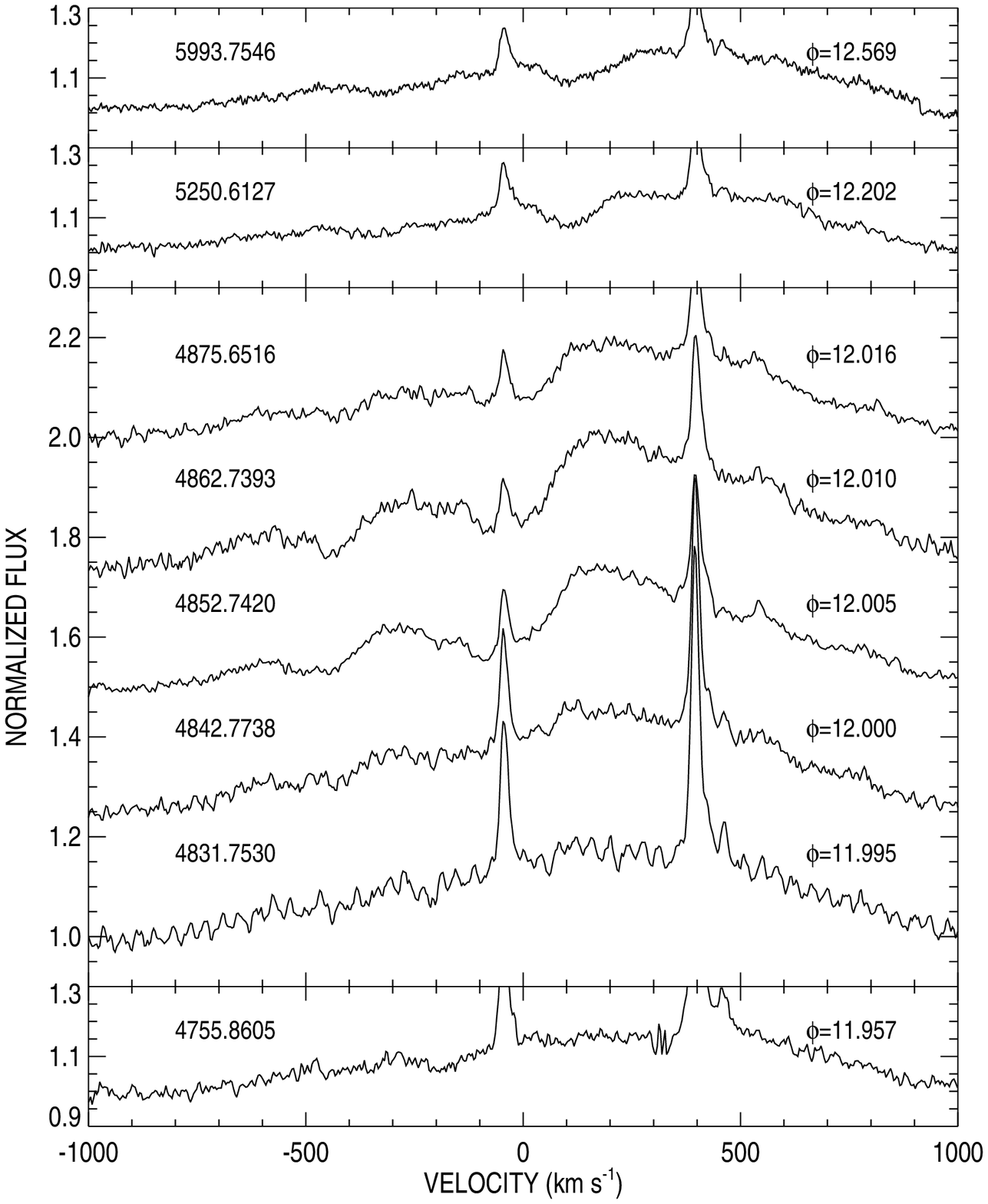}
\figsetgrpnote{Dynamical (left) and line plots (right) of \ion{Fe}{2} $\lambda$6238 during the 2009 event. The white horizontal bars on the dynamical representation indicate phases 11.99, 12.00, and 12.01.}
\figsetgrpend

\figsetgrpstart
\figsetgrpnum{1.20}
\figsetgrptitle{\ion{Fe}{2} $\lambda$6248 dynamical representation (left) and line plots (right). The white horizontal bars on the dynamical representation indicate phases 11.99, 12.00, and 12.01.}
\figsetplot{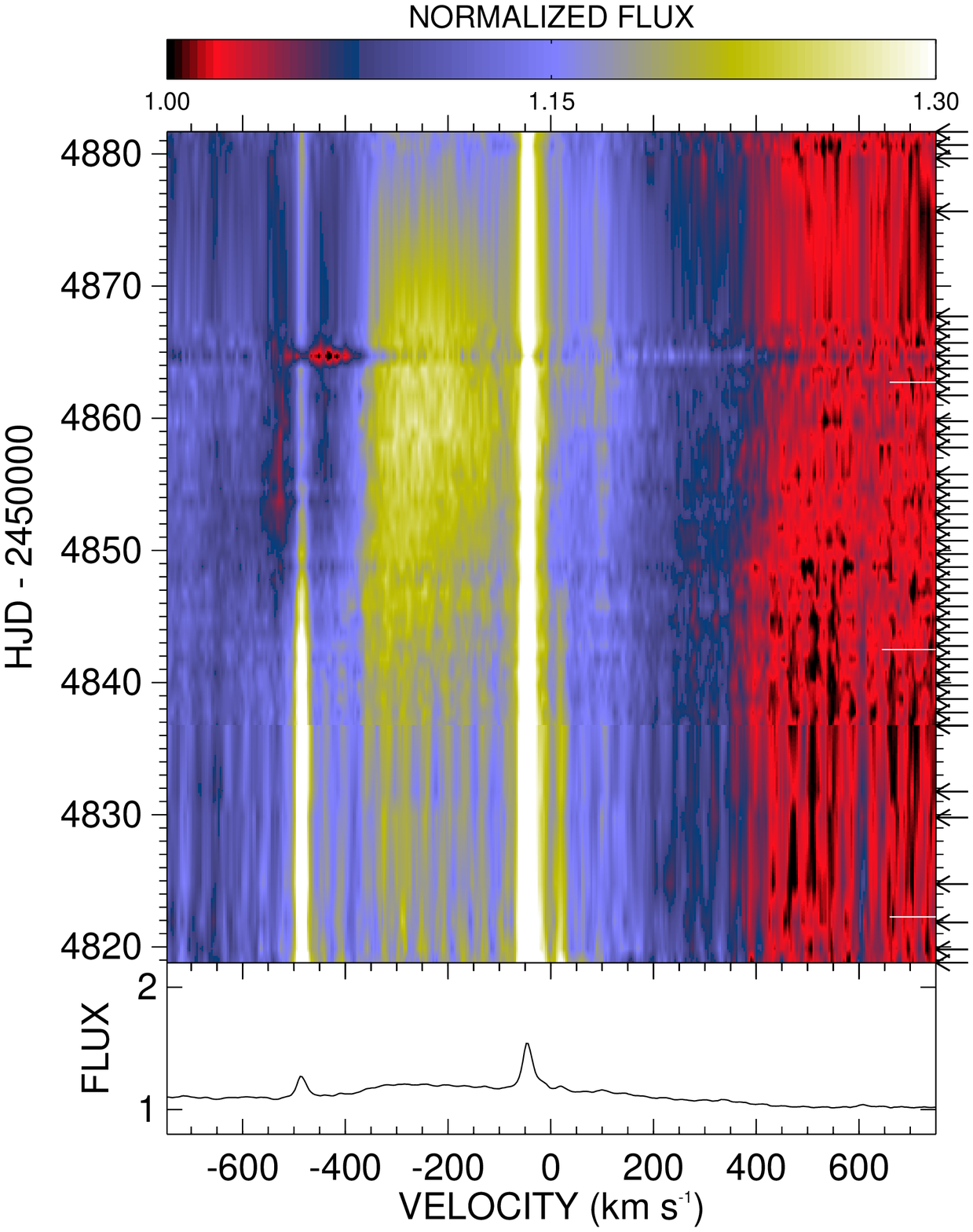}{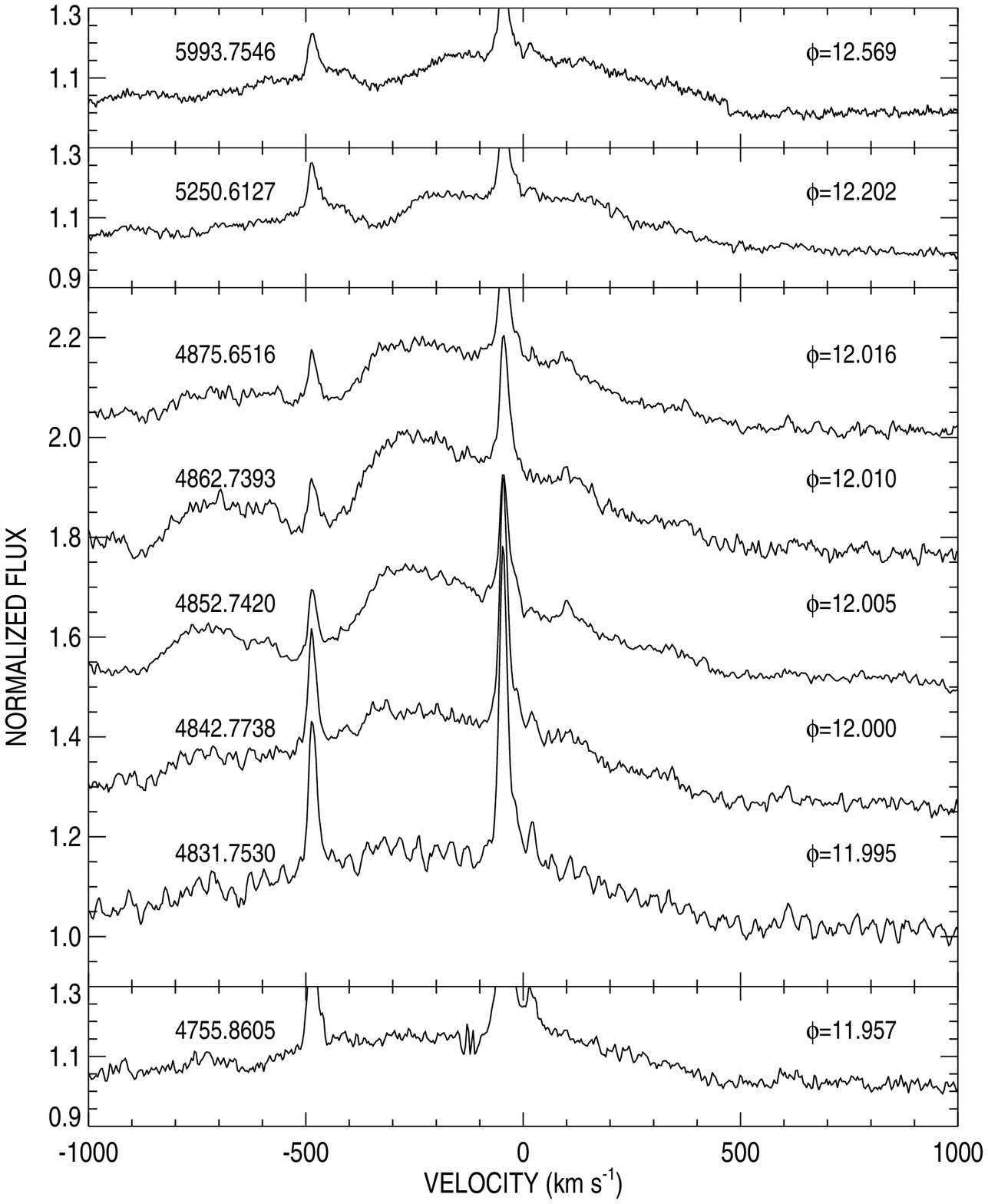}
\figsetgrpnote{Dynamical (left) and line plots (right) of \ion{Fe}{2} $\lambda$6248 during the 2009 event. The white horizontal bars on the dynamical representation indicate phases 11.99, 12.00, and 12.01.}
\figsetgrpend

\figsetgrpstart
\figsetgrpnum{1.21}
\figsetgrptitle{\ion{Fe}{2} $\lambda$6456 dynamical representation (left) and line plots (right). The white horizontal bars on the dynamical representation indicate phases 11.99, 12.00, and 12.01.}
\figsetplot{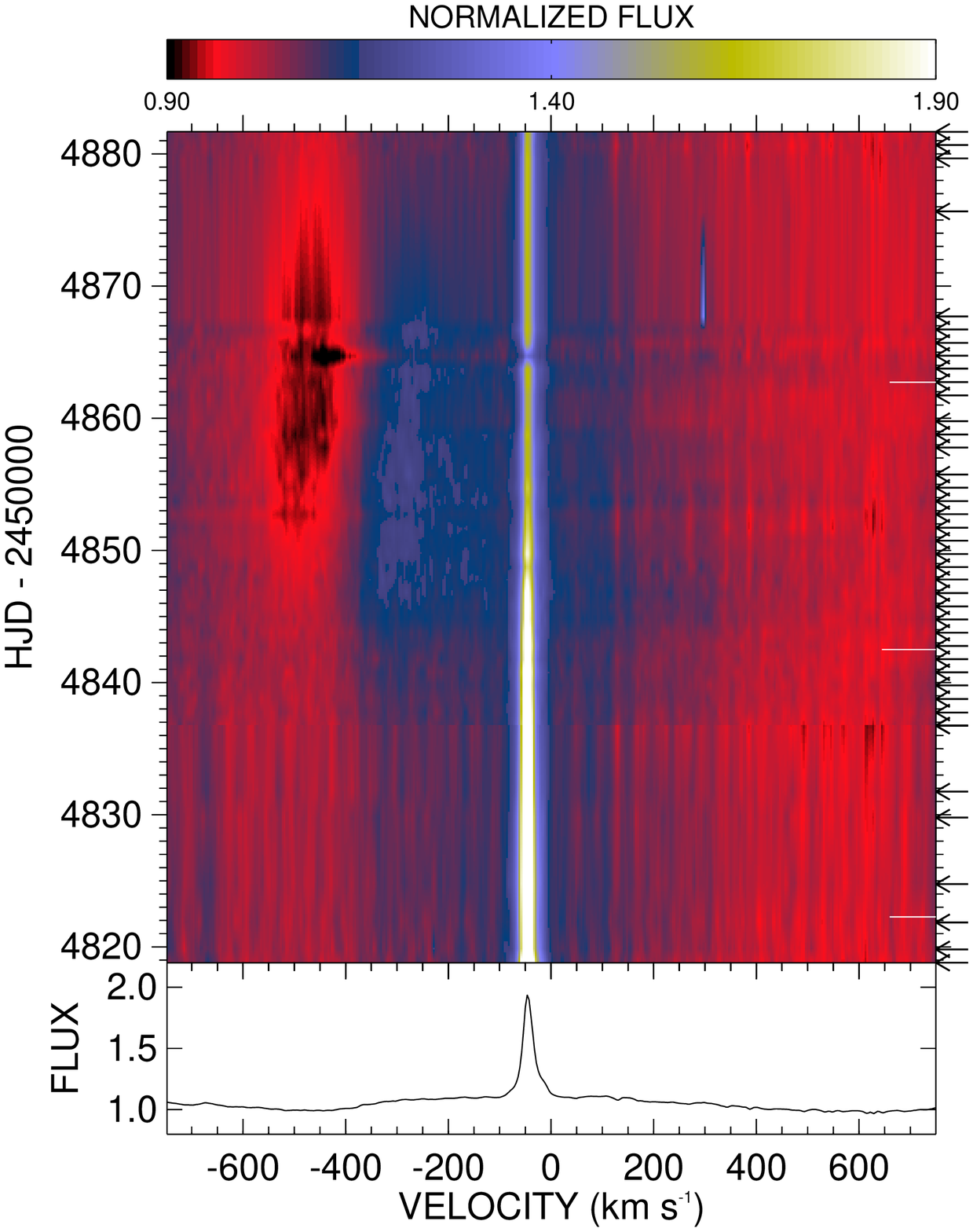}{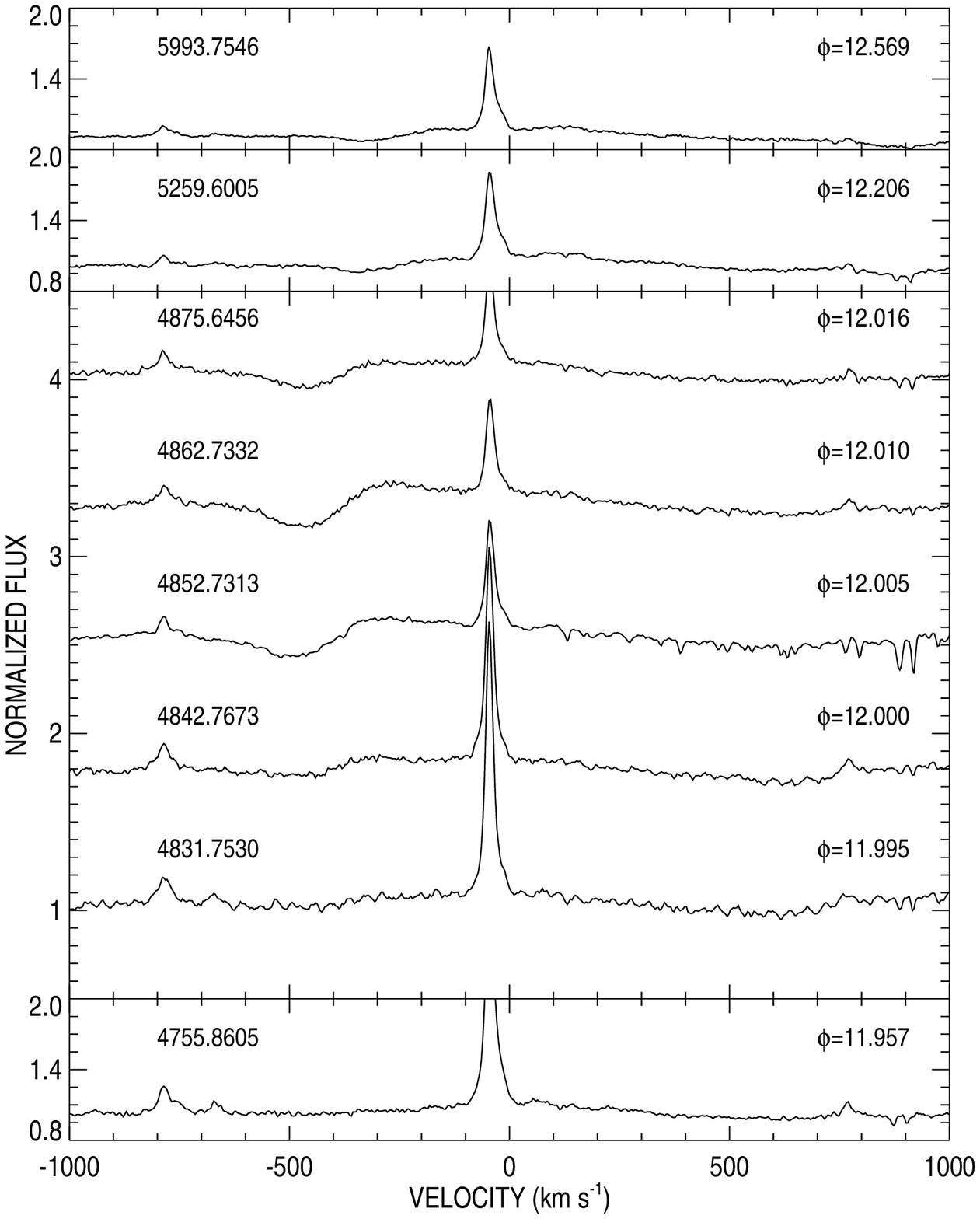}
\figsetgrpnote{Dynamical (left) and line plots (right) of \ion{Fe}{2} $\lambda$6456 during the 2009 event. The white horizontal bars on the dynamical representation indicate phases 11.99, 12.00, and 12.01.}
\figsetgrpend

\figsetend

\clearpage
\setcounter{figure}{1}
\renewcommand{\thefigure}{\arabic{figure}.1}
\begin{figure}
\figurenum{1.1}
\plottwo{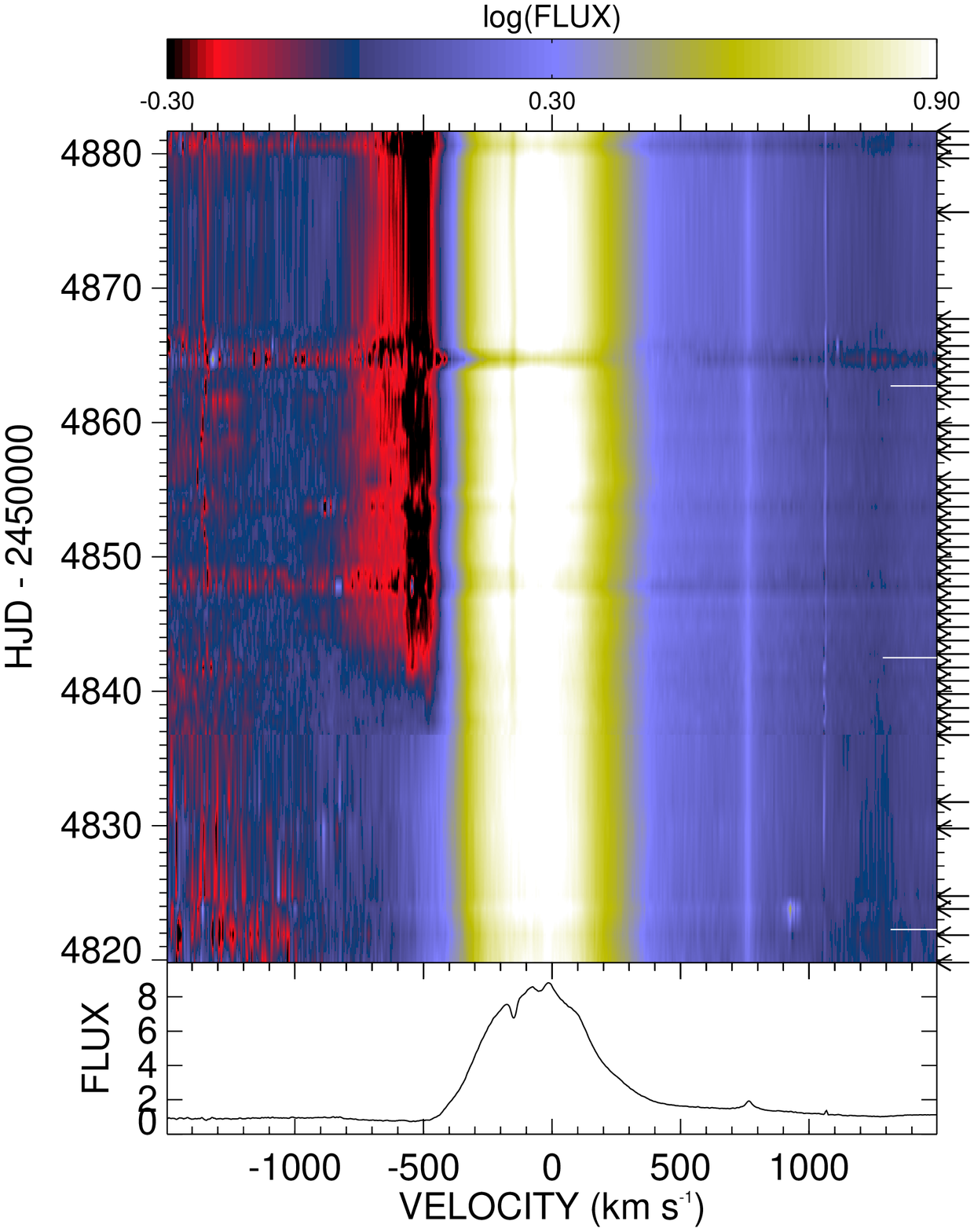}{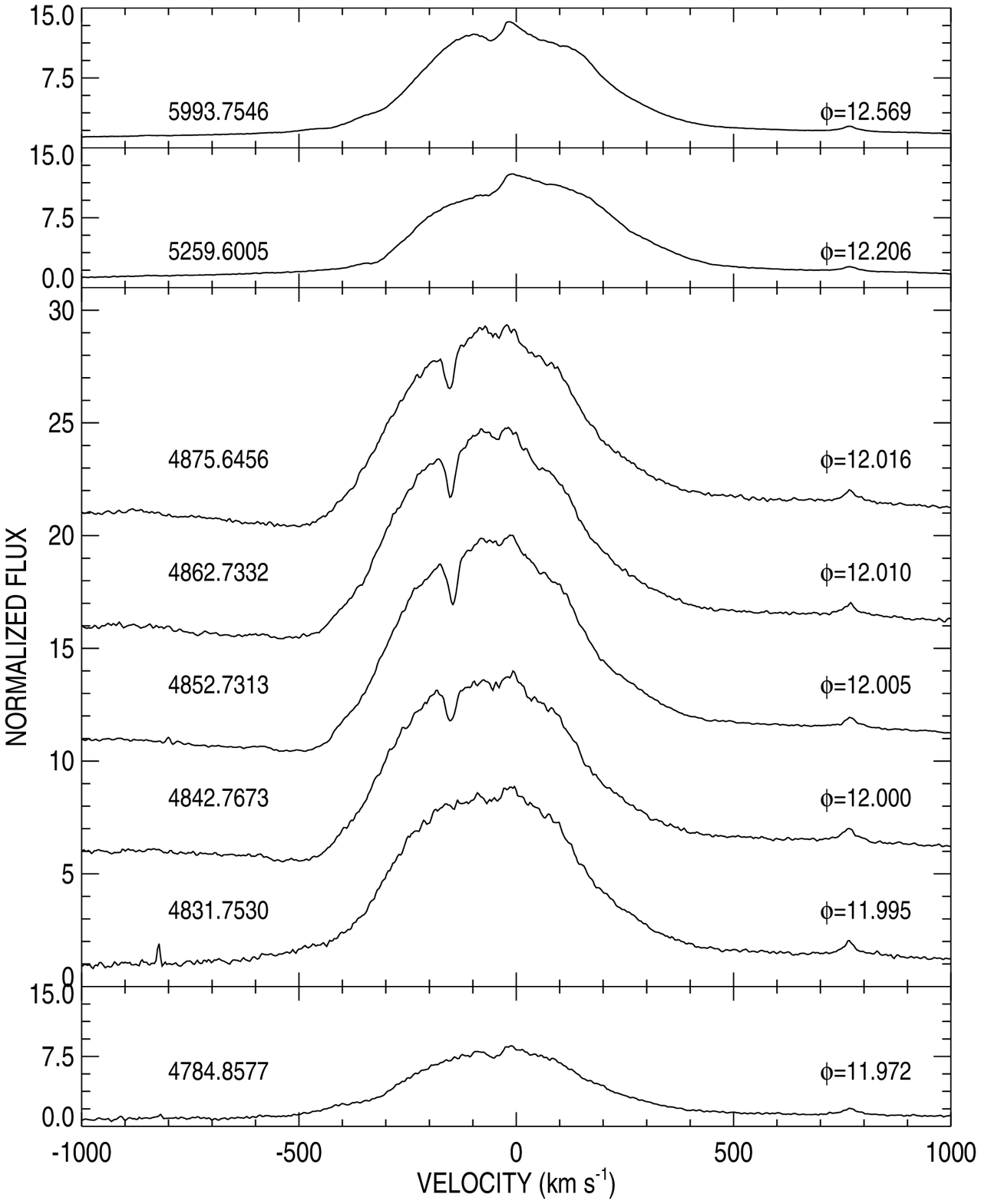}
\caption{Dynamical (left) and line plots (right) of H$\beta$ during the 2009 event. The white horizontal bars on the dynamical representation indicate phases 11.99, 12.00, and 12.01.}
\end{figure}

\clearpage
\setcounter{figure}{1}
\renewcommand{\thefigure}{\arabic{figure}.2}
\begin{figure}
\figurenum{1.2}
\plottwo{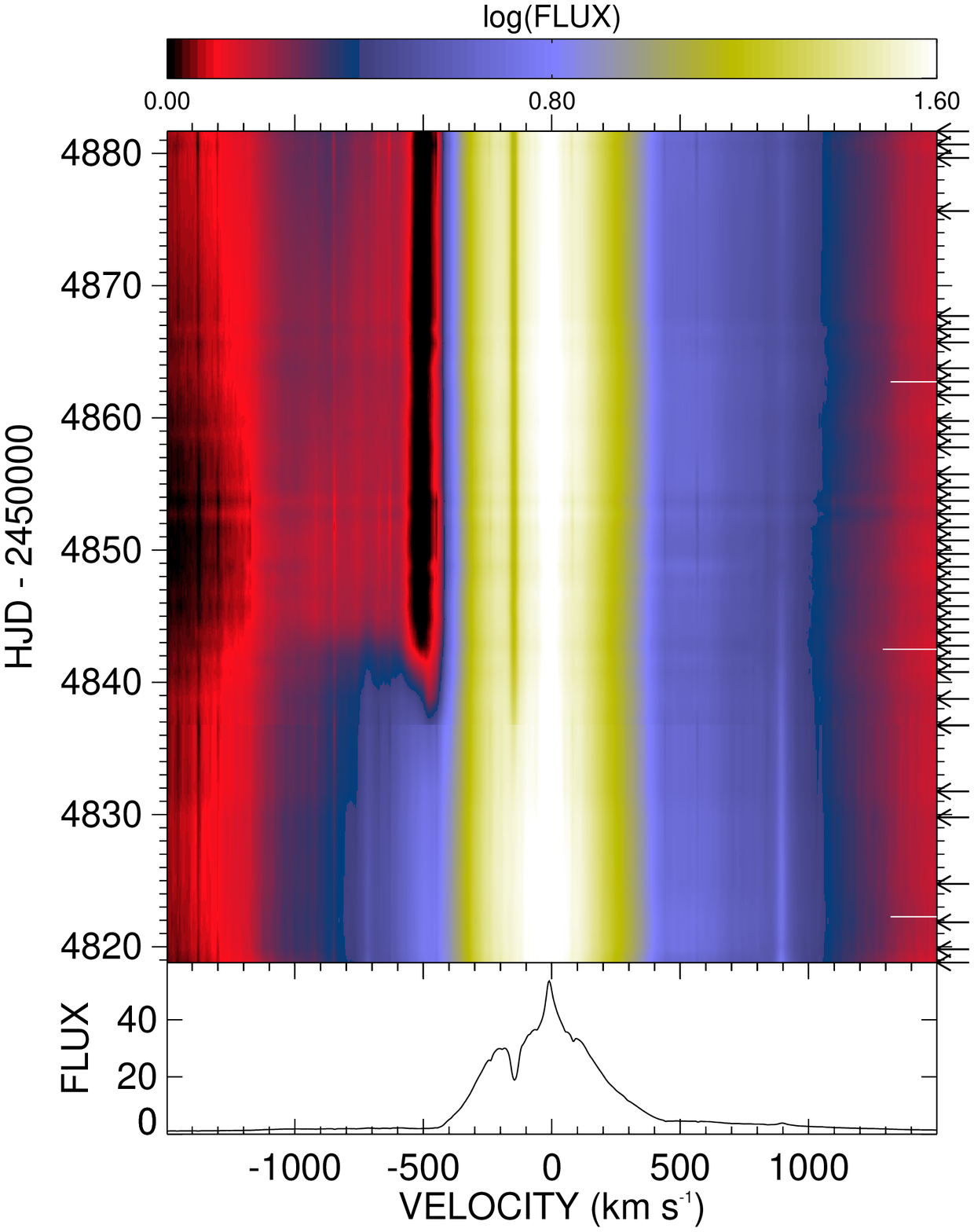}{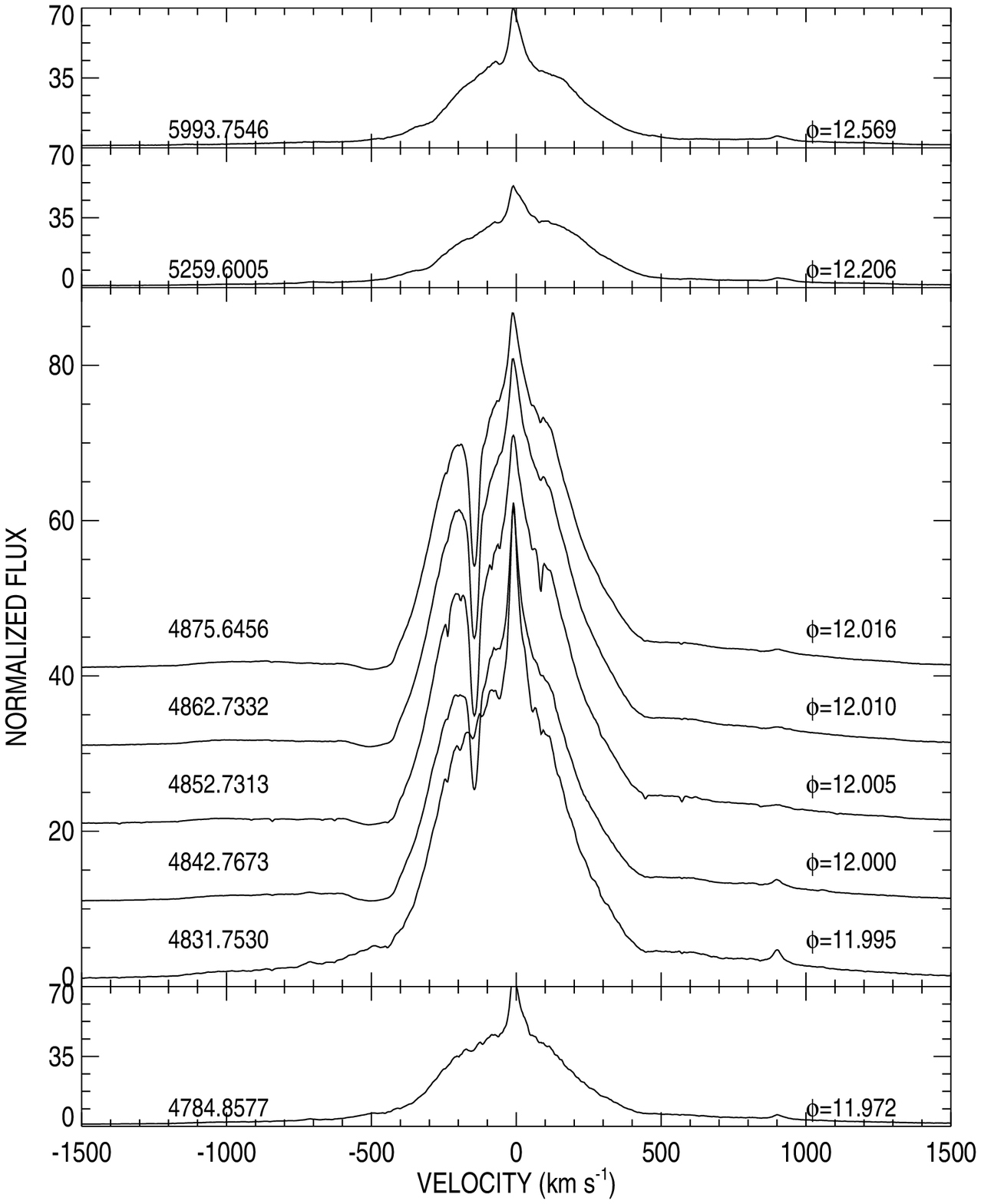}
\caption{Dynamical (left) and line plots (right) of H$\alpha$ during the 2009 event. The white horizontal bars on the dynamical representation indicate phases 11.99, 12.00, and 12.01.}
\end{figure}

\clearpage
\setcounter{figure}{1}
\renewcommand{\thefigure}{\arabic{figure}.3}
\begin{figure}
\figurenum{1.3}
\plottwo{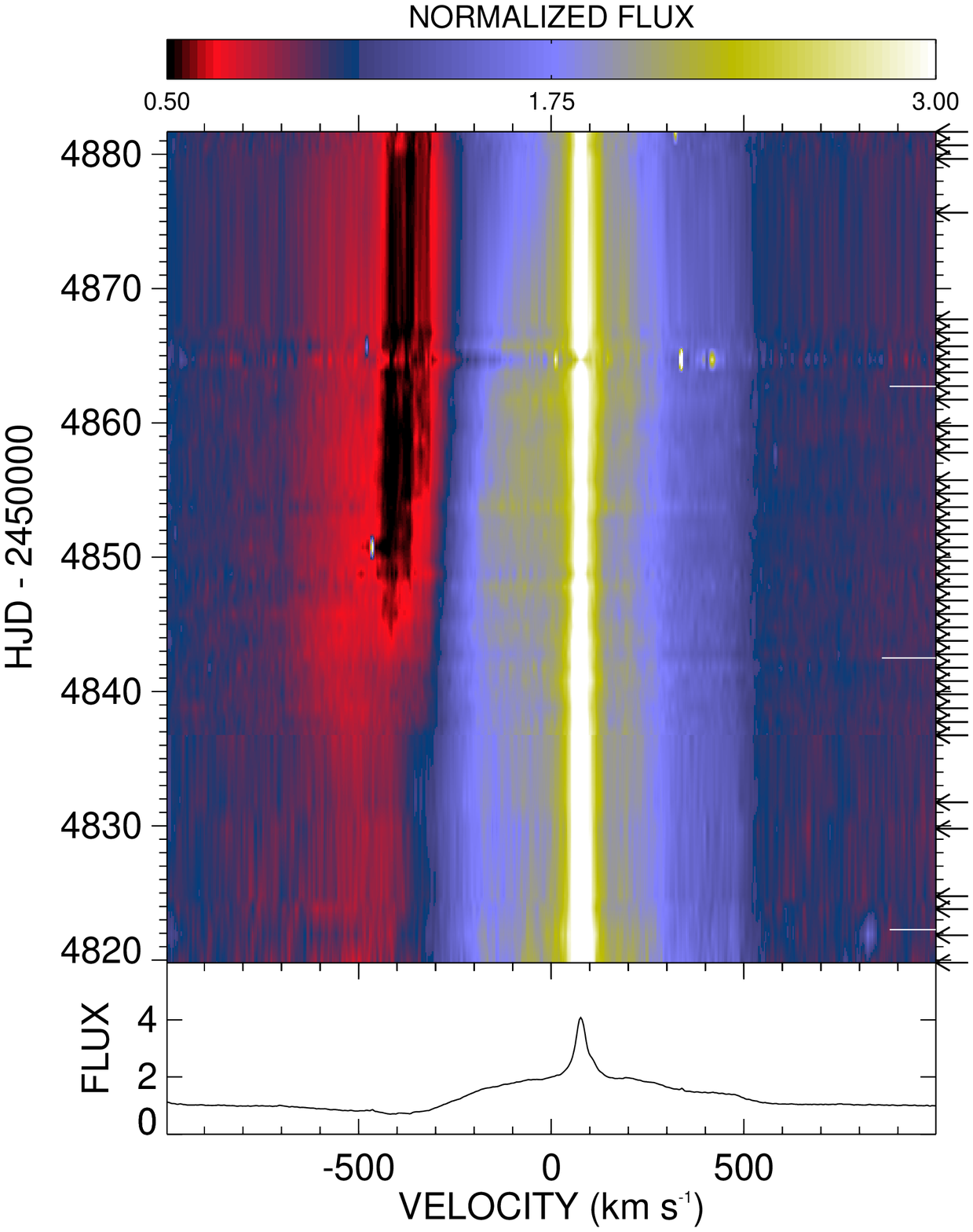}{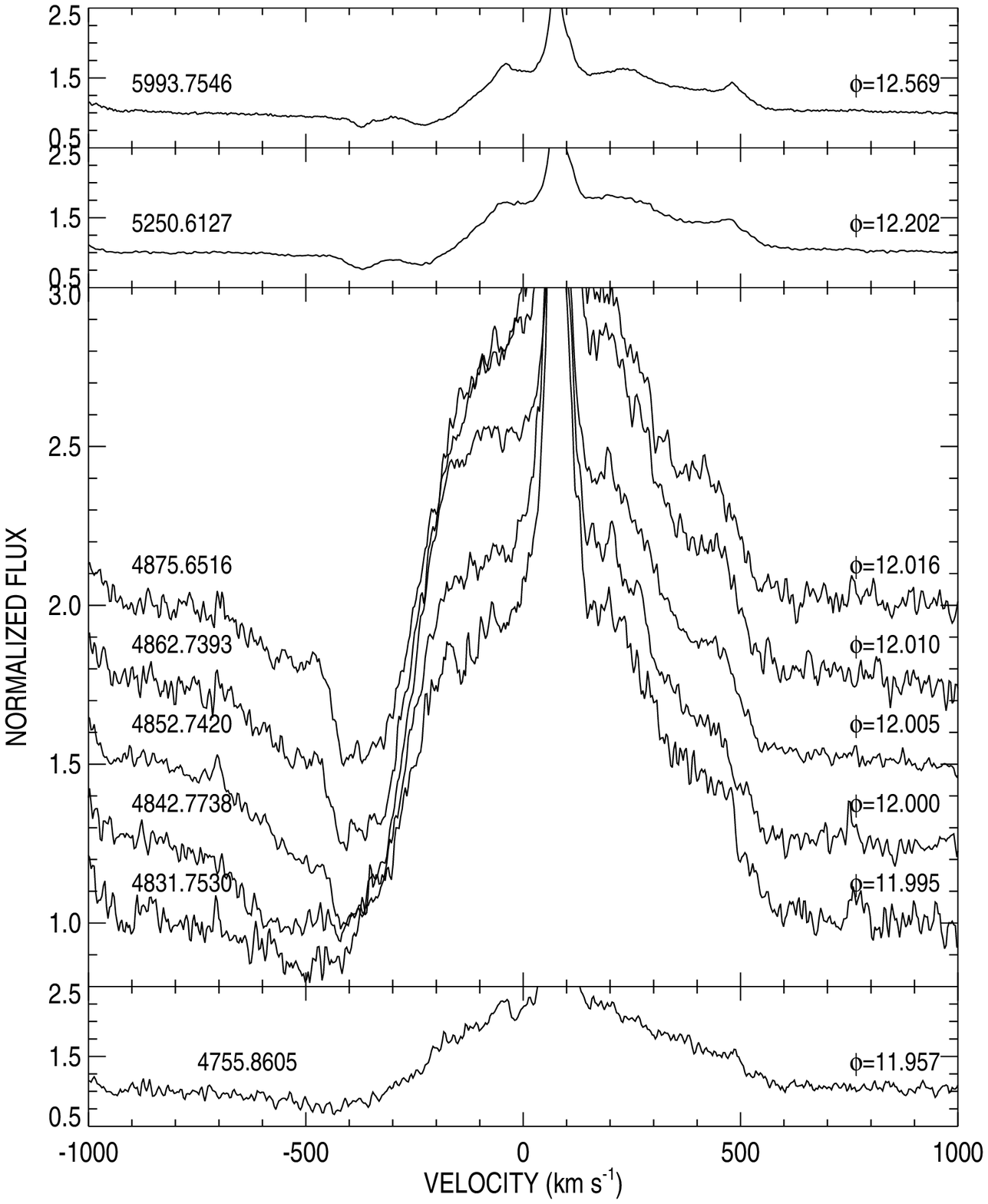}
\caption{Dynamical (left) and line plots (right) of \ion{He}{1} $\lambda$4922 during the 2009 event. The white horizontal bars on the dynamical representation indicate phases 11.99, 12.00, and 12.01.}
\end{figure}

\clearpage
\setcounter{figure}{1}
\renewcommand{\thefigure}{\arabic{figure}.1}
\begin{figure}
\figurenum{1.4}
\plottwo{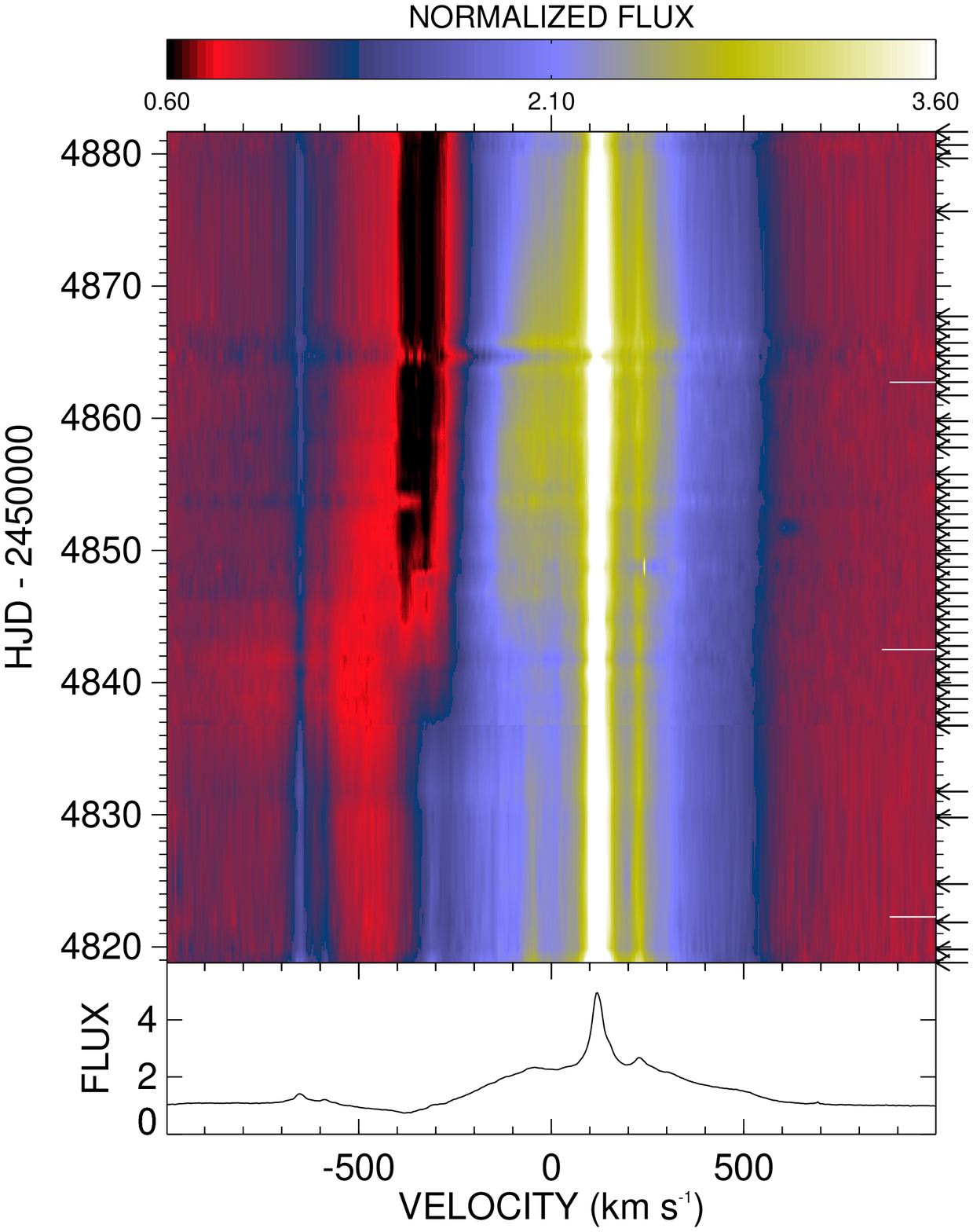}{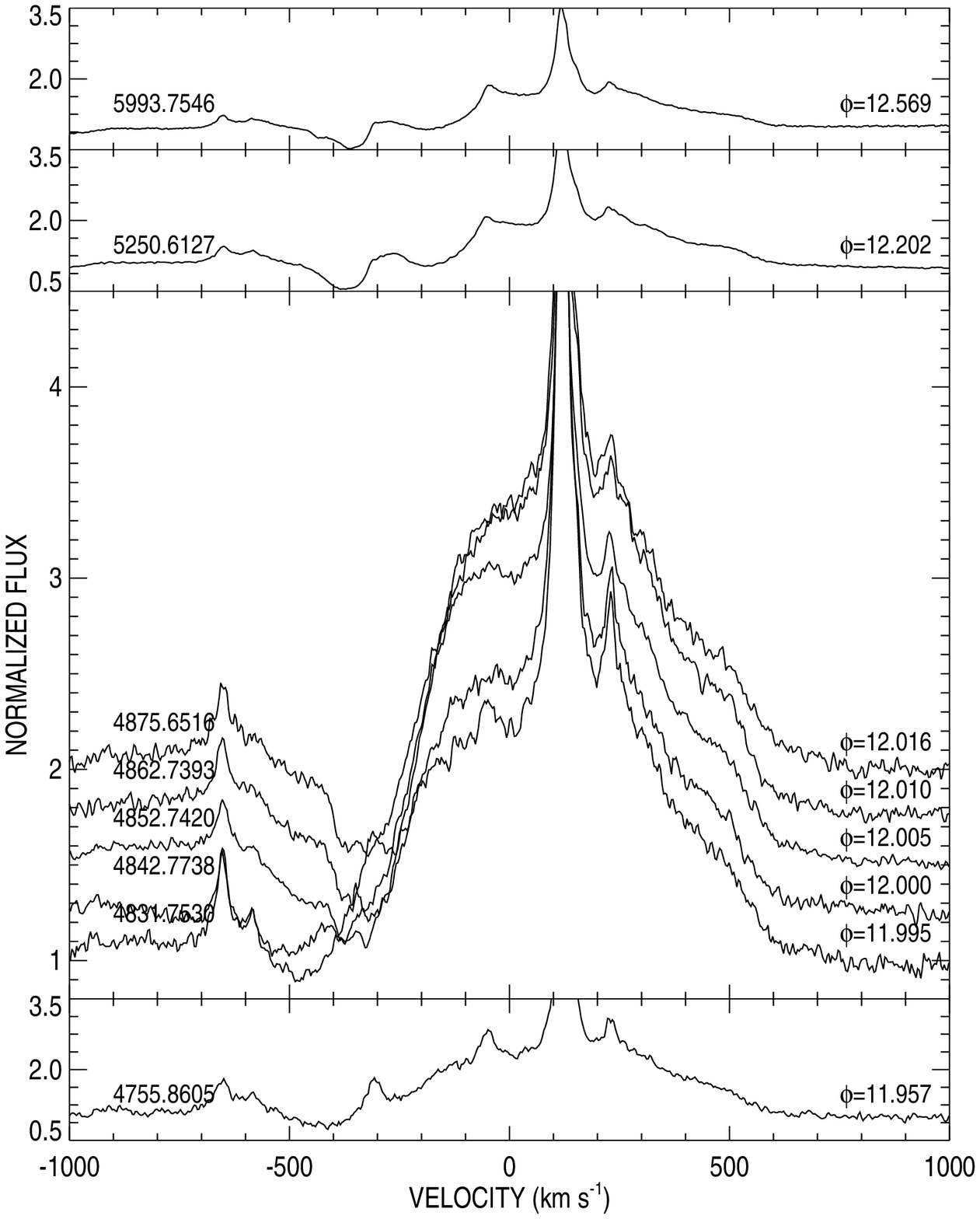}
\caption{Dynamical (left) and line plots (right) of \ion{He}{1} $\lambda$5015 during the 2009 event. The white horizontal bars on the dynamical representation indicate phases 11.99, 12.00, and 12.01.}
\end{figure}

\clearpage
\setcounter{figure}{1}
\renewcommand{\thefigure}{\arabic{figure}.1}
\begin{figure}
\figurenum{1.5}
\plottwo{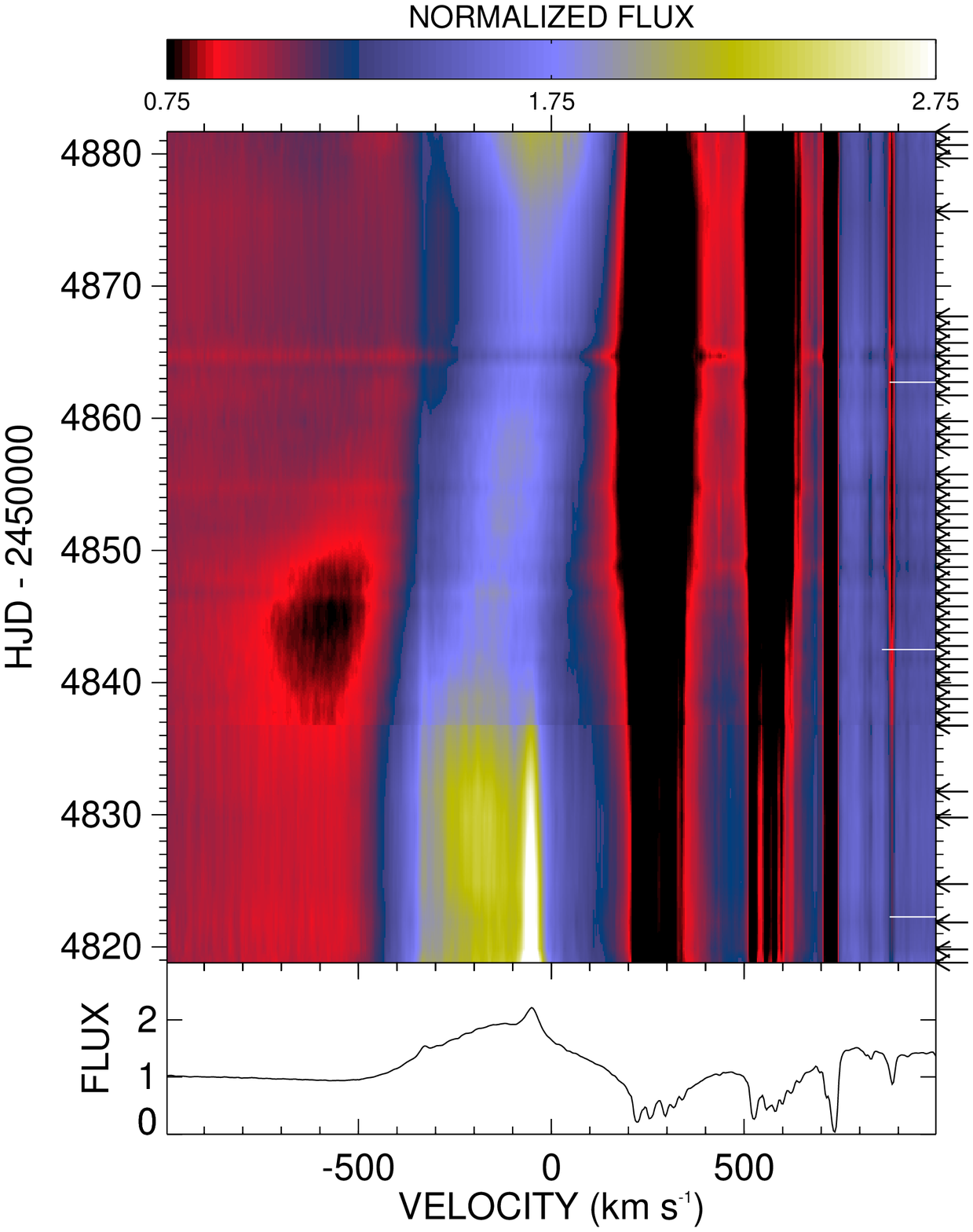}{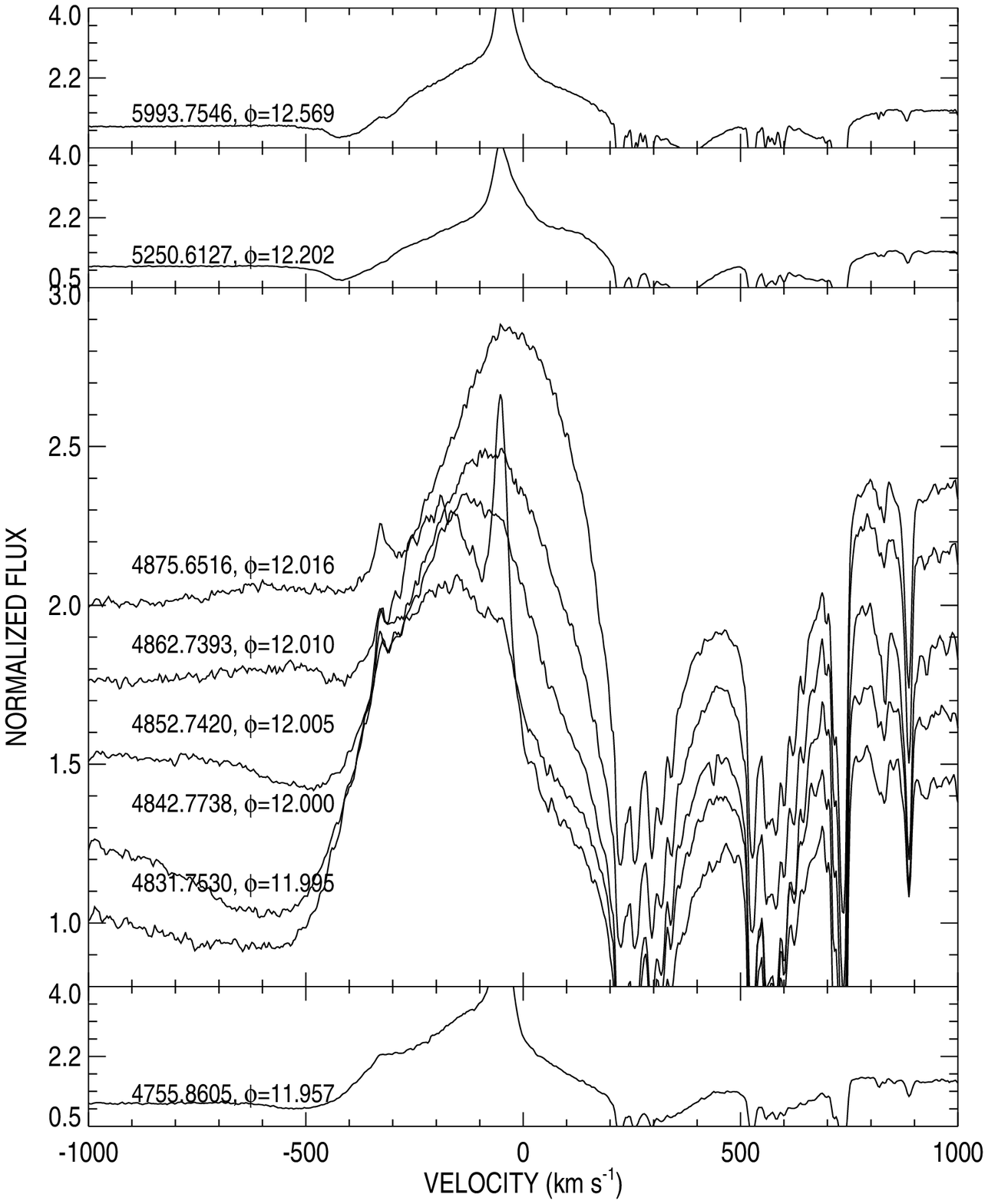}
\caption{Dynamical (left) and line plots (right) of \ion{He}{1} $\lambda$5876 during the 2009 event. The white horizontal bars on the dynamical representation indicate phases 11.99, 12.00, and 12.01.}
\end{figure}

\clearpage
\setcounter{figure}{1}
\renewcommand{\thefigure}{\arabic{figure}.1}
\begin{figure}
\figurenum{1.6}
\plottwo{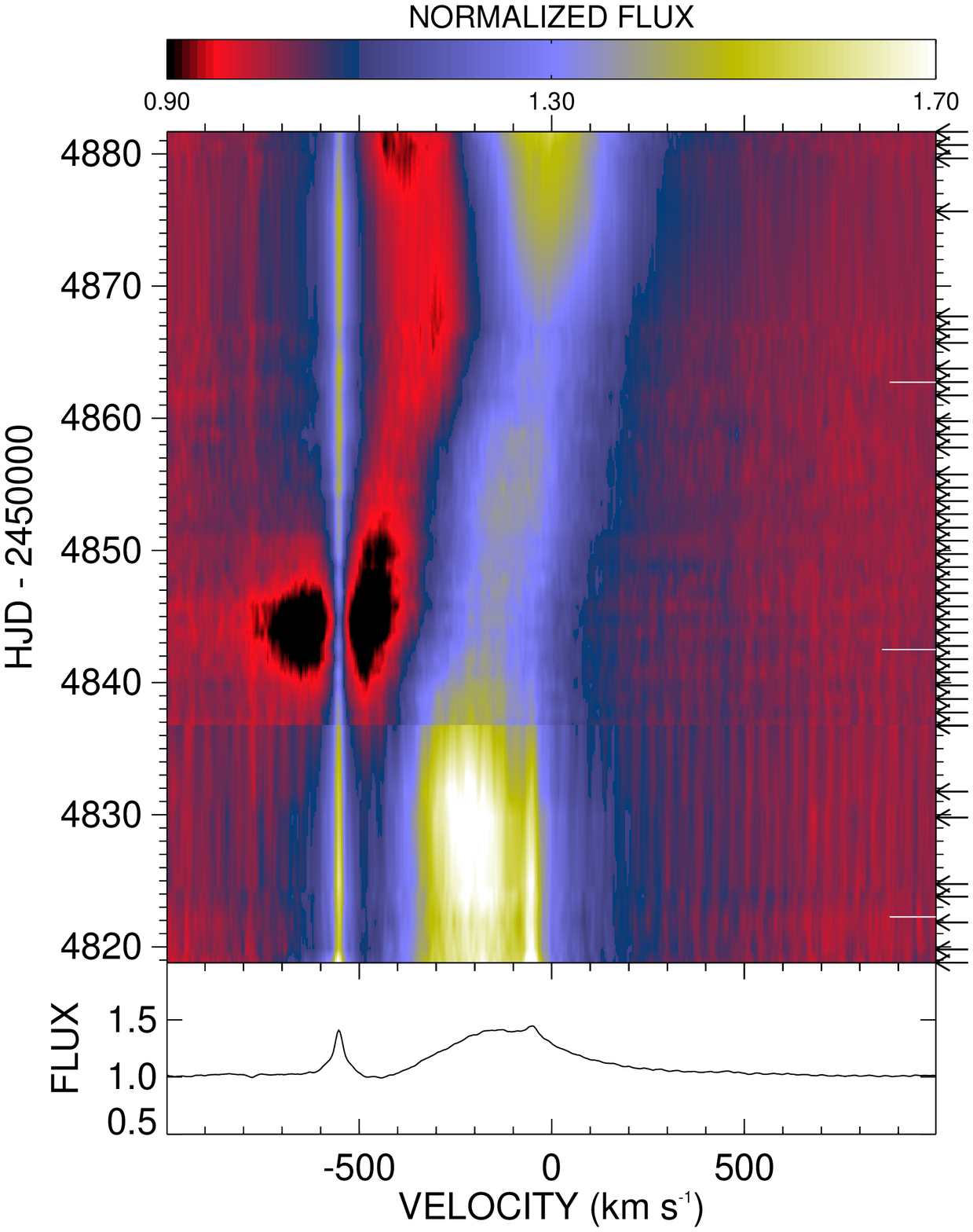}{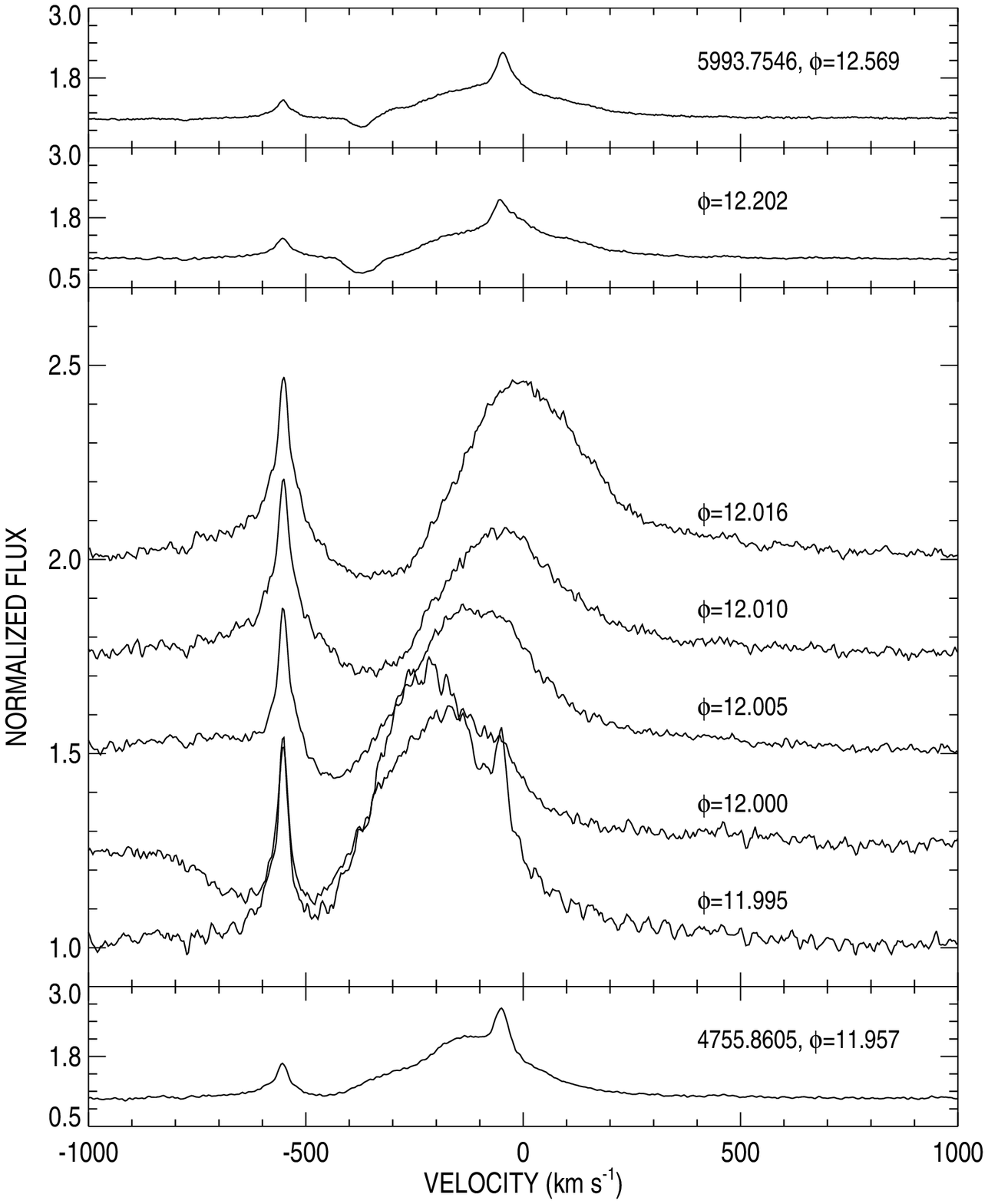}
\caption{Dynamical (left) and line plots (right) of \ion{He}{1} $\lambda$6678 during the 2009 event. The white horizontal bars on the dynamical representation indicate phases 11.99, 12.00, and 12.01.}
\end{figure}

\clearpage
\setcounter{figure}{1}
\renewcommand{\thefigure}{\arabic{figure}.1}
\begin{figure}
\figurenum{1.7}
\plottwo{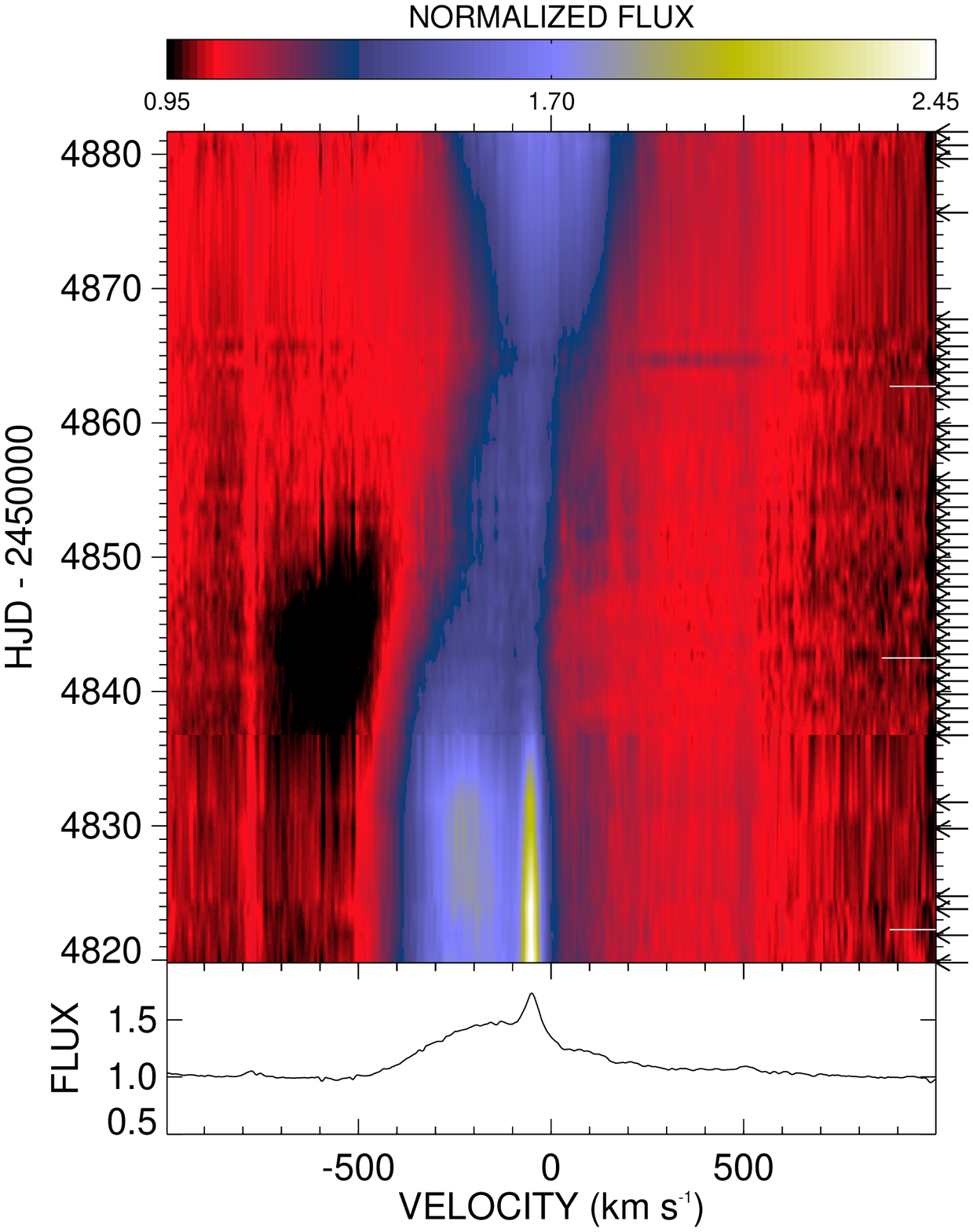}{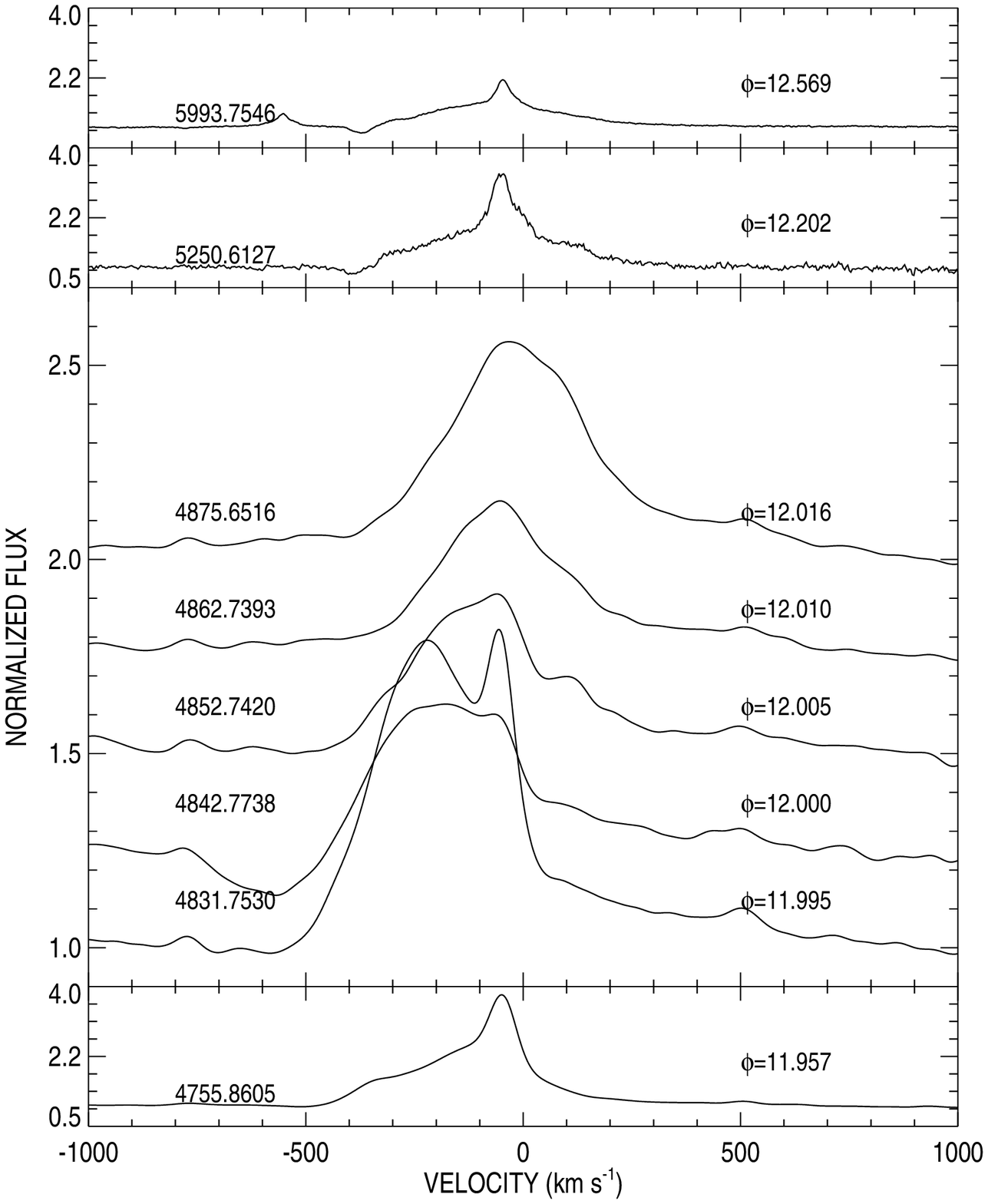}
\caption{Dynamical (left) and line plots (right) of \ion{He}{1} $\lambda$7065 during the 2009 event. The white horizontal bars on the dynamical representation indicate phases 11.99, 12.00, and 12.01.}
\end{figure}

\clearpage
\setcounter{figure}{1}
\renewcommand{\thefigure}{\arabic{figure}.1}
\begin{figure}
\figurenum{1.8}
\plottwo{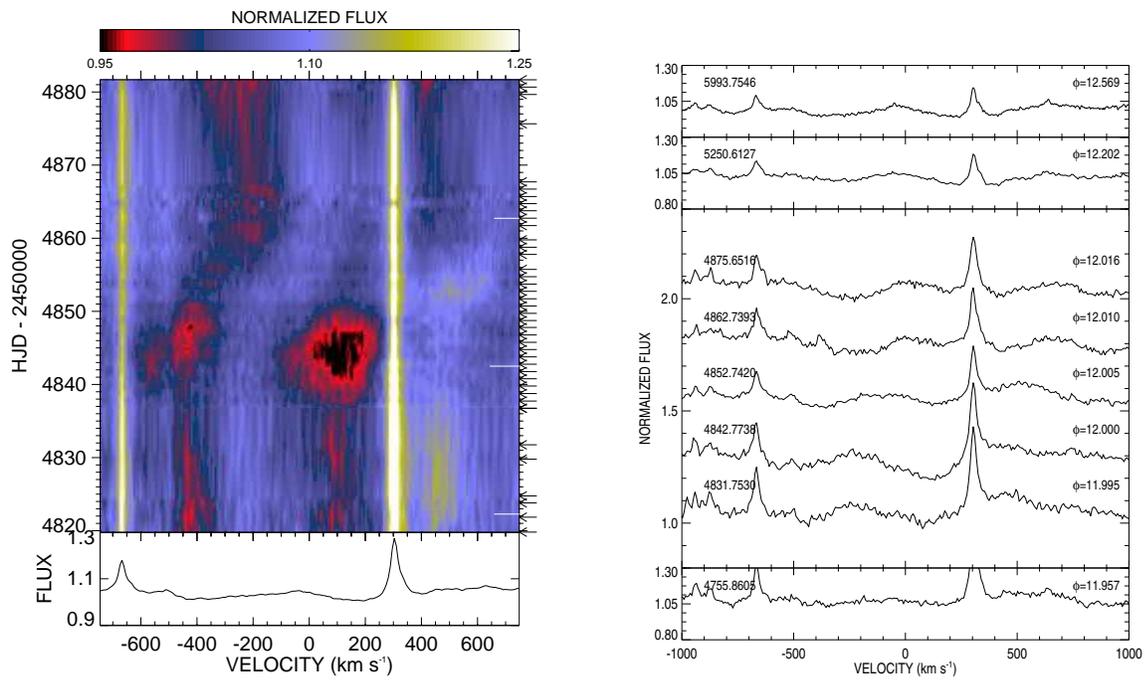}{f1_08b.eps}
\caption{Dynamical (left) and line plots (right) of \ion{N}{2} $\lambda$5666 during the 2009 event. The white horizontal bars on the dynamical representation indicate phases 11.99, 12.00, and 12.01. Note that \ion{N}{2} $\lambda$5676 is visible on the right.}
\end{figure}

\clearpage
\setcounter{figure}{1}
\renewcommand{\thefigure}{\arabic{figure}.1}
\begin{figure}
\figurenum{1.9}
\plottwo{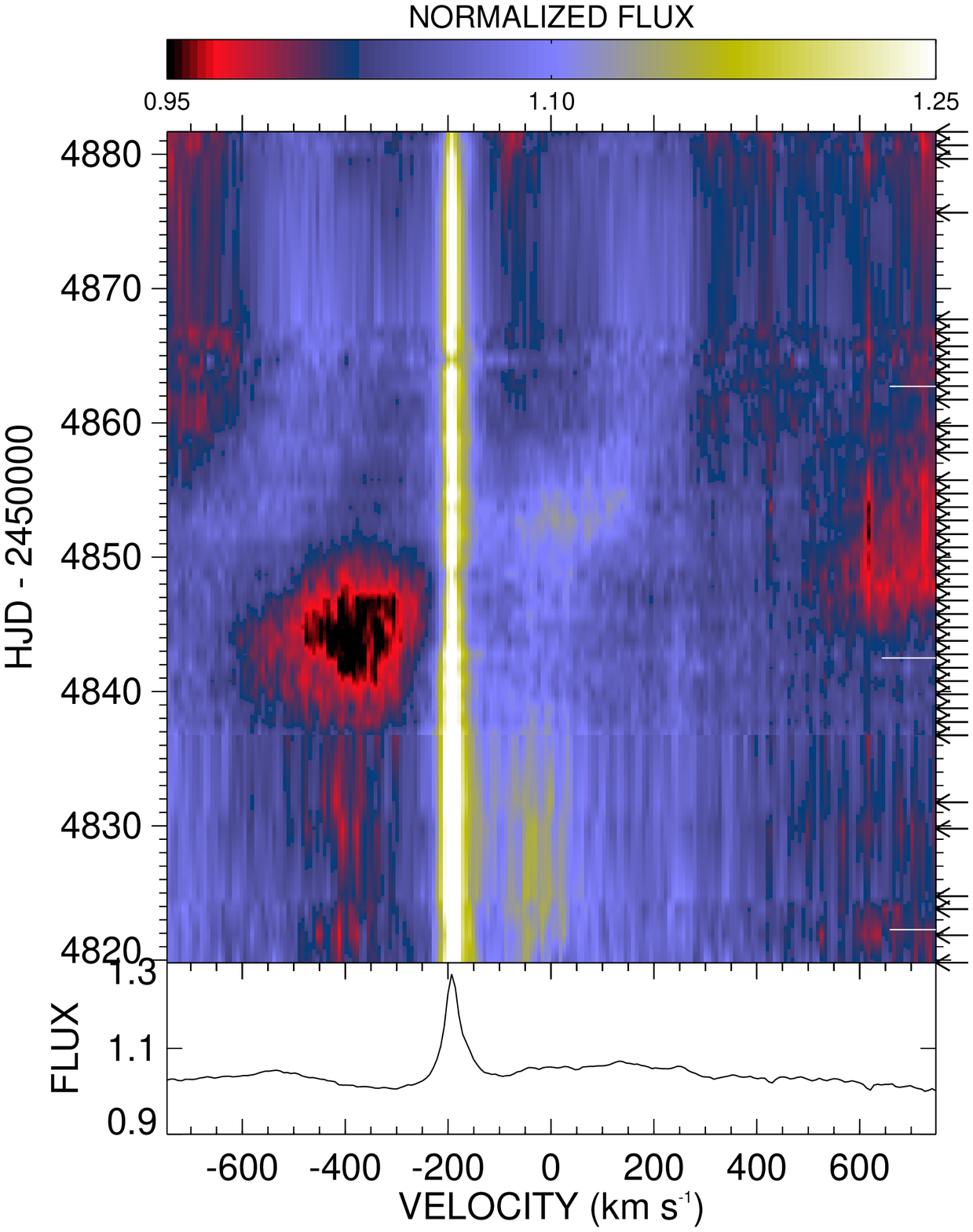}{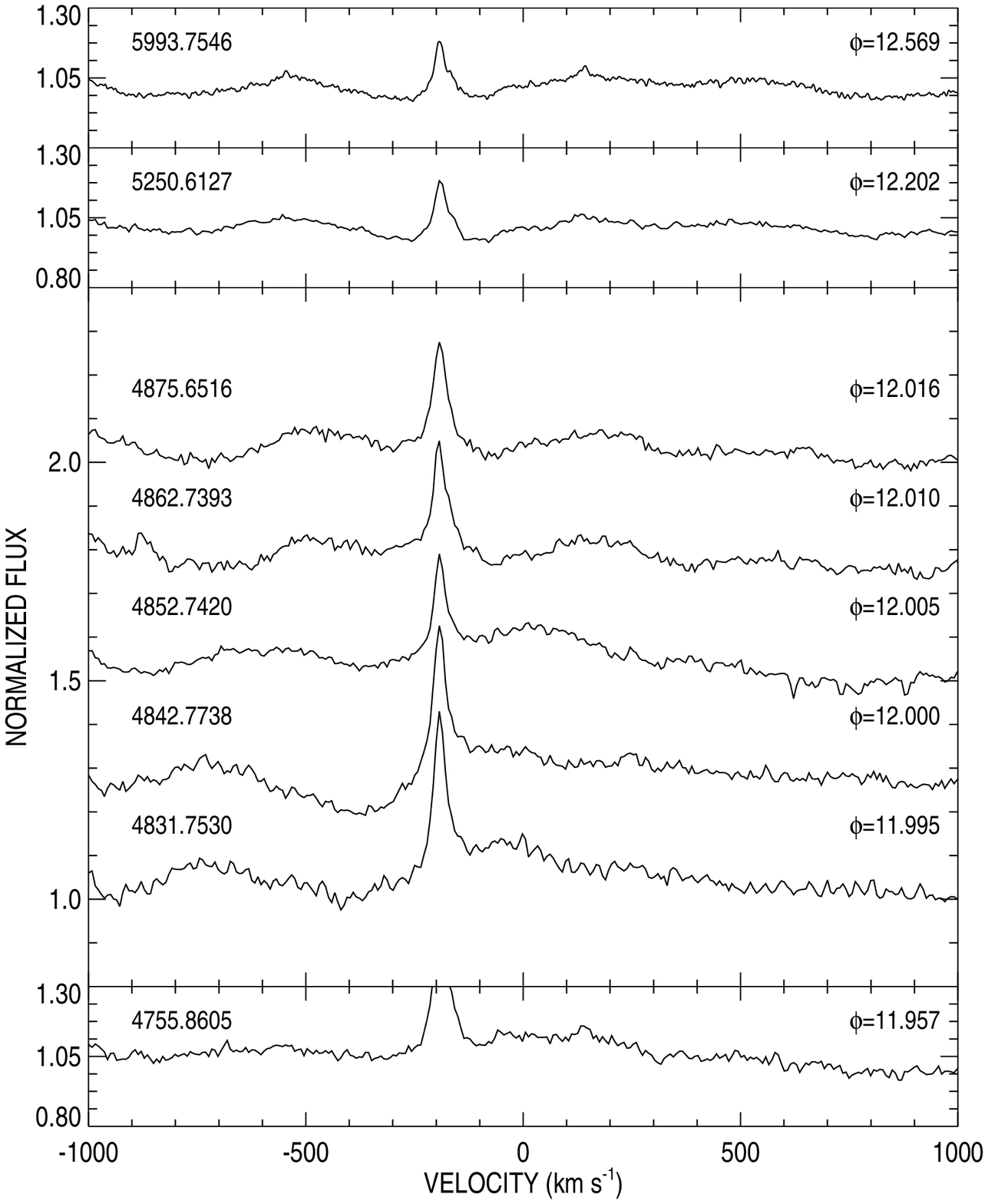}
\caption{Dynamical (left) and line plots (right) of \ion{N}{2} $\lambda$5676 during the 2009 event. The white horizontal bars on the dynamical representation indicate phases 11.99, 12.00, and 12.01.}
\end{figure}

\clearpage
\setcounter{figure}{1}
\renewcommand{\thefigure}{\arabic{figure}.1}
\begin{figure}
\figurenum{1.10}
\plottwo{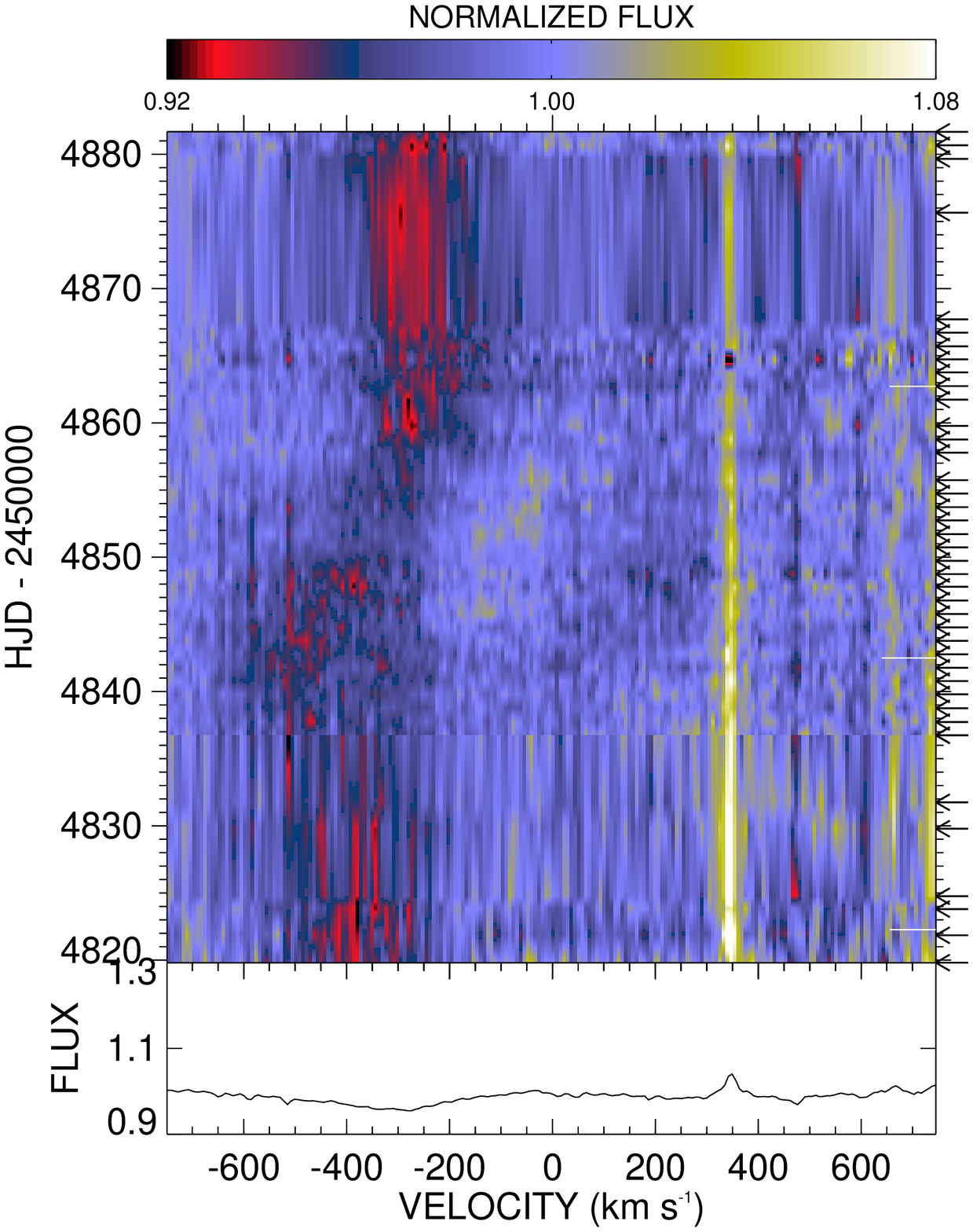}{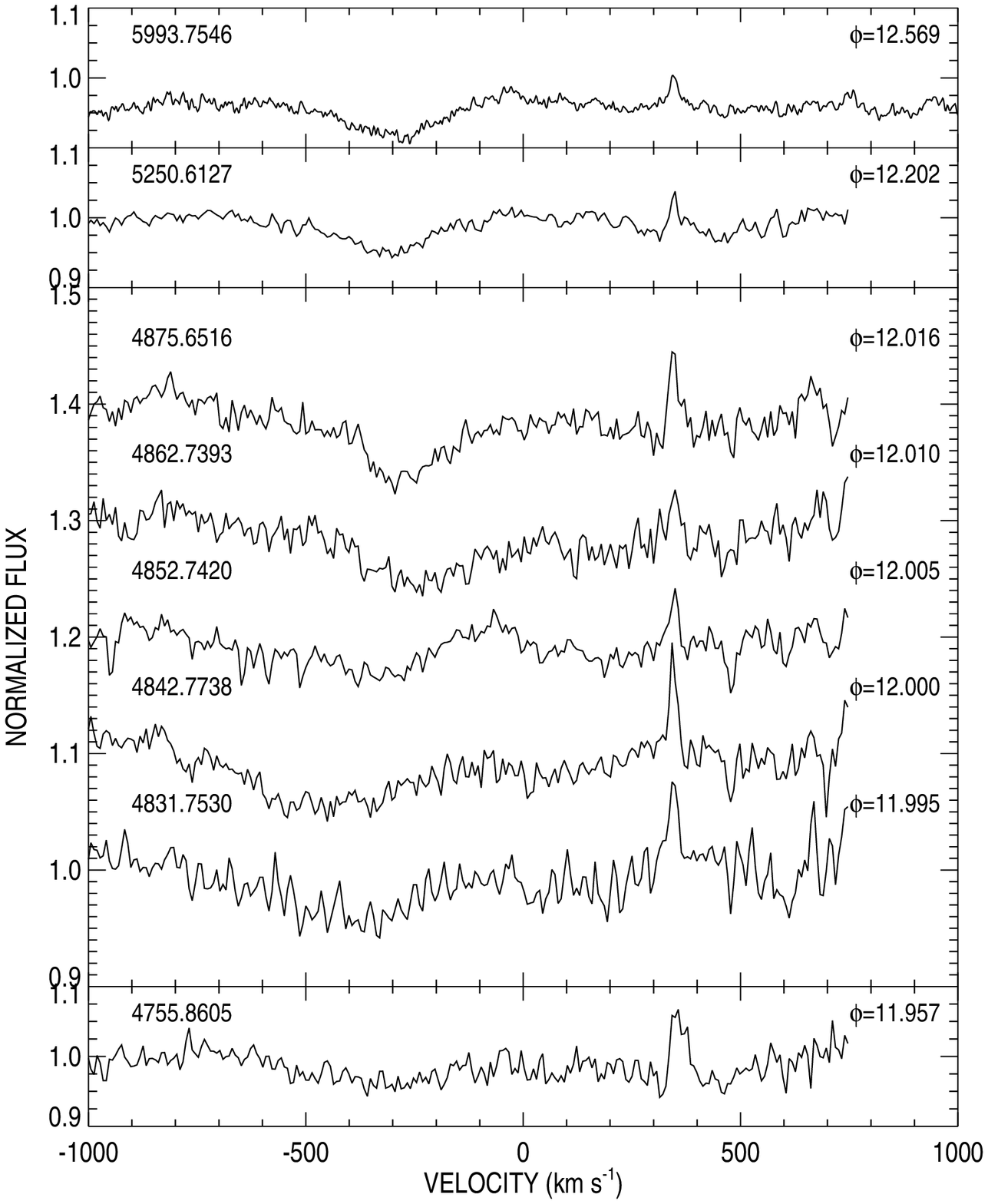}
\caption{Dynamical (left) and line plots (right) of \ion{N}{2} $\lambda$5711 during the 2009 event. The white horizontal bars on the dynamical representation indicate phases 11.99, 12.00, and 12.01.}
\end{figure}

\clearpage
\setcounter{figure}{1}
\renewcommand{\thefigure}{\arabic{figure}.1}
\begin{figure}
\figurenum{1.11}
\plottwo{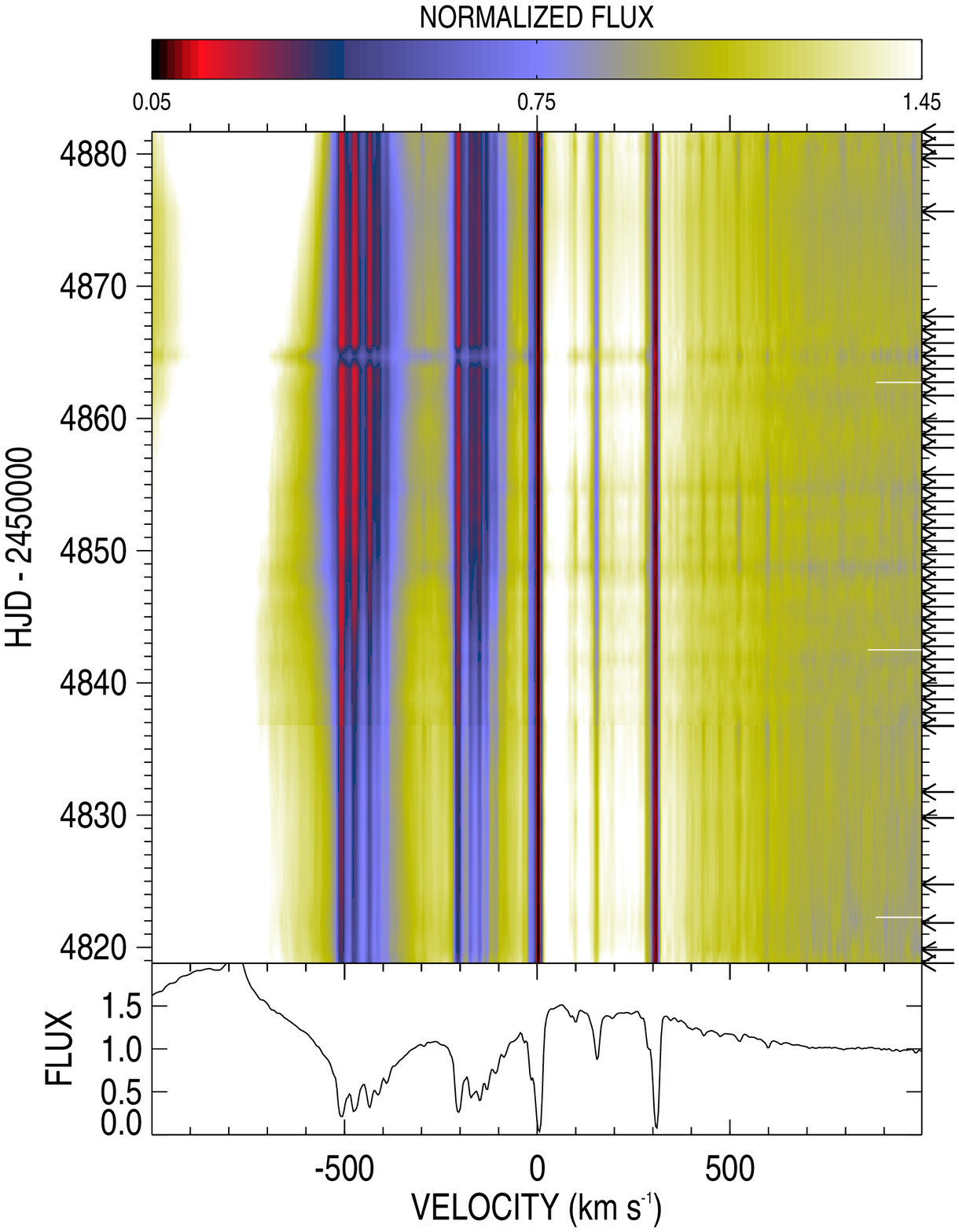}{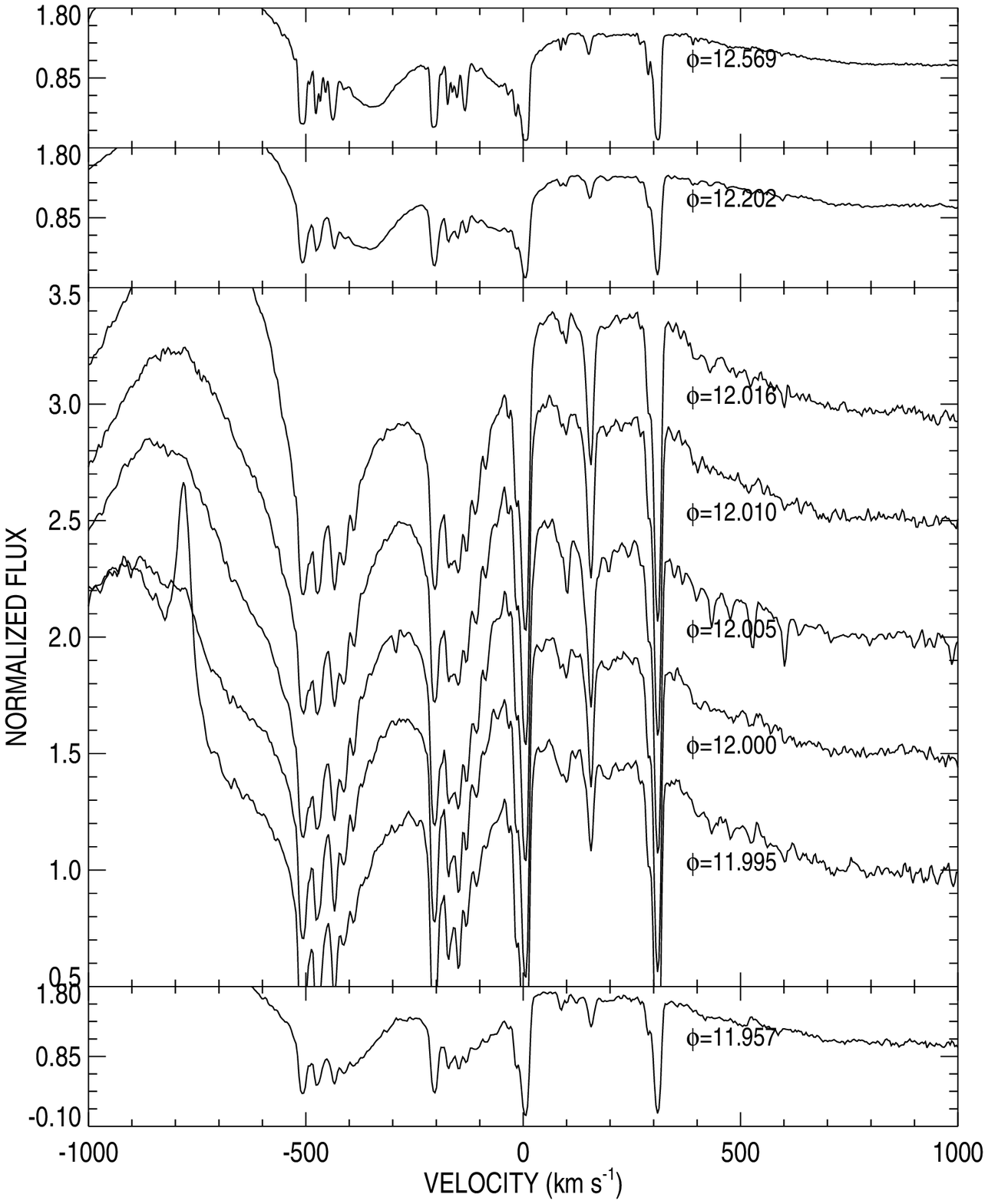}
\caption{Dynamical (left) and line plots (right) of \ion{Na}{1} D$_2 \lambda 5890$ during the 2009 event. The white horizontal bars on the dynamical representation indicate phases 11.99, 12.00, and 12.01.}
\end{figure}

\clearpage
\setcounter{figure}{1}
\renewcommand{\thefigure}{\arabic{figure}.1}
\begin{figure}
\figurenum{1.12}
\plottwo{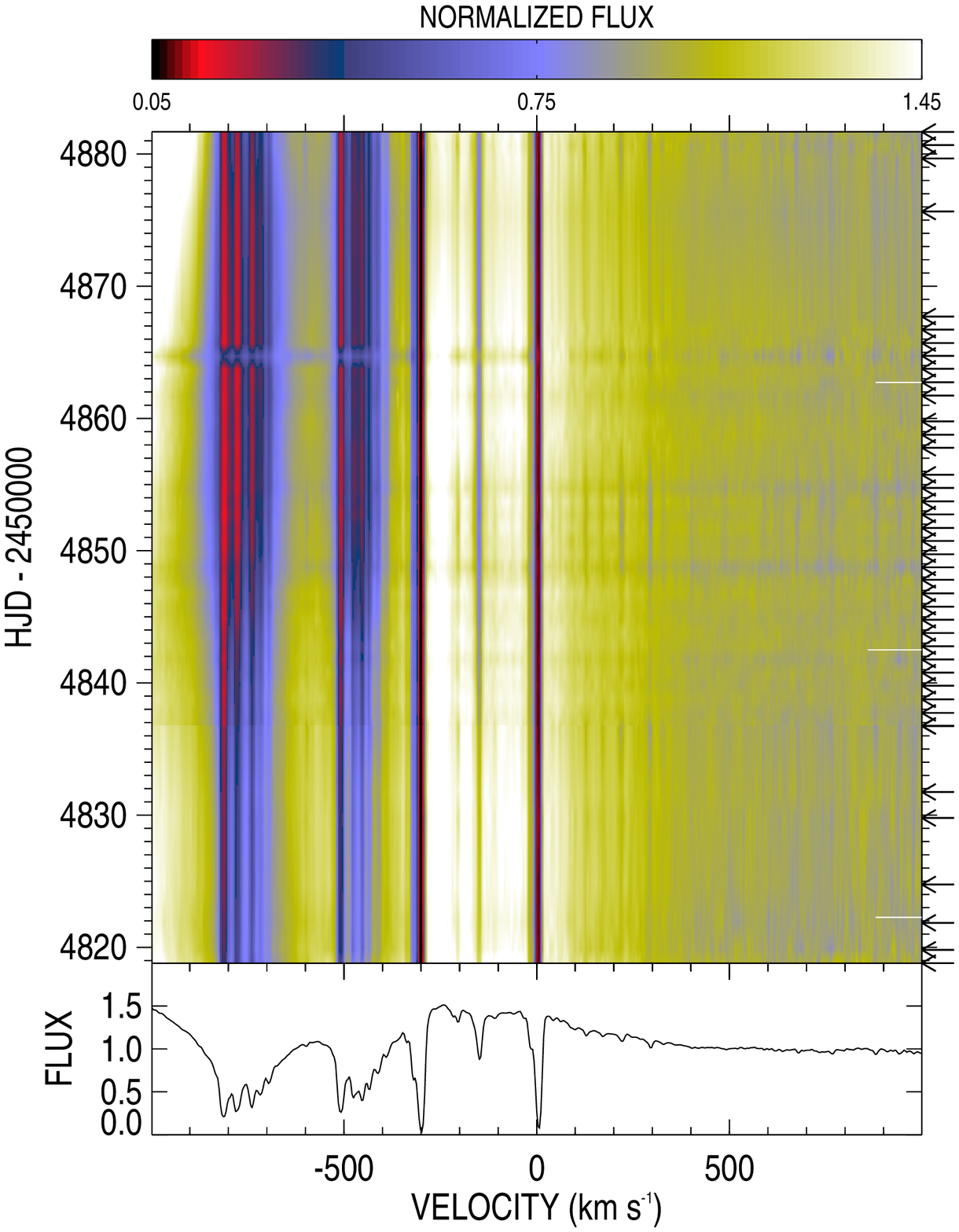}{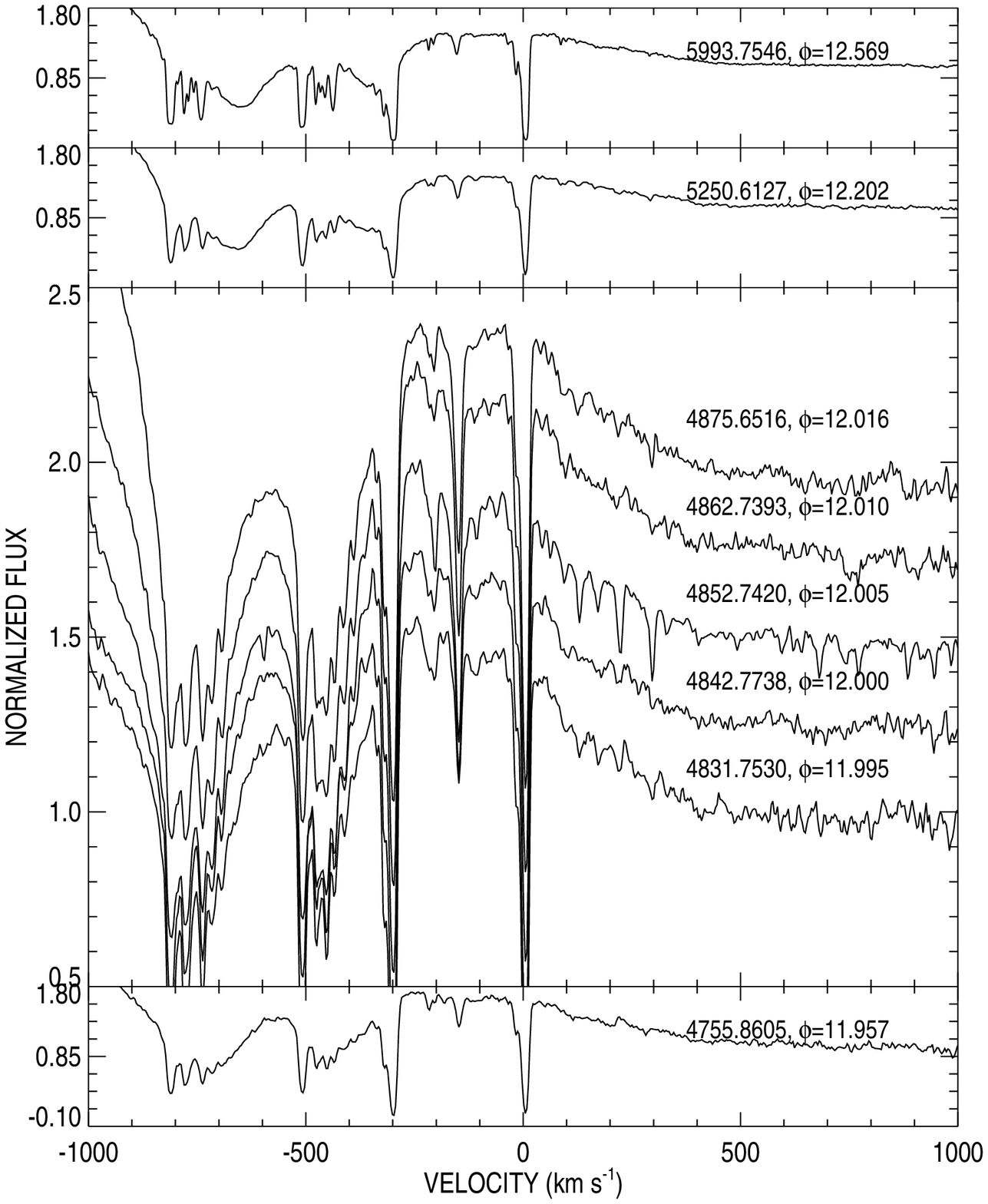}
\caption{Dynamical (left) and line plots (right) of \ion{Na}{1} D$_1 \lambda 5896$ during the 2009 event. The white horizontal bars on the dynamical representation indicate phases 11.99, 12.00, and 12.01.}
\end{figure}

\clearpage
\setcounter{figure}{1}
\renewcommand{\thefigure}{\arabic{figure}.1}
\begin{figure}
\figurenum{1.13}
\plottwo{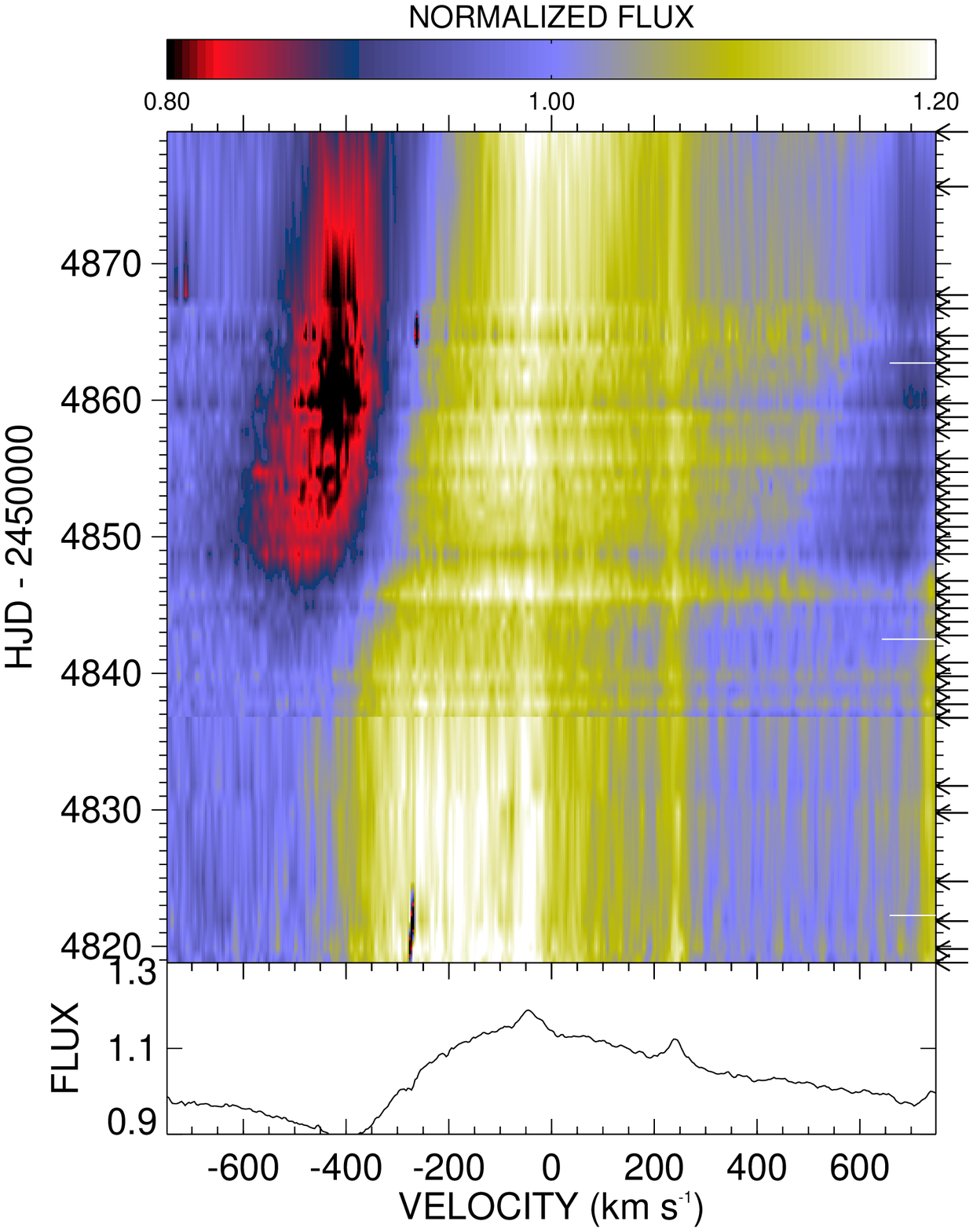}{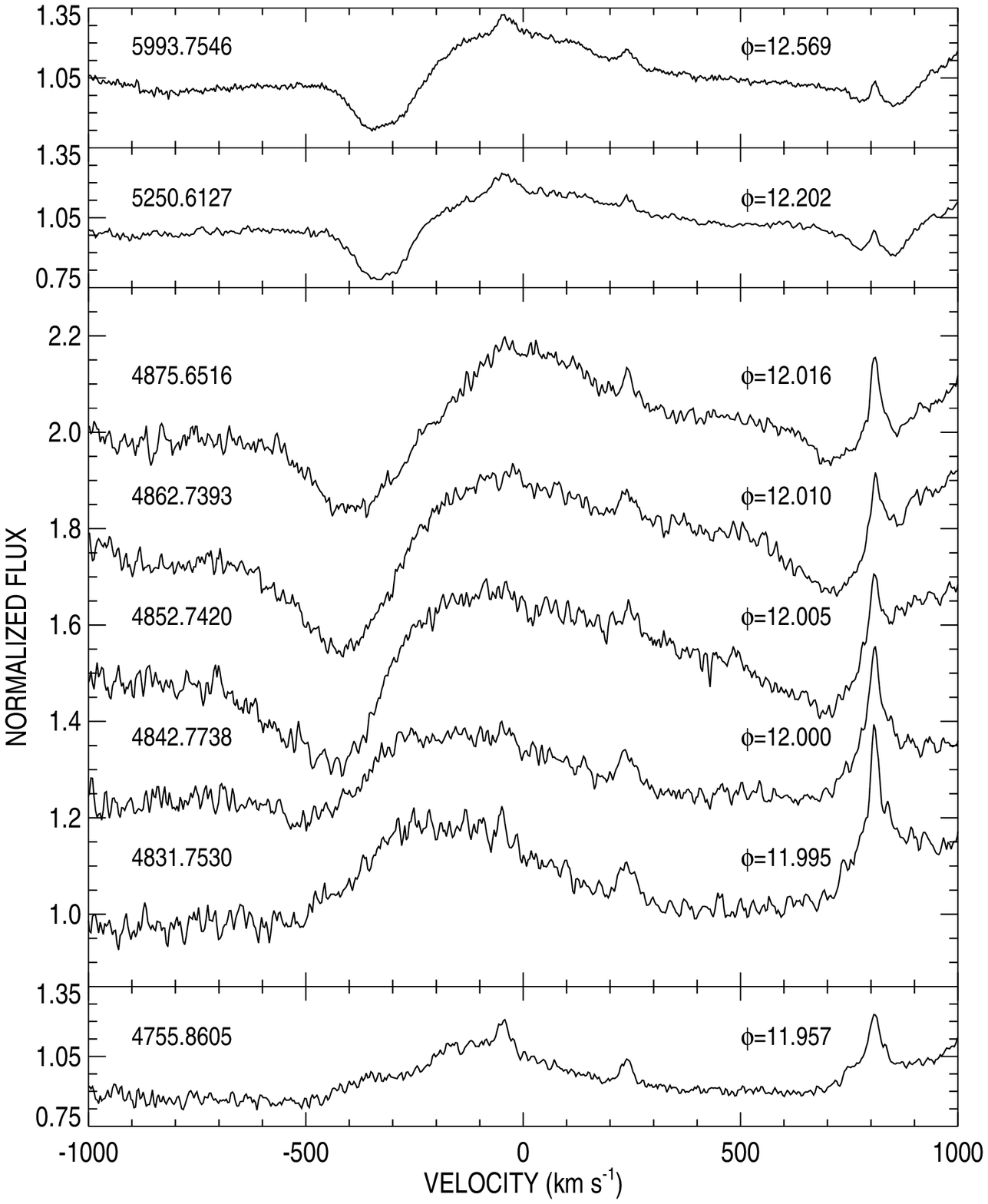}
\caption{Dynamical (left) and line plots (right) of \ion{Si}{2} $\lambda$6347 during the 2009 event. The white horizontal bars on the dynamical representation indicate phases 11.99, 12.00, and 12.01.}
\end{figure}

\clearpage
\setcounter{figure}{1}
\renewcommand{\thefigure}{\arabic{figure}.1}
\begin{figure}
\figurenum{1.14}
\plottwo{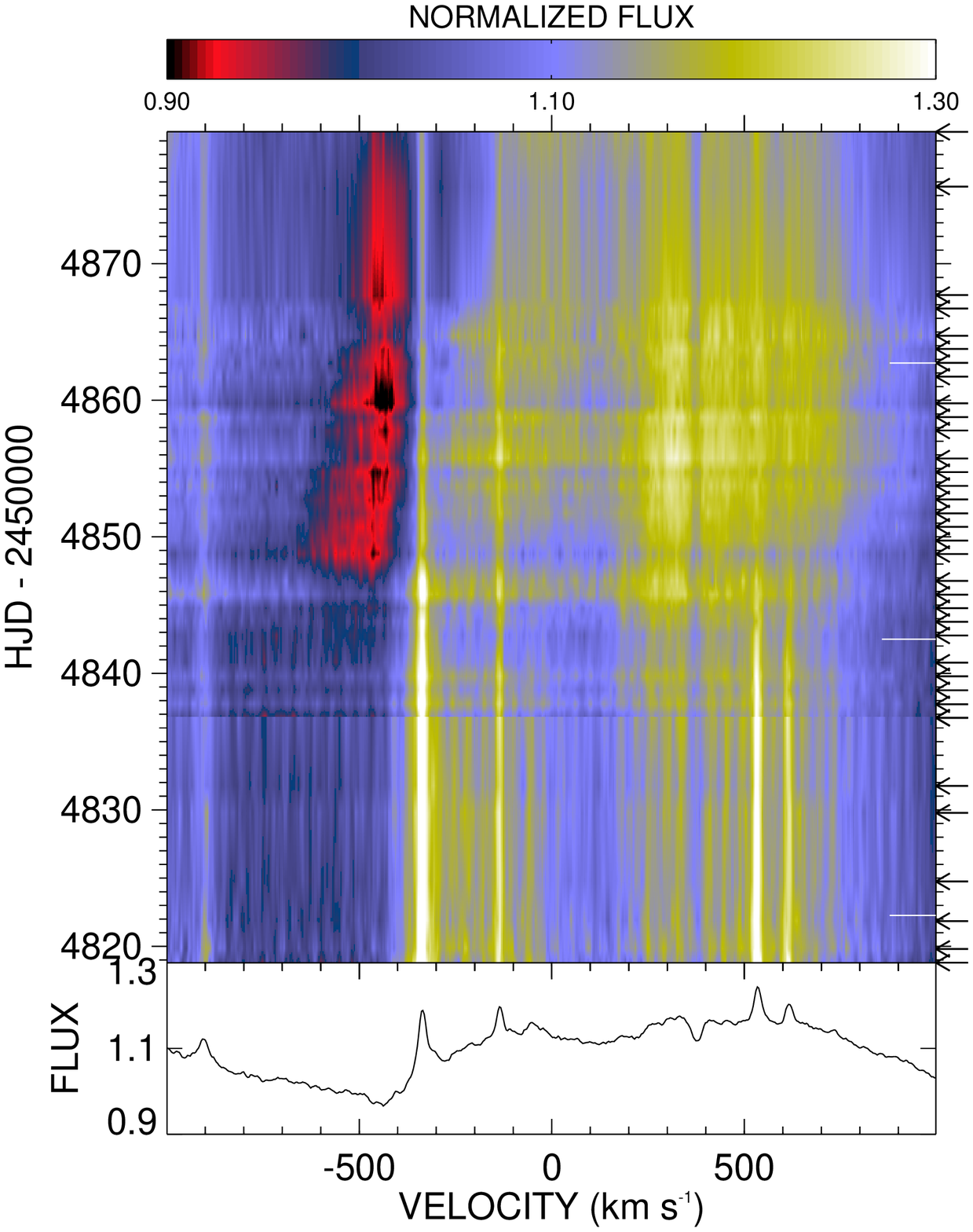}{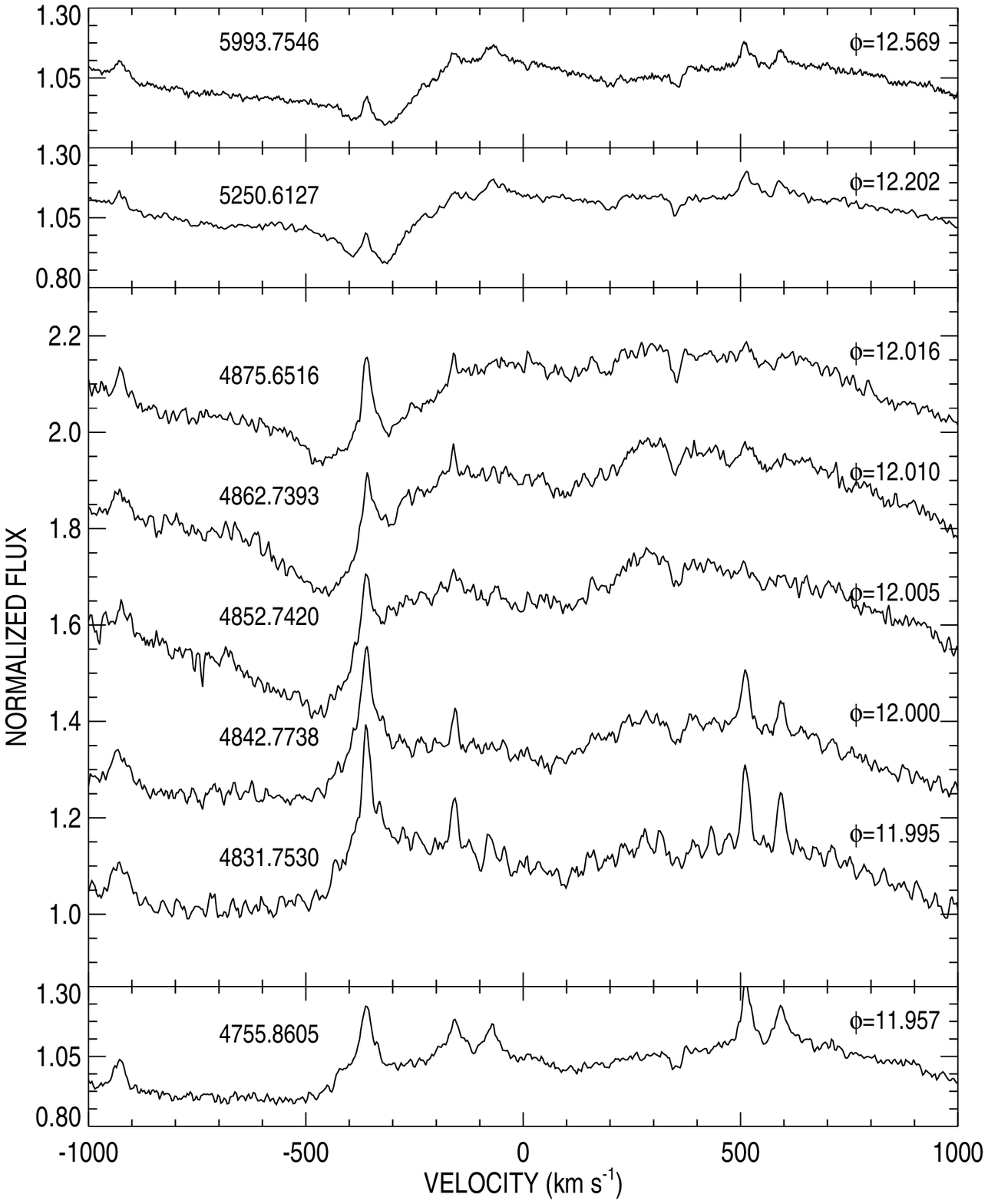}
\caption{Dynamical (left) and line plots (right) of \ion{Si}{2} $\lambda$6371 during the 2009 event. The white horizontal bars on the dynamical representation indicate phases 11.99, 12.00, and 12.01.}
\end{figure}

\clearpage
\setcounter{figure}{1}
\renewcommand{\thefigure}{\arabic{figure}.1}
\begin{figure}
\figurenum{1.15}
\plottwo{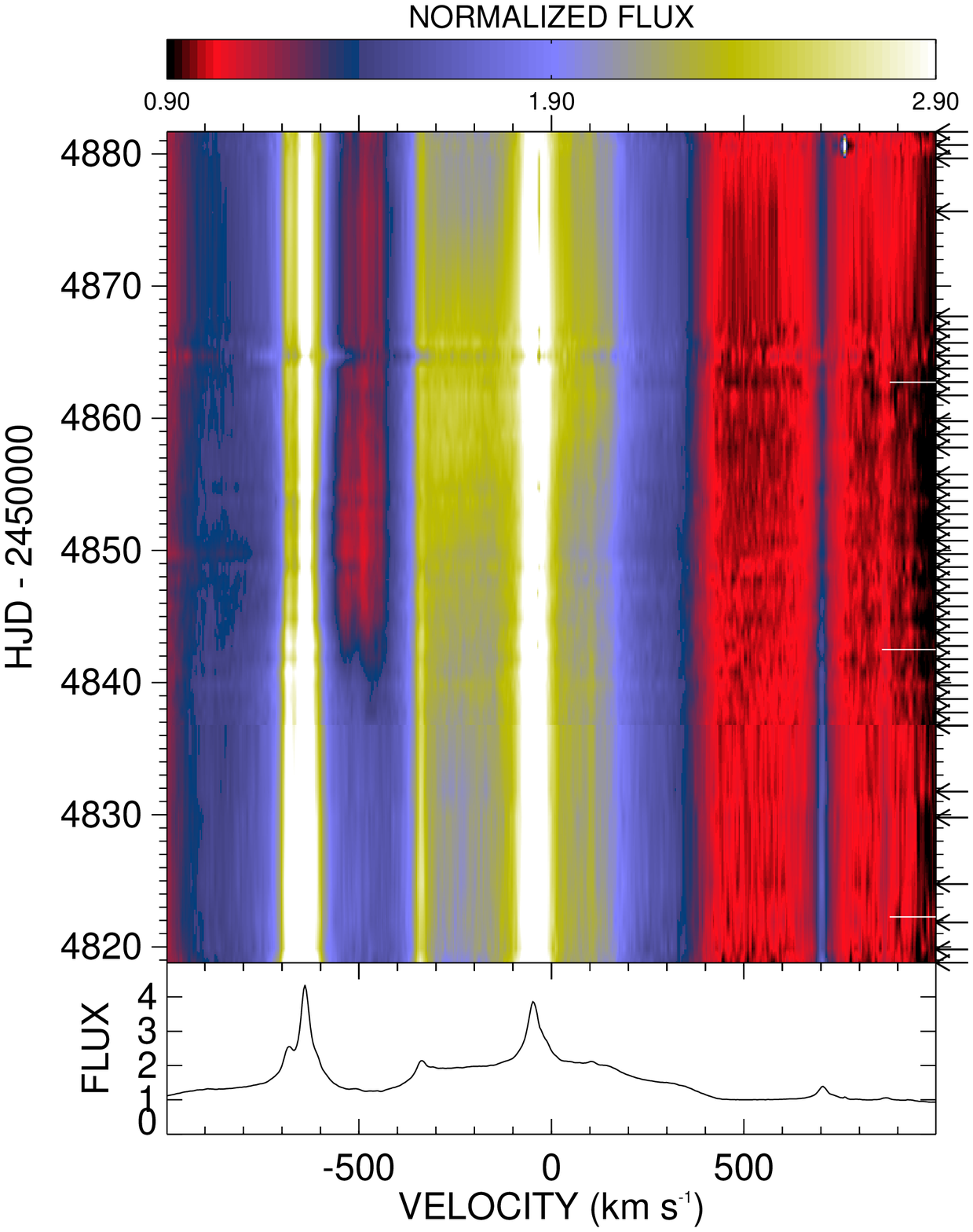}{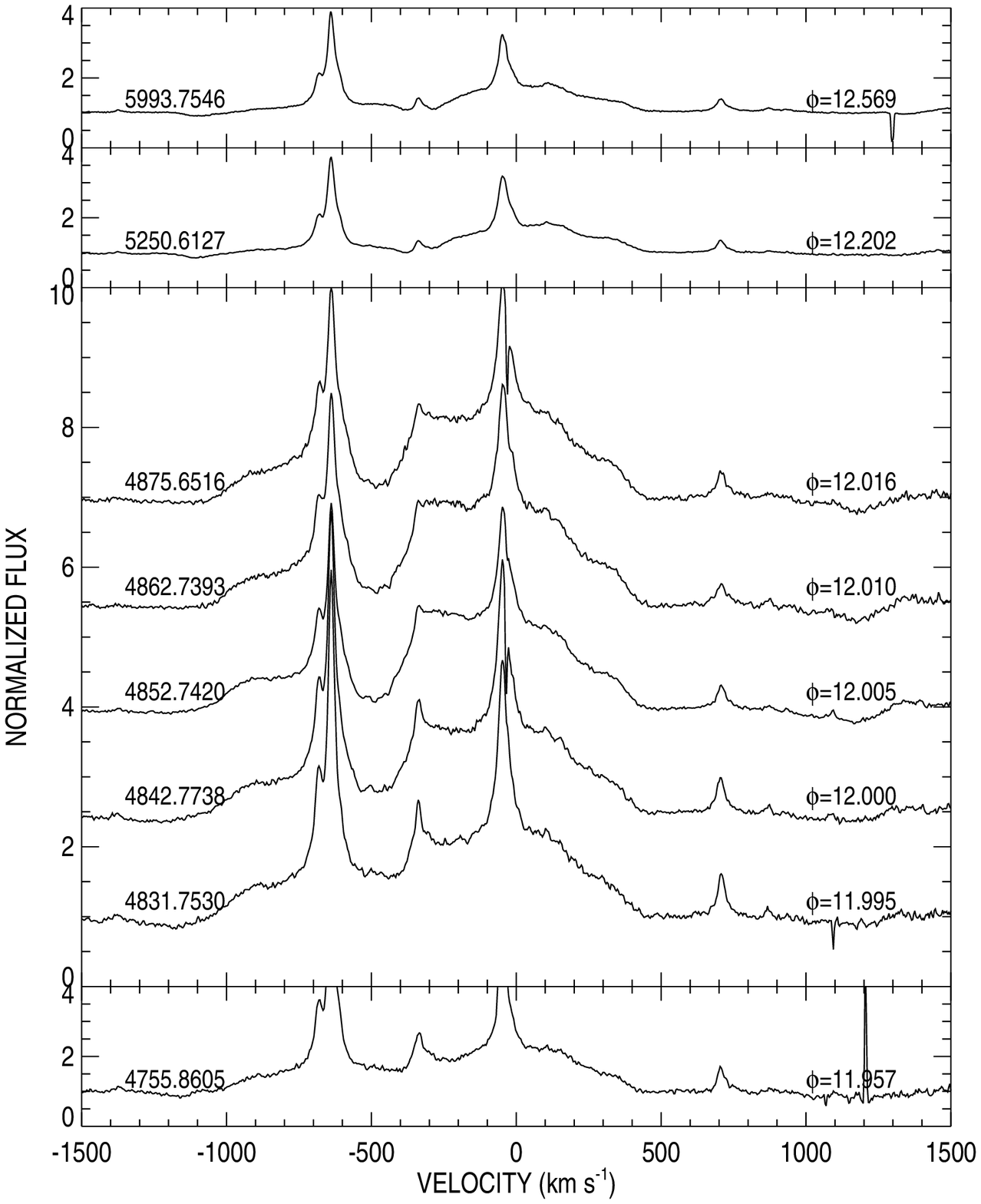}
\caption{Dynamical (left) and line plots (right) of \ion{Fe}{2} $\lambda$5169 during the 2009 event. The white horizontal bars on the dynamical representation indicate phases 11.99, 12.00, and 12.01.}
\end{figure}

\clearpage
\setcounter{figure}{1}
\renewcommand{\thefigure}{\arabic{figure}.1}
\begin{figure}
\figurenum{1.16}
\plottwo{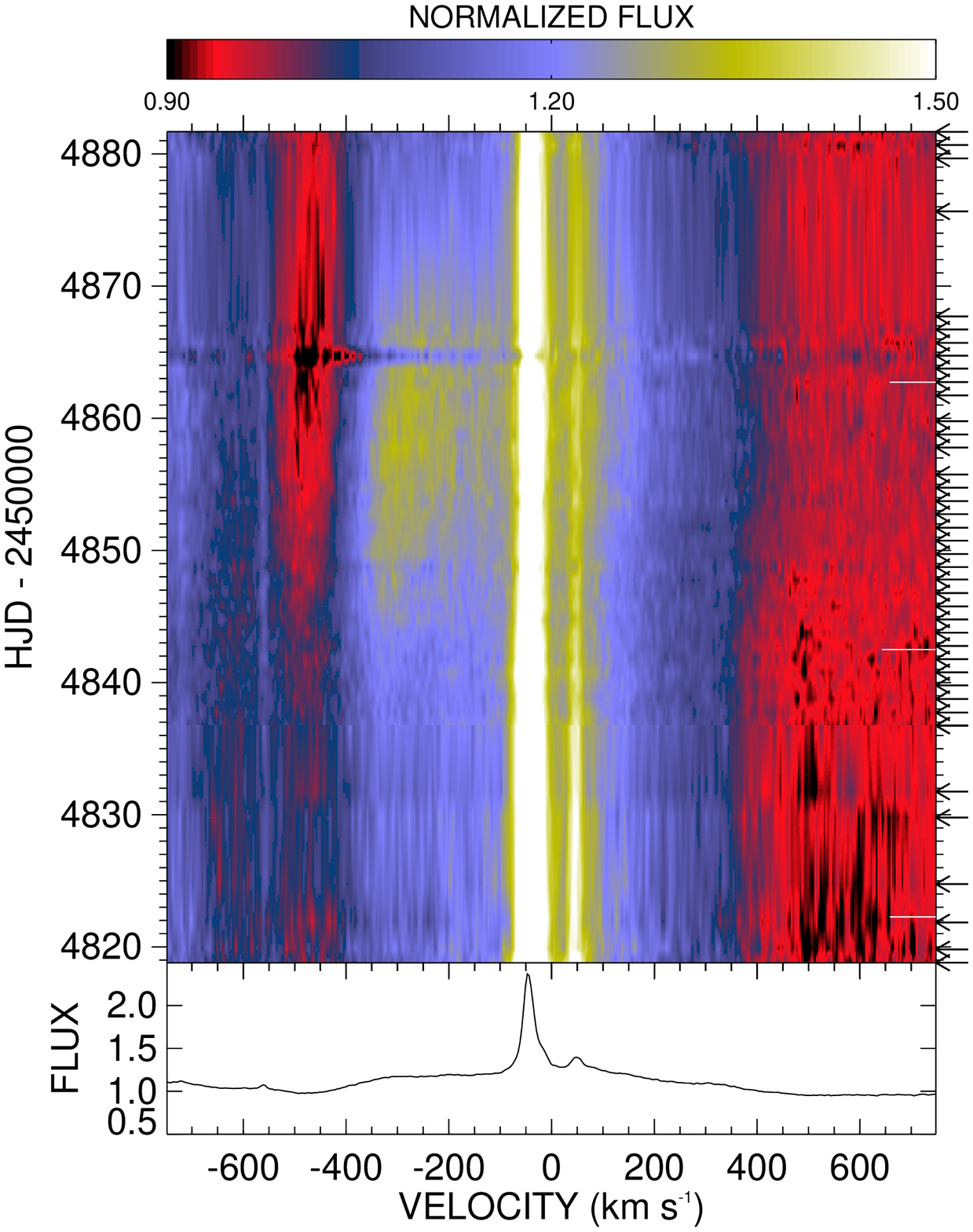}{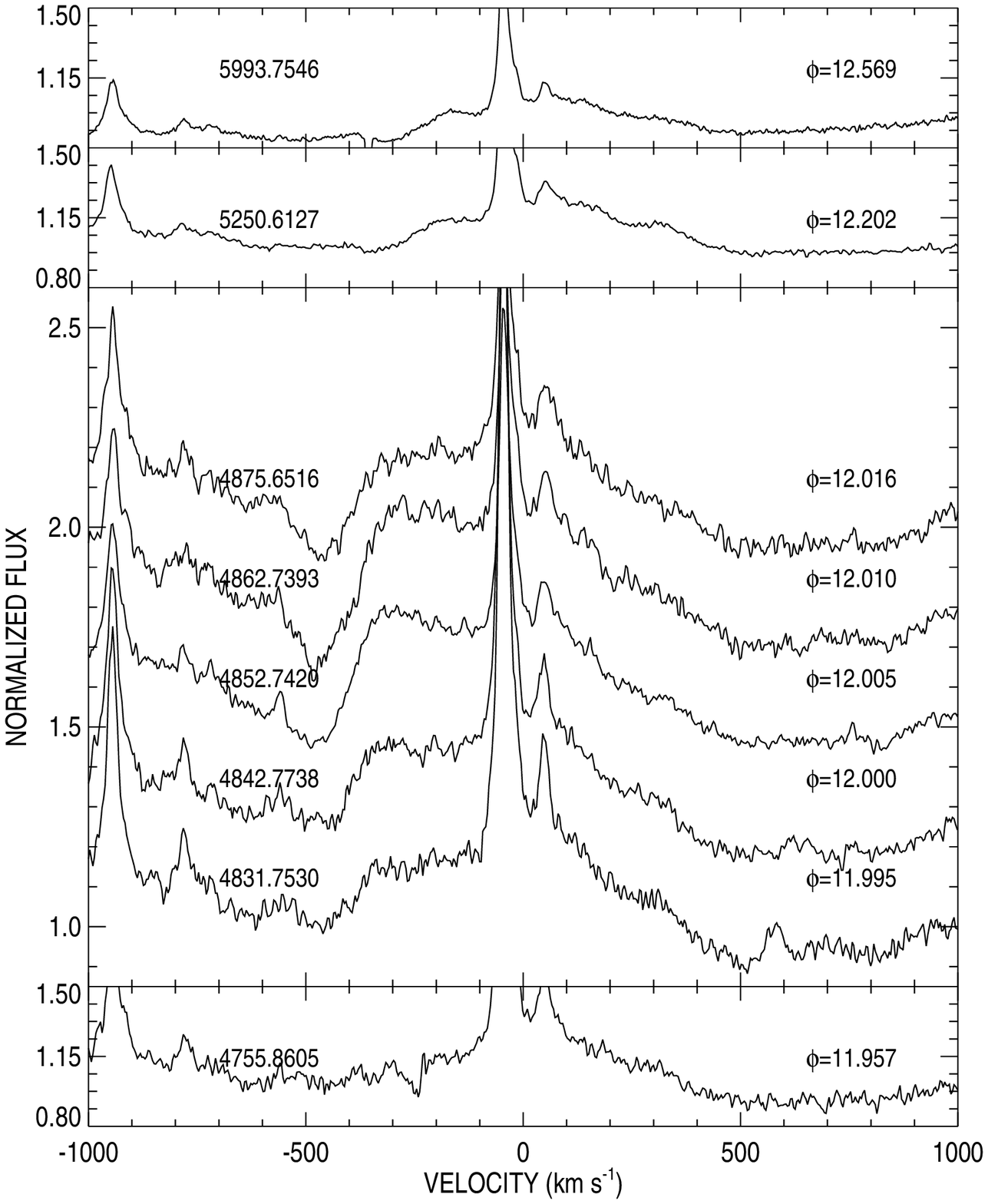}
\caption{Dynamical (left) and line plots (right) of \ion{Fe}{2} $\lambda$5197 during the 2009 event. The white horizontal bars on the dynamical representation indicate phases 11.99, 12.00, and 12.01.}
\end{figure}

\clearpage
\setcounter{figure}{1}
\renewcommand{\thefigure}{\arabic{figure}.1}
\begin{figure}
\figurenum{1.17}
\plottwo{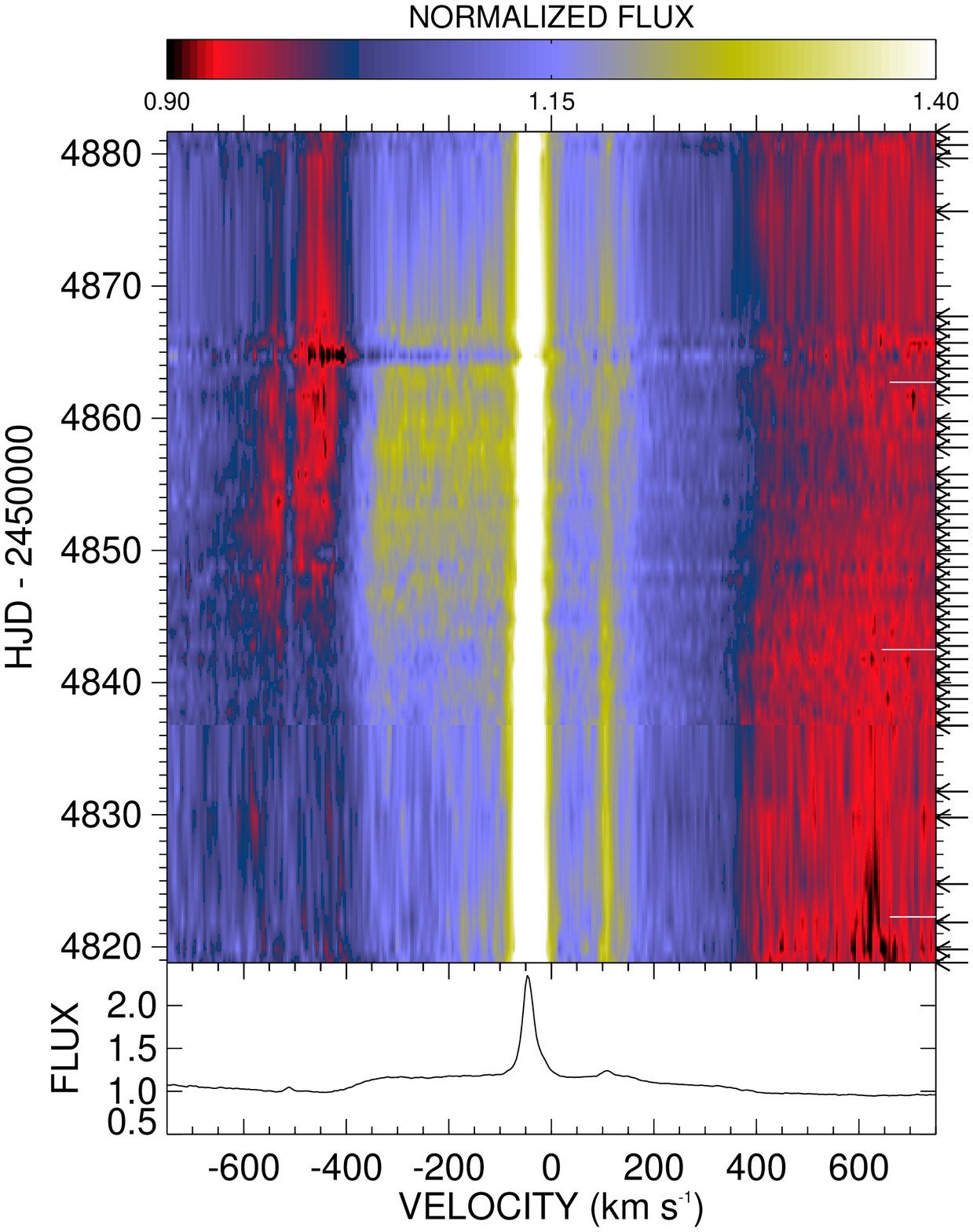}{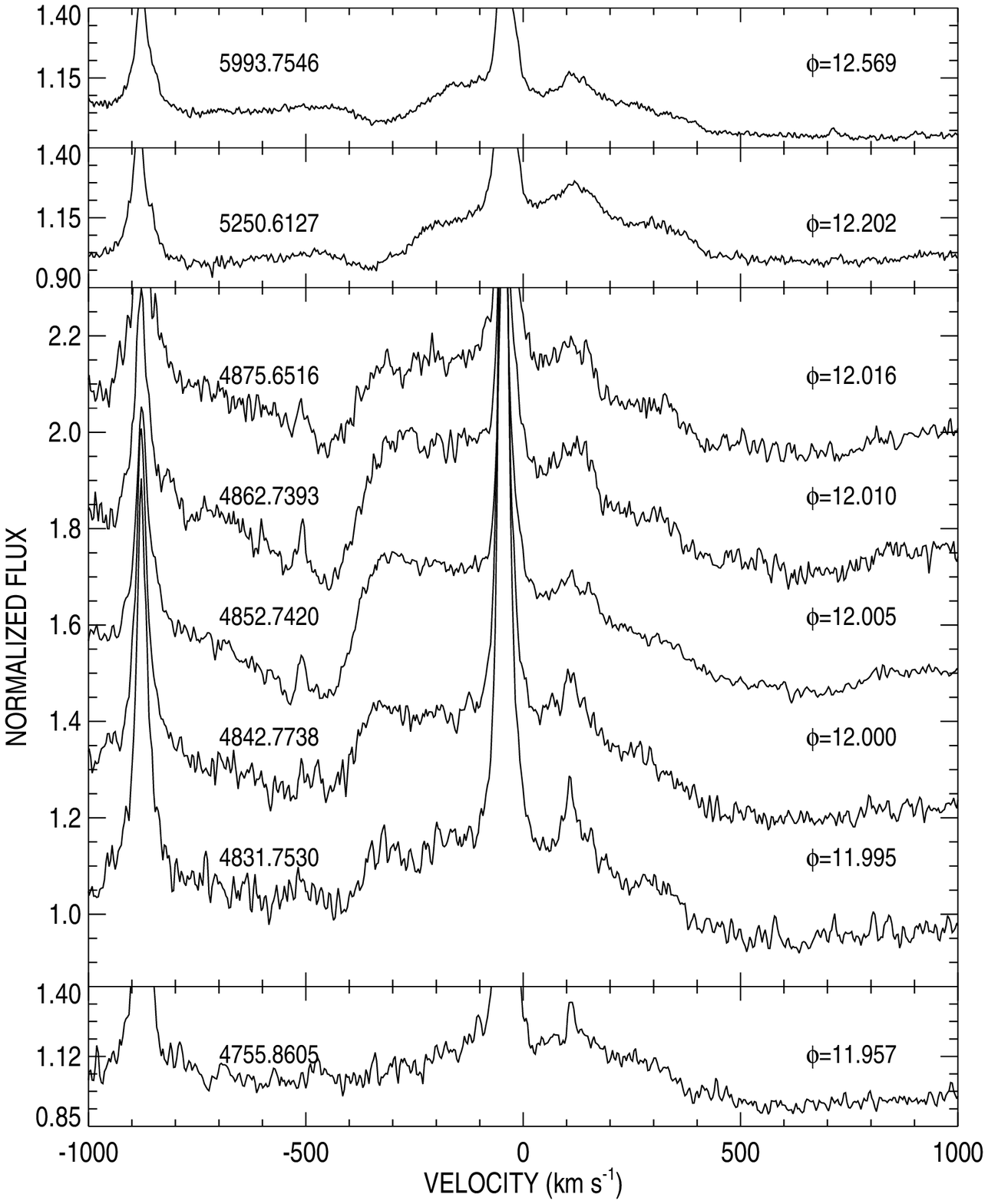}
\caption{Dynamical (left) and line plots (right) of \ion{Fe}{2} $\lambda$5234 during the 2009 event. The white horizontal bars on the dynamical representation indicate phases 11.99, 12.00, and 12.01.}
\end{figure}

\clearpage
\setcounter{figure}{1}
\renewcommand{\thefigure}{\arabic{figure}.1}
\begin{figure}
\figurenum{1.18}
\plottwo{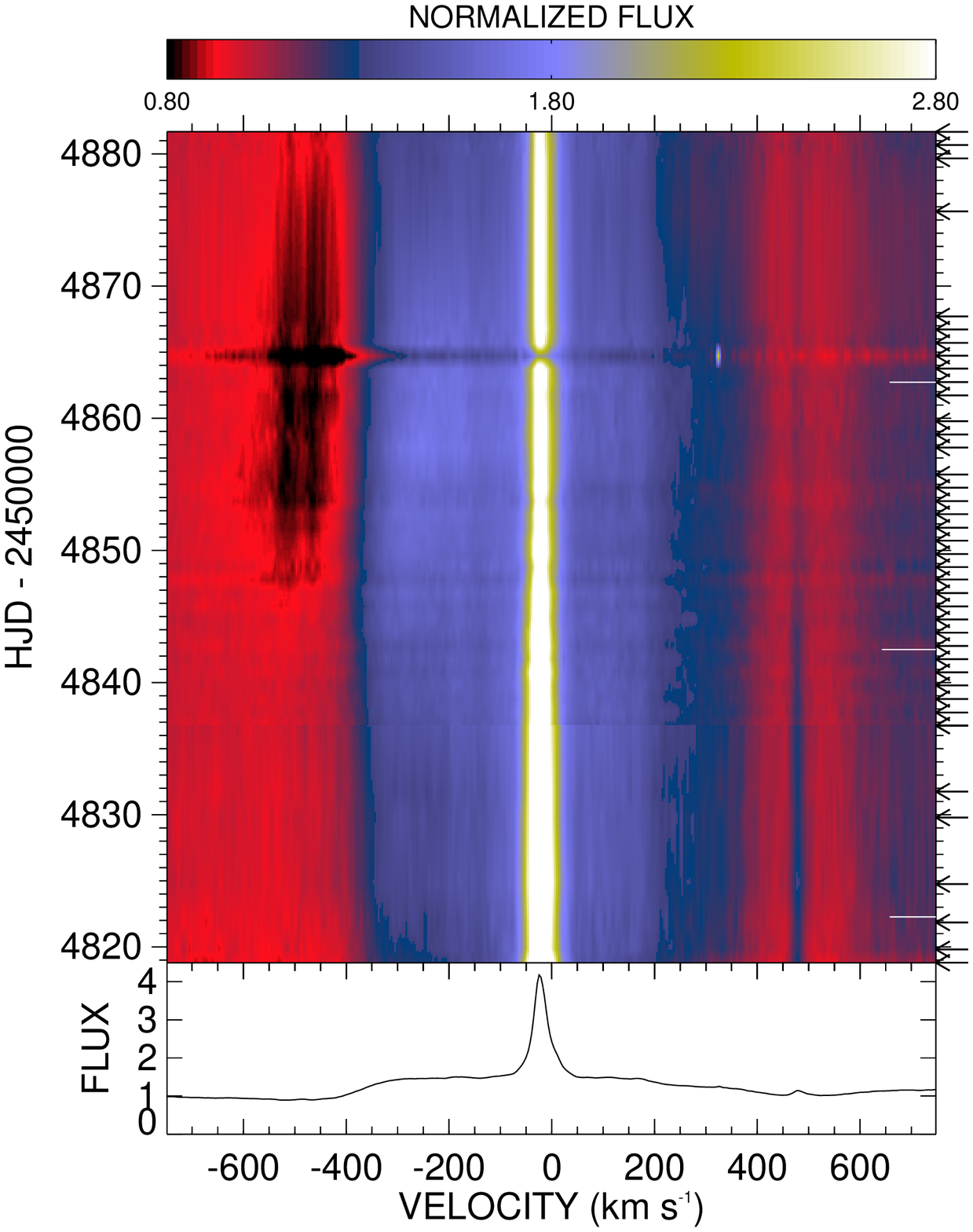}{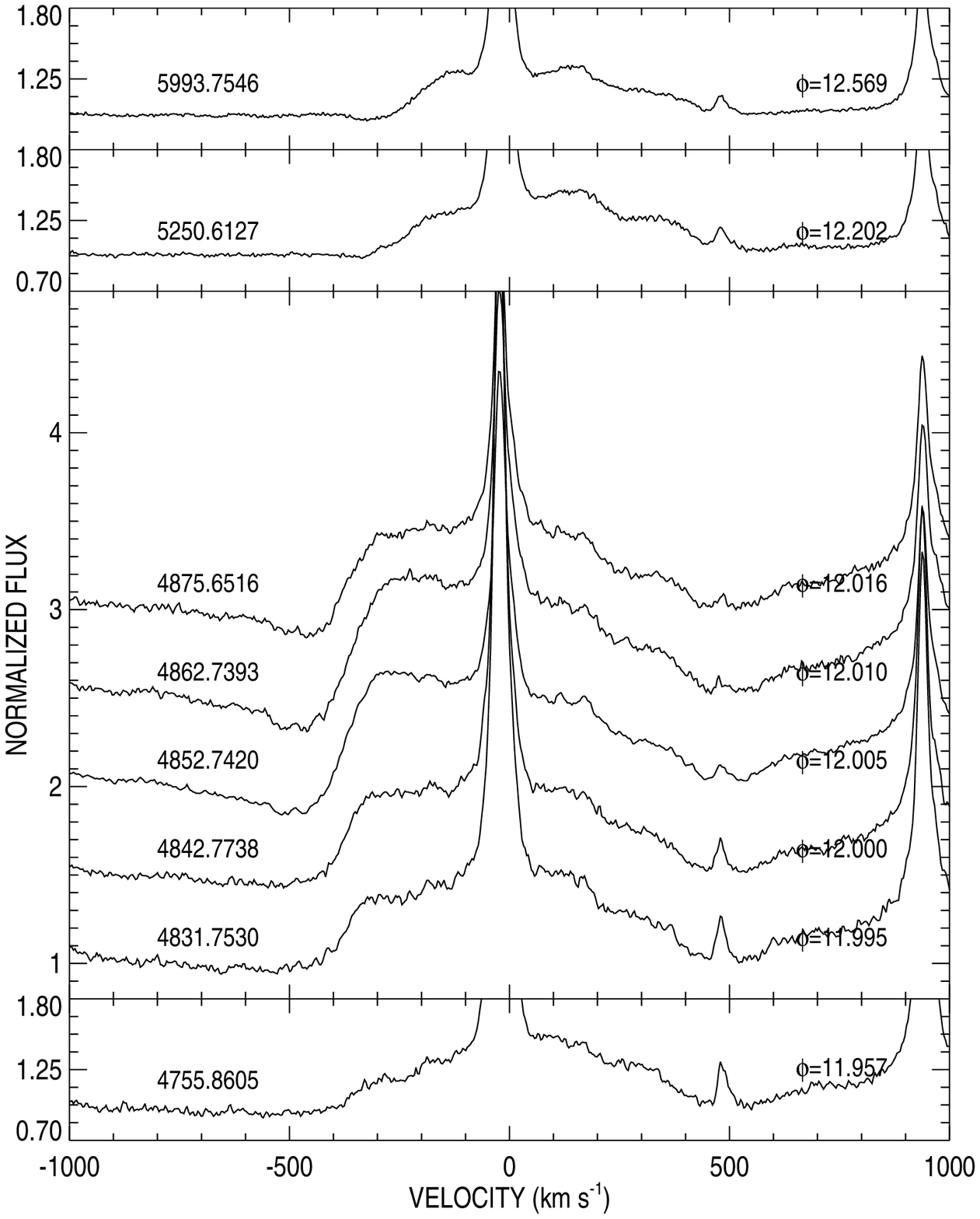}
\caption{Dynamical (left) and line plots (right) of \ion{Fe}{2} $\lambda$5316 during the 2009 event. The white horizontal bars on the dynamical representation indicate phases 11.99, 12.00, and 12.01.}
\end{figure}

\clearpage
\setcounter{figure}{1}
\renewcommand{\thefigure}{\arabic{figure}.1}
\begin{figure}
\figurenum{1.19}
\plottwo{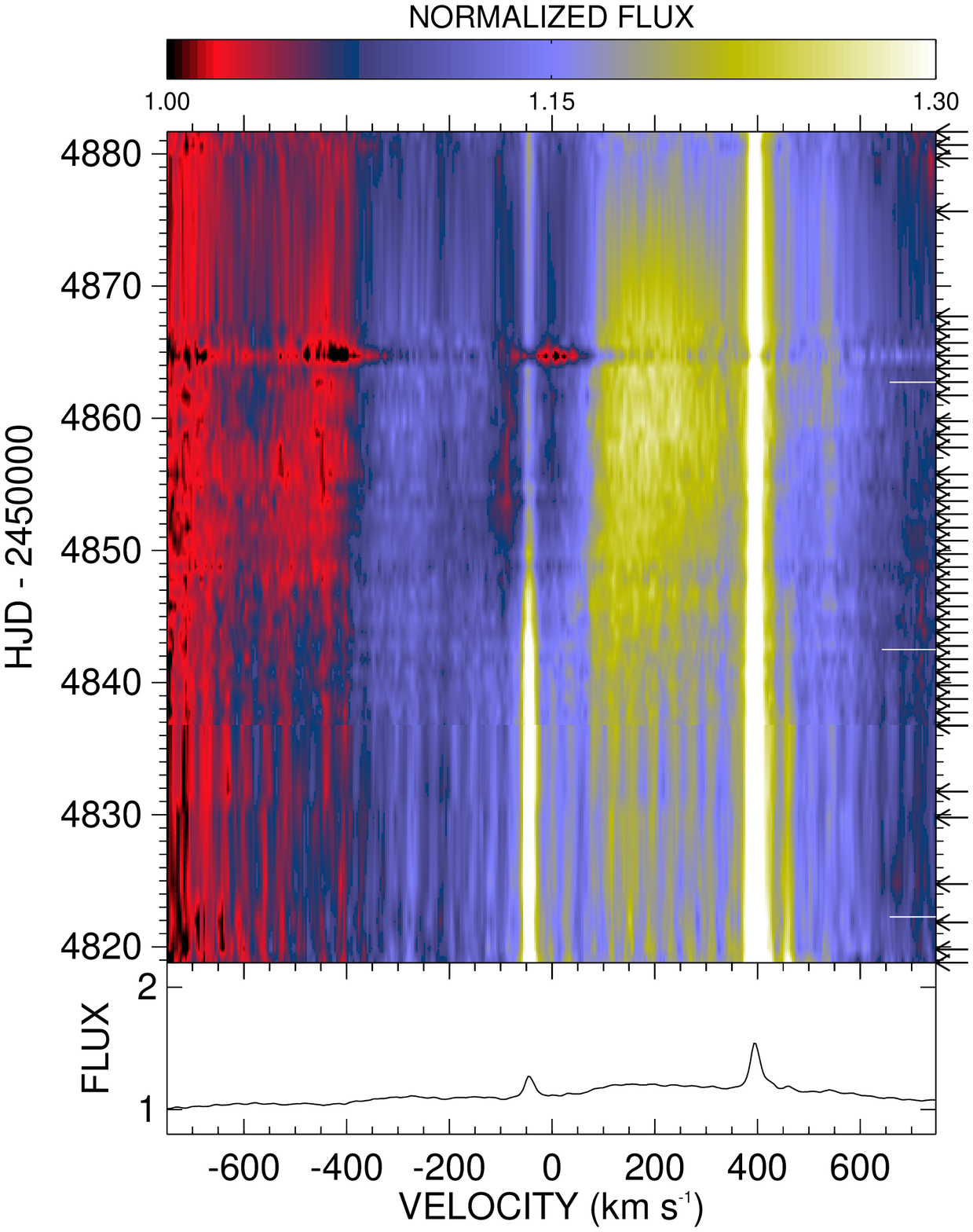}{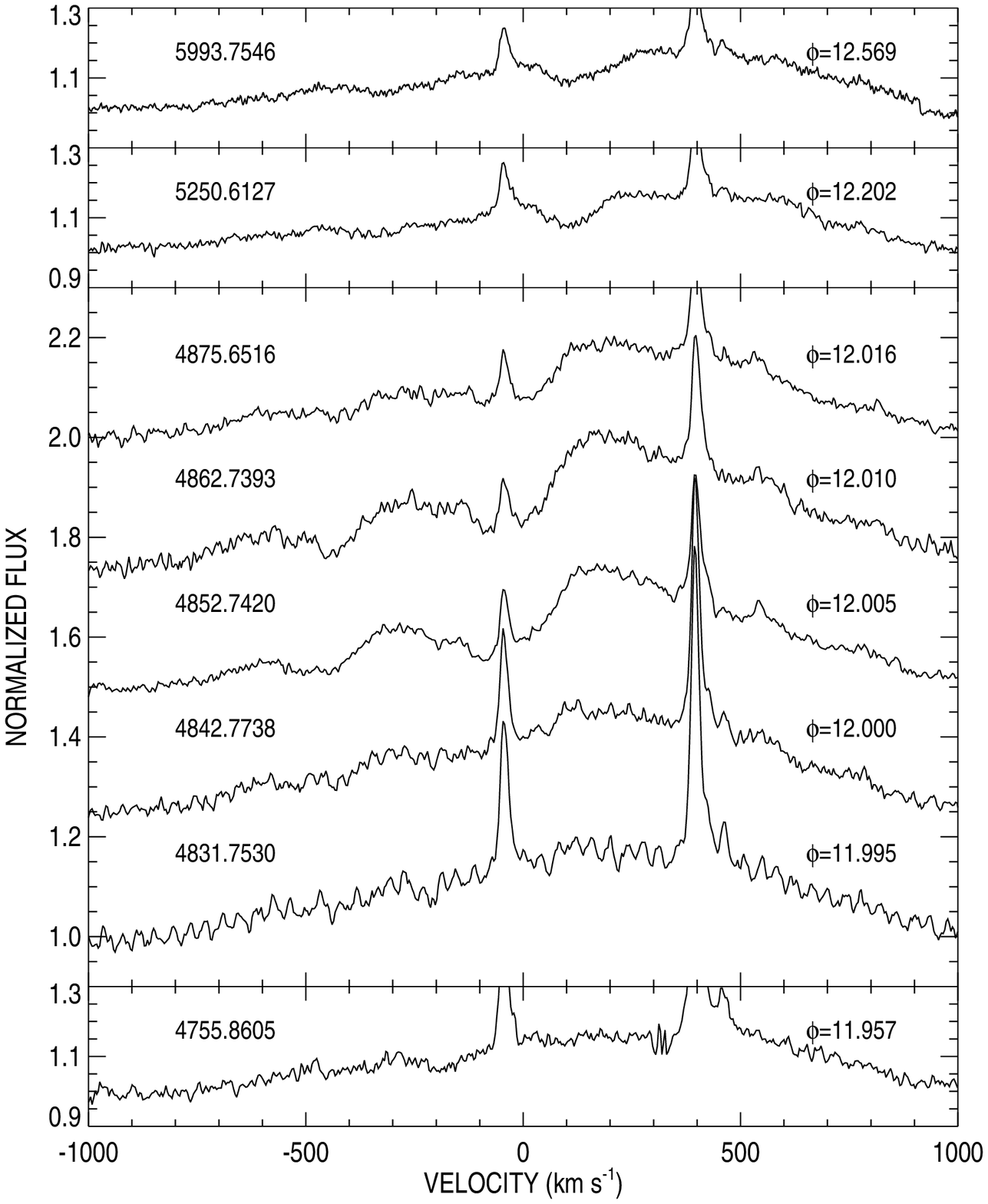}
\caption{Dynamical (left) and line plots (right) of \ion{Fe}{2} $\lambda$6238 during the 2009 event. The white horizontal bars on the dynamical representation indicate phases 11.99, 12.00, and 12.01.}
\end{figure}

\clearpage
\setcounter{figure}{1}
\renewcommand{\thefigure}{\arabic{figure}.1}
\begin{figure}
\figurenum{1.20}
\plottwo{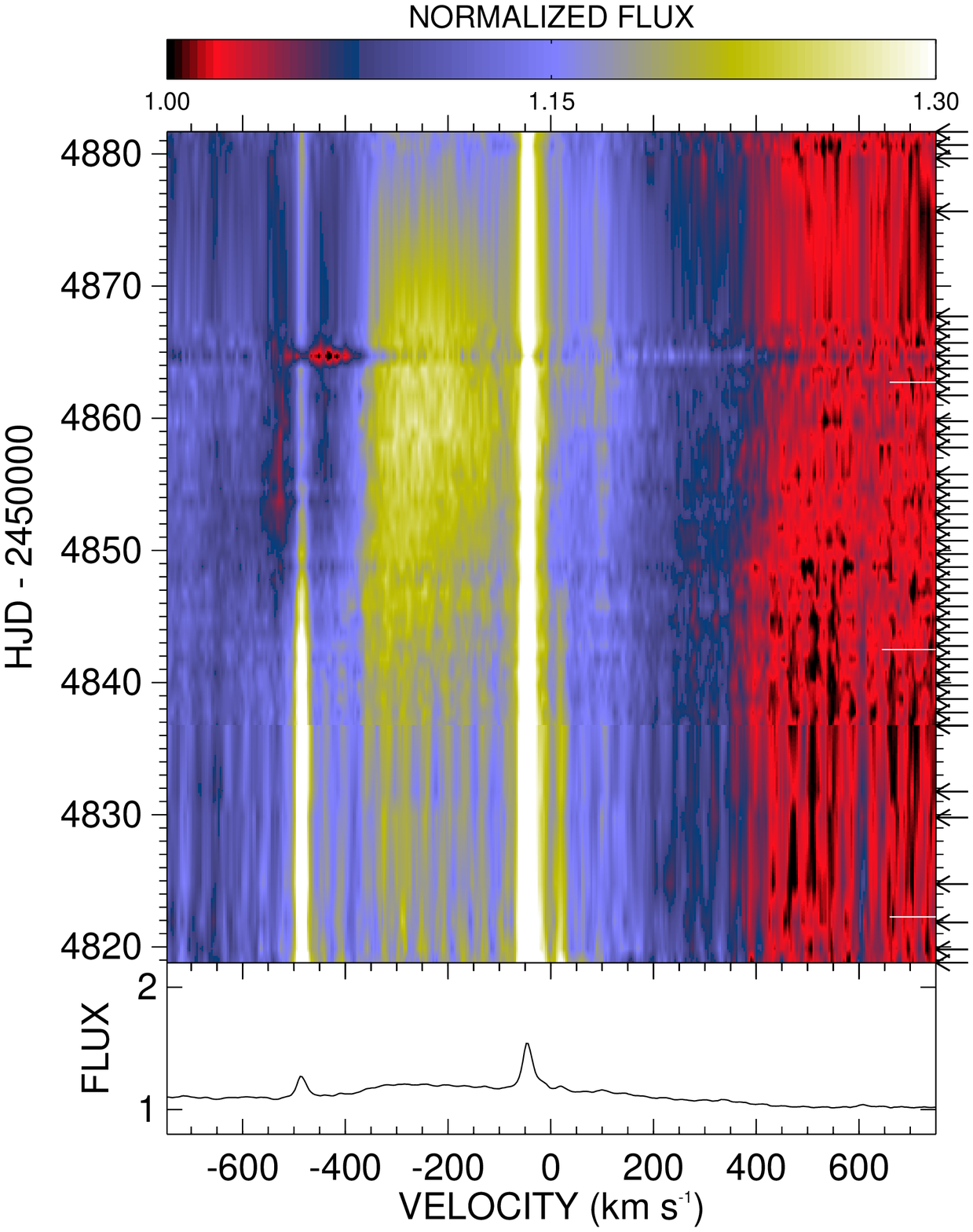}{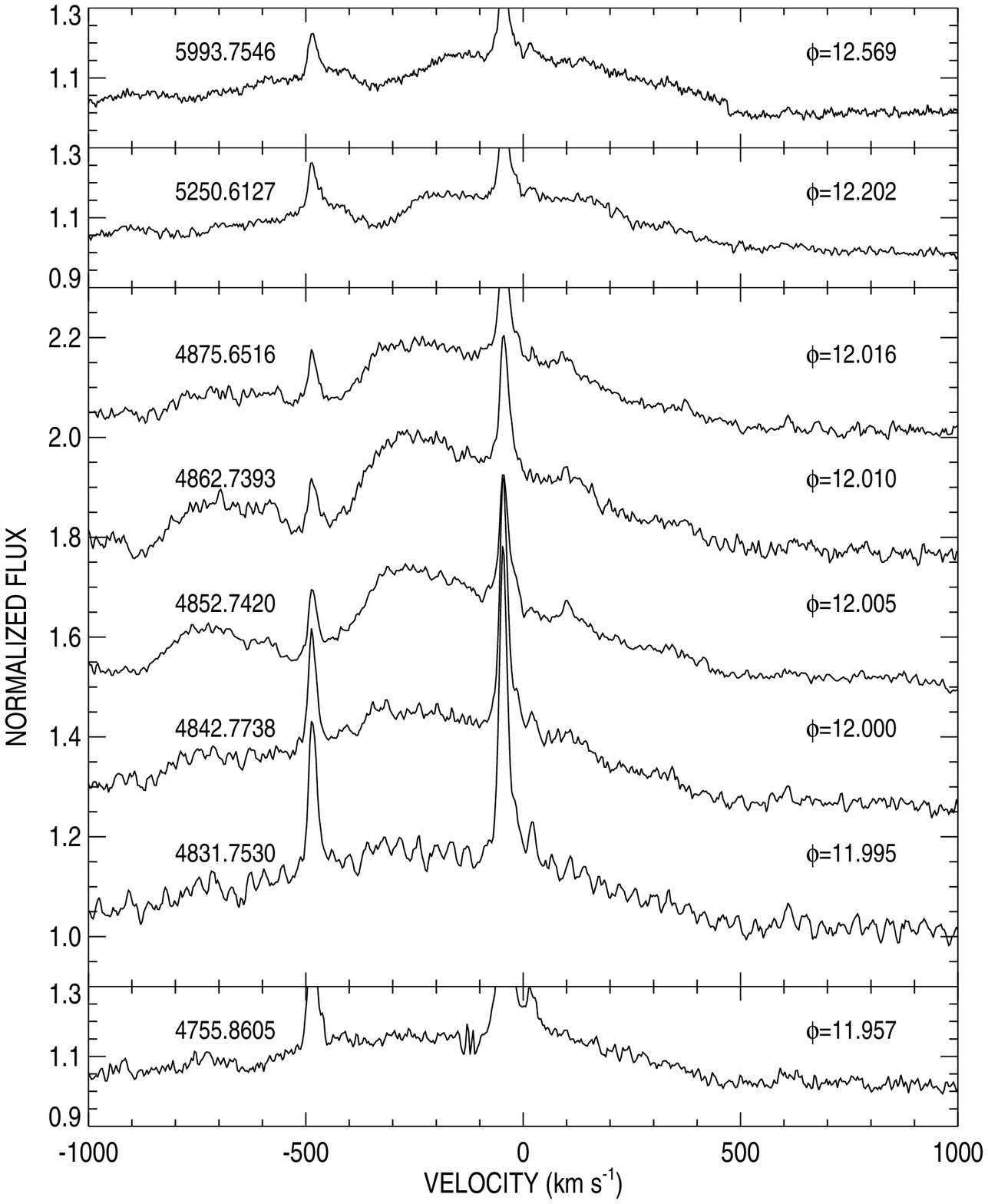}
\caption{Dynamical (left) and line plots (right) of \ion{Fe}{2} $\lambda$6248 during the 2009 event. The white horizontal bars on the dynamical representation indicate phases 11.99, 12.00, and 12.01.}
\end{figure}

\clearpage
\setcounter{figure}{1}
\renewcommand{\thefigure}{\arabic{figure}.21}
\begin{figure}
\figurenum{1.21}
\plottwo{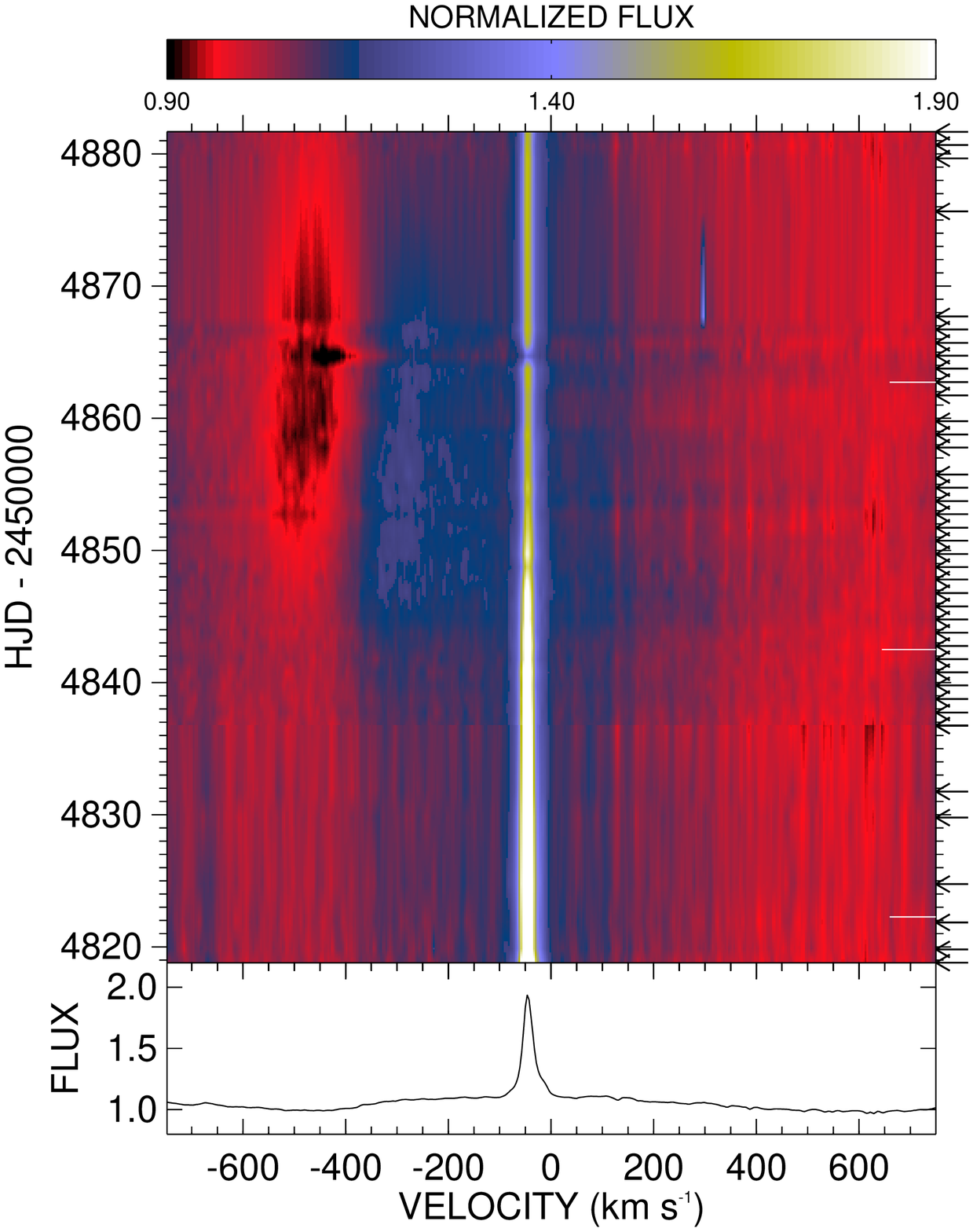}{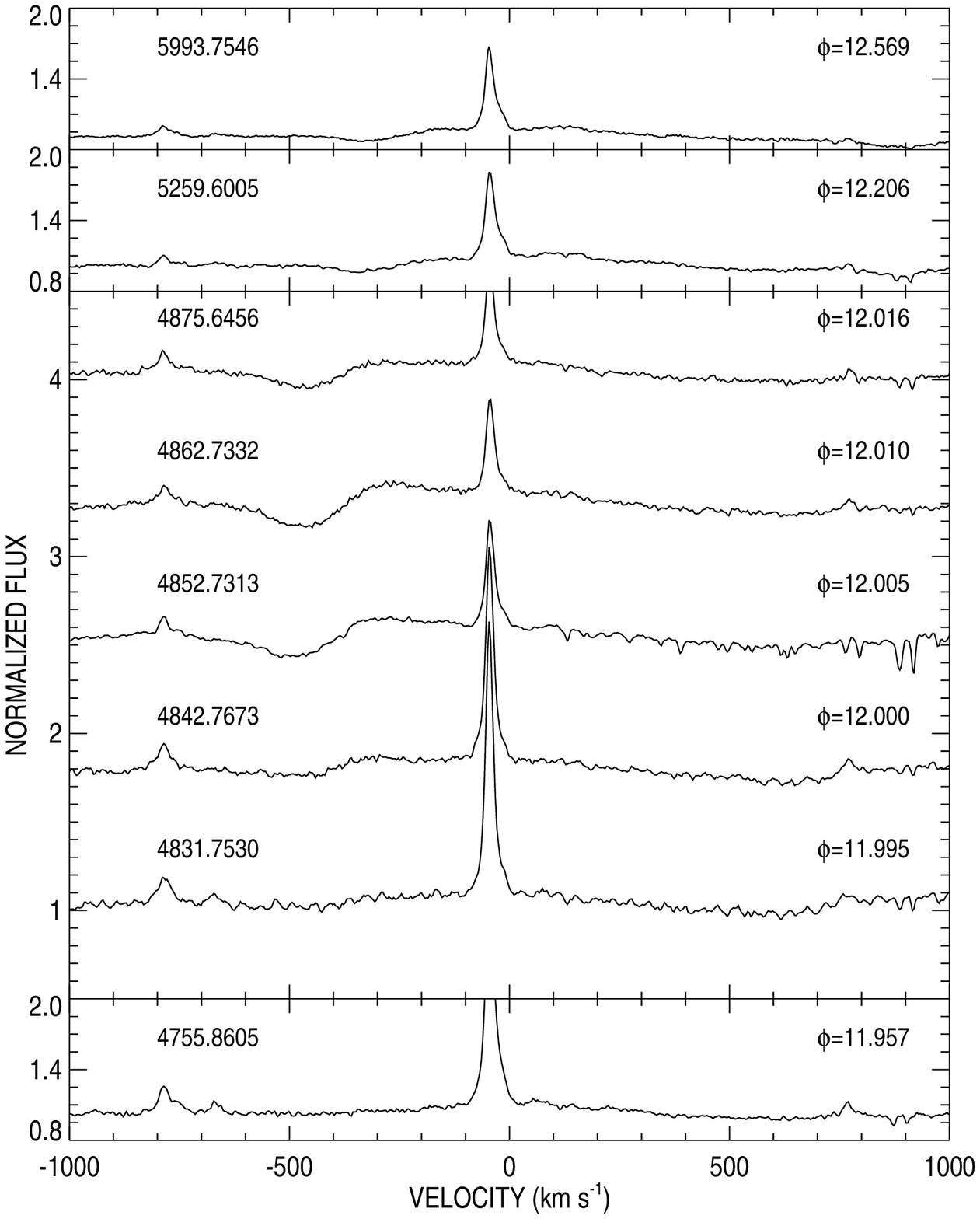}
\caption{Dynamical (left) and line plots (right) of \ion{Fe}{2} $\lambda$6456 during the 2009 event. The white horizontal bars on the dynamical representation indicate phases 11.99, 12.00, and 12.01.}
\end{figure}


\clearpage
\begin{figure}
\figurenum{2}
\epsscale{0.85}
\label{fig10}
\includegraphics[angle=0,width=5.5in]{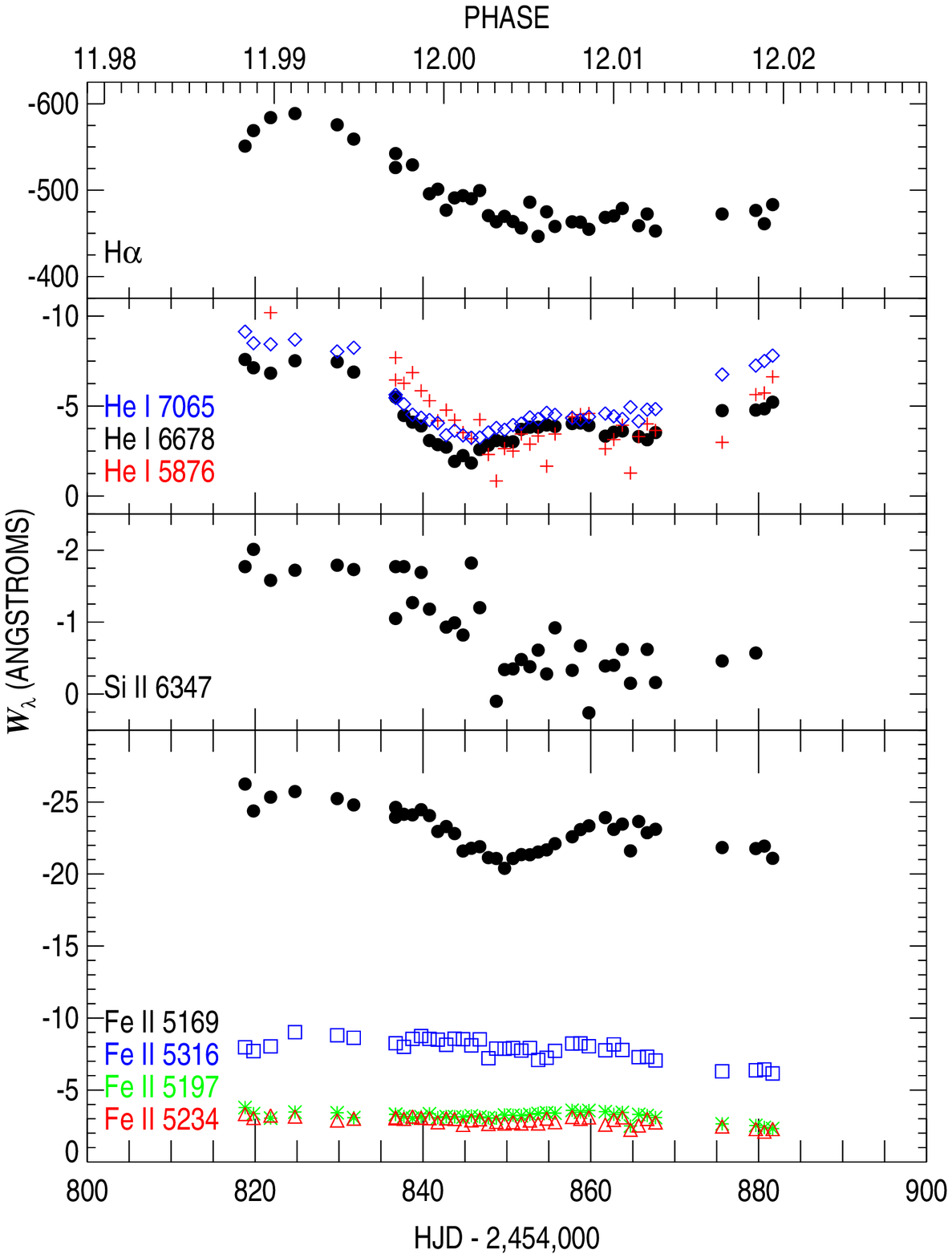}
\caption{Equivalent width measurements from our echelle data across the 2009 event. }
\end{figure}

\clearpage
\begin{figure}

\figurenum{3}
\epsscale{0.85}
\label{fig12}
\includegraphics[angle=0,width=5in]{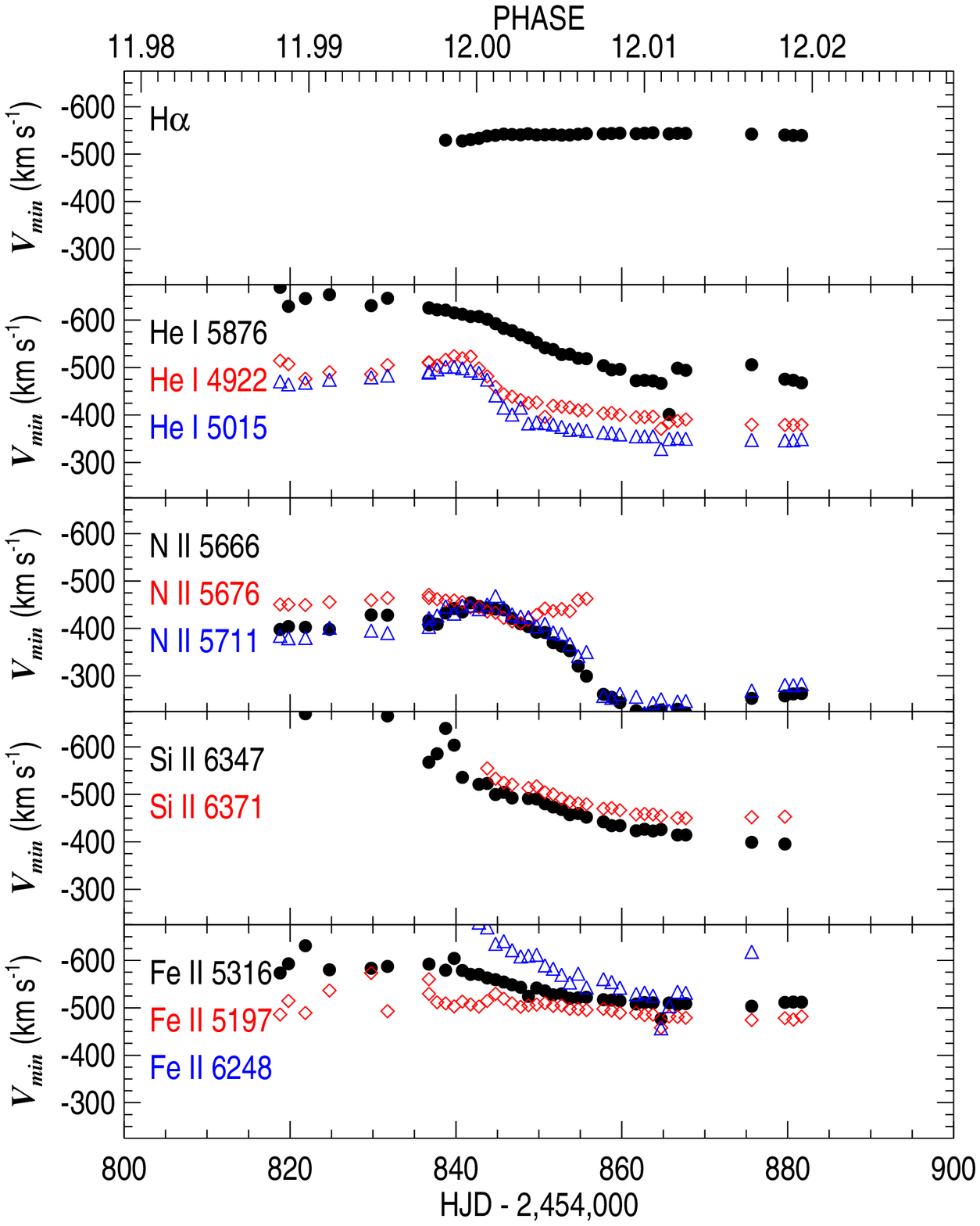}
\caption{Measurements of radial velocities of the P Cygni absorption components, $V_{\rm min}$. }
\end{figure}

\clearpage
\begin{figure}
\figurenum{4}
\label{fig11}
\includegraphics[angle=0,width=5in]{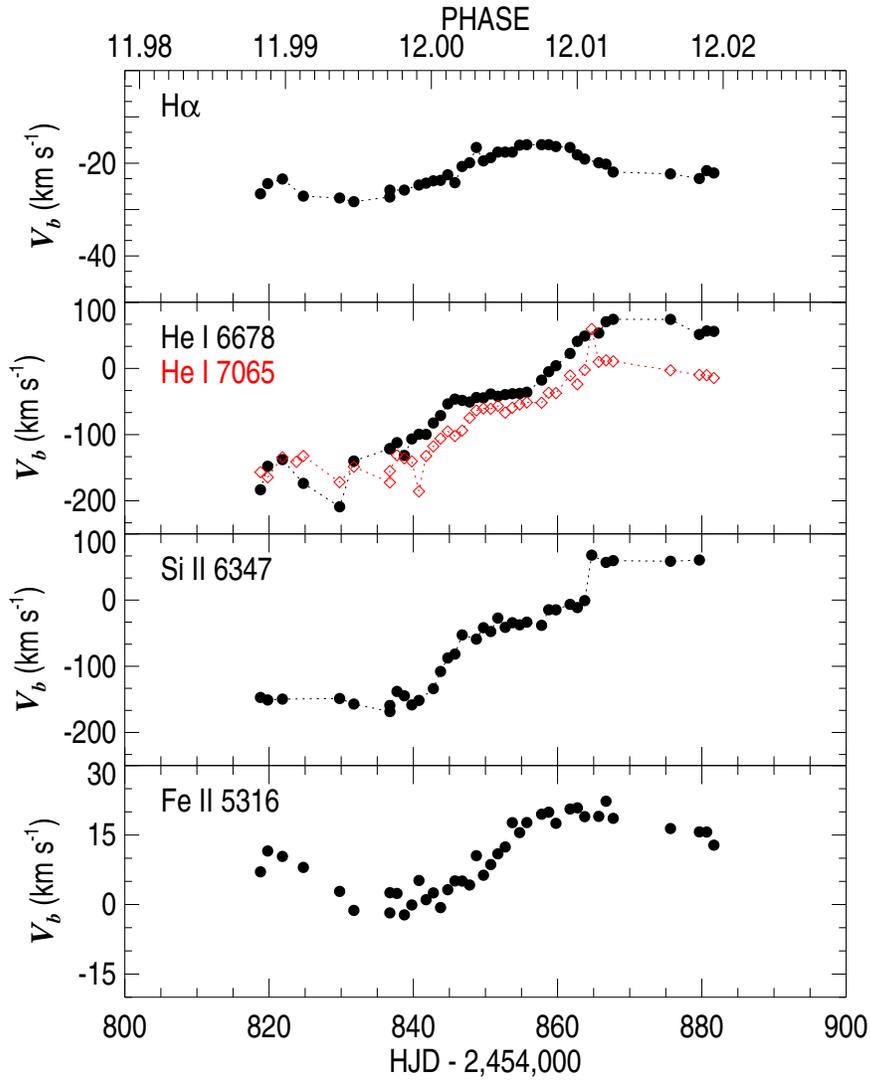}
\caption{Measurements of emission line bisector radial velocities, $V_b$. The bisector velocities for H$\alpha$ from Paper 1 are shown for comparison, and each emission line is marked.}
\end{figure}

\clearpage
\begin{figure}
\figurenum{5}
\epsscale{0.85}
\label{fig14}
 \scalebox{0.7}{\includegraphics{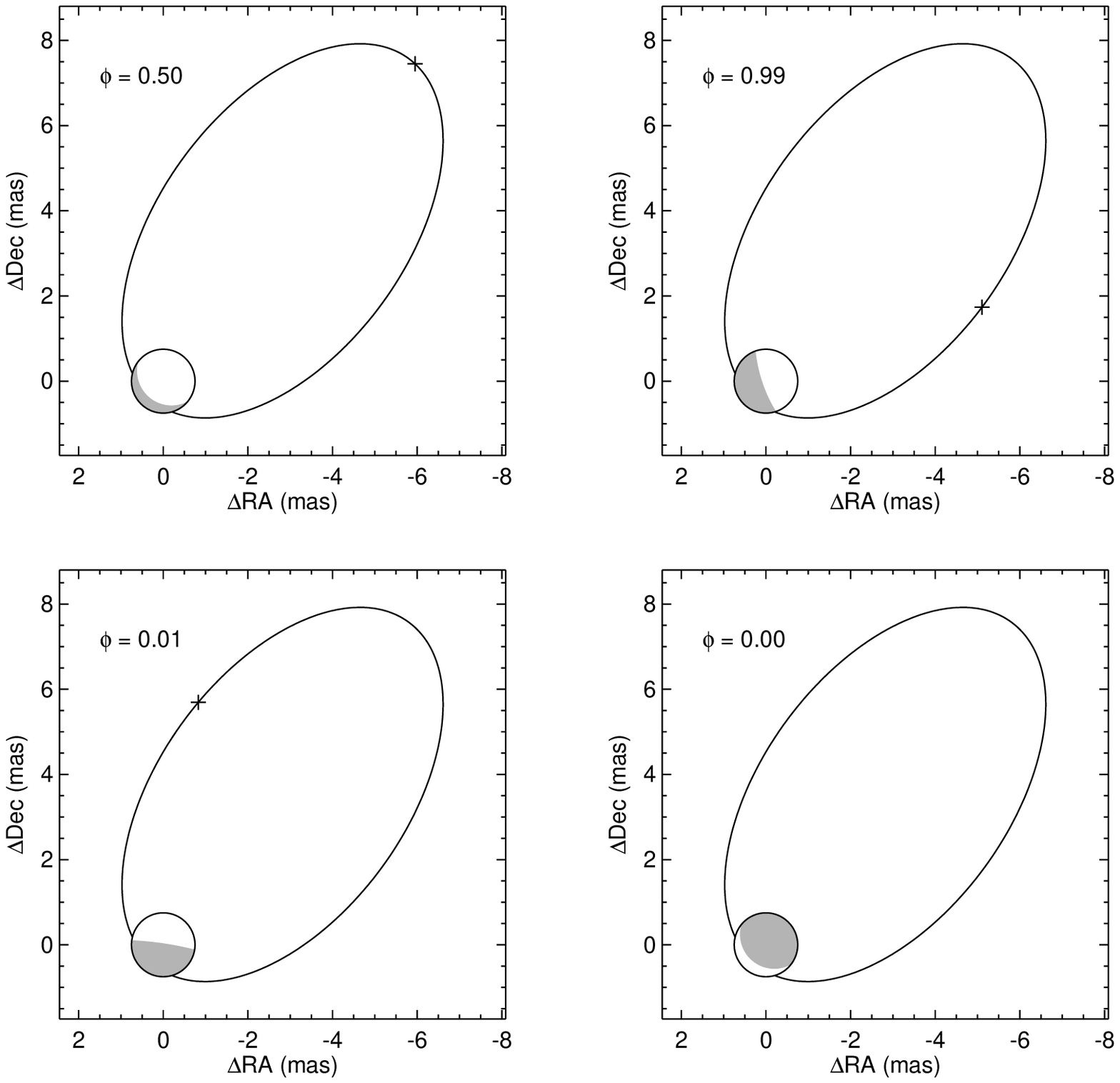}}
\caption{A depiction of the relative astrometric orbit of the companion in the plane of the sky.  The four panels show the orientation of the primary and secondary stars at different orbital phases (arranged in a clockwise order), where the primary star's wind is illuminated by the secondary (white is illuminated, grey is not). }
\end{figure}

\clearpage
\begin{figure}
\figurenum{6}
\epsscale{0.85}
\label{fig14}

 \scalebox{0.3}{\includegraphics[angle=90]{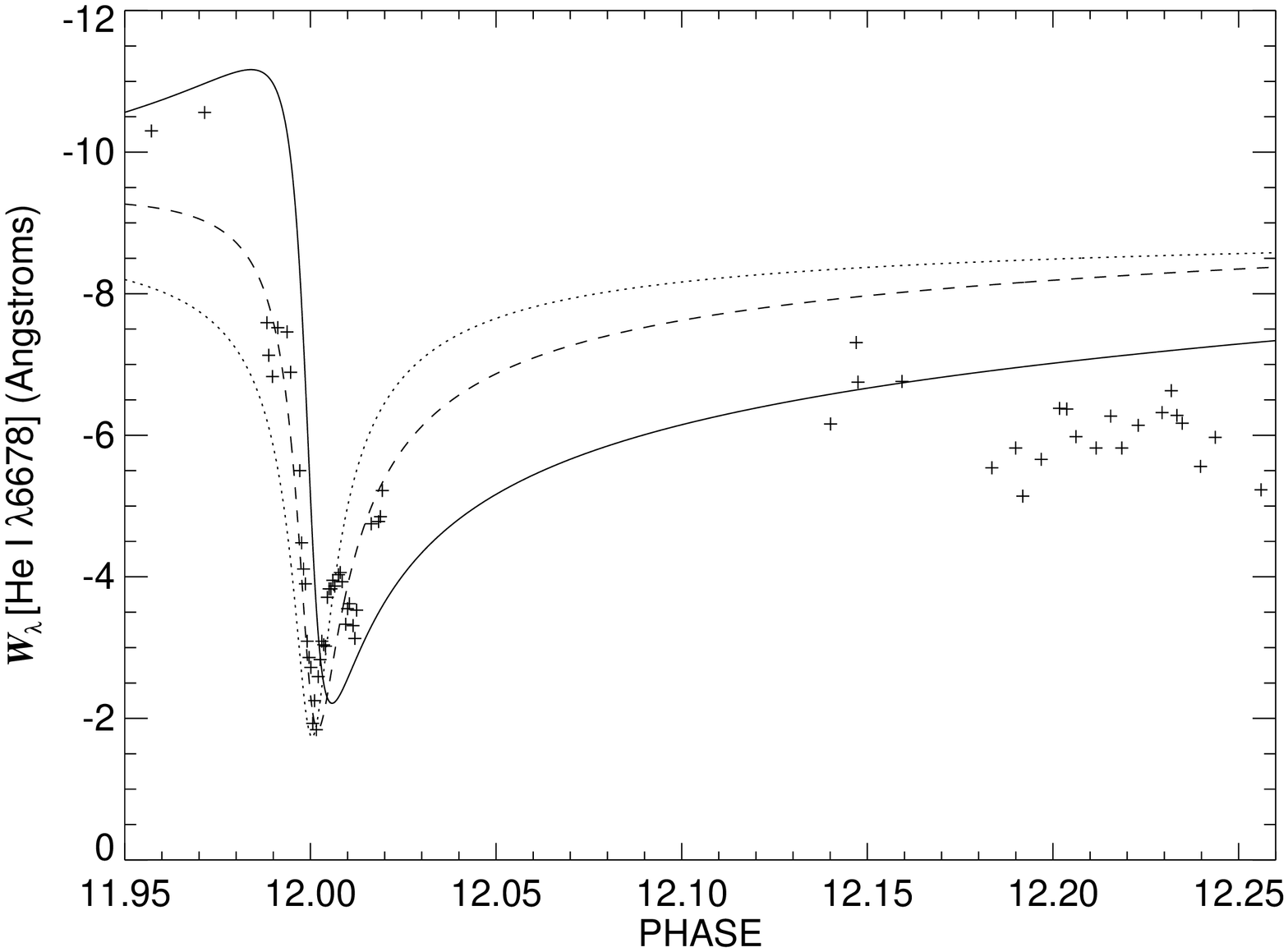}}
 \scalebox{0.3}{\includegraphics[angle=90]{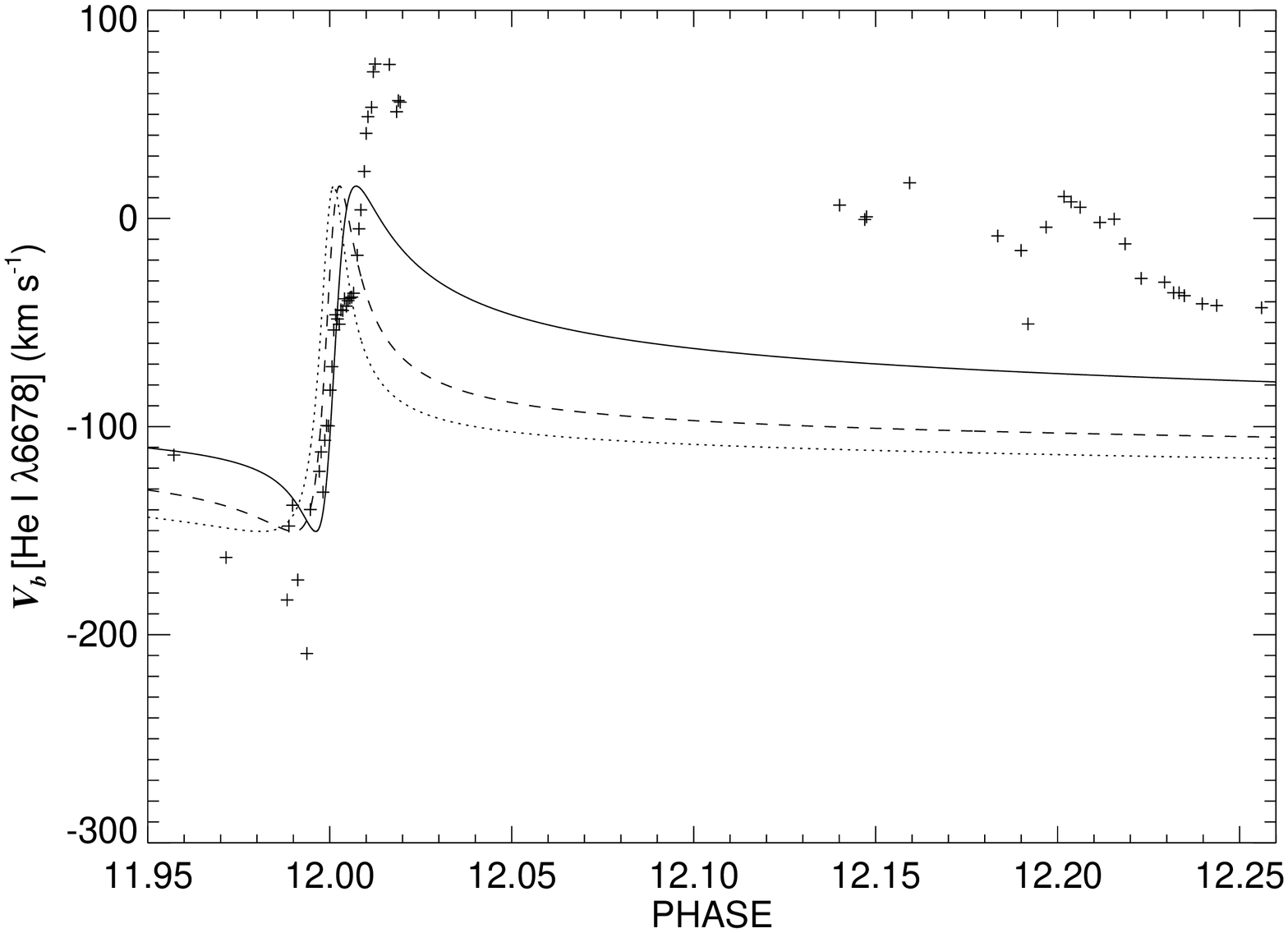}}
\caption{The temporal variations in the \ion{He}{1} $\lambda 6678$ equivalent width
(left panel) and emission peak bisector velocity (right panel) compared
with predictions of the illuminated hemisphere model.  The three lines
correspond to trial models with a longitude of periastron $\omega_p$
equal to $200^\circ$ (dotted line), $240^\circ$ (dashed line), and
$263^\circ$ (solid line).  Deviations from the fits at later time (phase 12.2)
may be related to the presence then of strong P~Cygni absorption which
reduces the emission equivalent width and shifts the emission line bisector
to more positive values.
 }
\end{figure}

\clearpage
\begin{deluxetable}{ccccccccc}
\rotate
\label{tab-atomic}
\tablecaption{Optical Wind Line Parameters\tablenotemark{a}}
\tablewidth{0pt}
\tablehead{
	\colhead{}	&
	\colhead{Rest}	&	
	\colhead{Ionization} &
	\colhead{Energy}	&
	\colhead{Energy}	&
	\colhead{Blue Int.}	&
	\colhead{Red Int.}	&
	\colhead{Figure} 		&
	\colhead{}		\\	

	\colhead{}	&
	\colhead{Wavelength}	&	
	\colhead{Energy} &
	\colhead{$E_{\rm lower}$}	&
	\colhead{$E_{\rm upper}$}		&
	\colhead{Limit}	&
	\colhead{Limit}	&
	\colhead{Set} 		&
	\colhead{Line}	\\	

	\colhead{Species}	&
	\colhead{(\AA)}	&	
	\colhead{(eV)}	&
	\colhead{(eV)}	&
	\colhead{(eV)}		&
	\colhead{(km s$^{-1}$)}	&
	\colhead{(km s$^{-1}$)}	&
	\colhead{Number} 		&
	\colhead{Blends\tablenotemark{b}}	}		
\startdata
\ion{H}{1}(H$\beta$)	&	4861.33		&	13.598	&	10.2028	&	12.7535 	&	\nodata	&	\nodata	&	1	&	W, LHA	\\
\ion{H}{1}(H$\alpha$) &	6562.82		&	13.598	& 	10.2028	 &	12.0922 	&	$-2500$	&	$+2500$	&	2	&	W, LHA, [N II] 6548, 6583	\\
\ion{He}{1}		&	4921.93 		&	24.587	&	21.2180	&	23.7357	&	\phn$-800$	&	\phn$+800$	&	3	& W, Fe II 4922			\\
\ion{He}{1}		&	5015.68		&	24.587	&	20.6152	&	23.0864	&	$-1000$	&	$+1000$	&	4	& W, Fe II 5015		\\
\ion{He}{1}		&	5875.62		&	24.587&	20.9723	&	23.0827 	&	$-1000$	&	$+2000$	&	5	& W, [Fe II], Na D1, D2	\\
\ion{He}{1}		&	6678.15		&	24.587	&	21.2263	&	23.0831  	&	$-1000$	&	$+1000$	&	6	& W, [Ni II]	\\
\ion{He}{1}		&	7065.19		&	24.587	&	21.2263	&	22.7273  	&	$-1000$	&	$+1000$	&	7	& W, telluric	\\
\ion{N}{2}			&	5666.63		&	29.601	&	18.4734	&	20.6616	&	\nodata	&	\nodata	&	8	& Fe II 5658	\\
\enddata
\tablenotetext{a}{Atomic data collected from NIST (http://nist.gov/pml/data/asd.cfm).}
\tablenotetext{b}{A `W' indicates a Weigelt knot contamination from the same transition, which dominates the narrow emission component. `LHA' indicates ``anomalous absorption", which is from intervening gas located in the Little Homunculus, while `HA' refers to absorption from the intervening gas of the Homunculus.}
\end{deluxetable}

\clearpage
\begin{deluxetable}{lccccc}
\label{Measurements}
\tablecaption{Wind Line Measurements.  - complete table to only be online} 
\tablewidth{0pt}
\tablehead{
	\colhead{Ion / }	&
	\colhead{HJD}	&	
	\colhead{Phase} 	&	
	\colhead{$W_\lambda$} &
	\colhead{$V_{\rm min}$} &
	\colhead{$V_{b}$}   \\	

	\colhead{Line}	&
	\colhead{-2,454,000}	&	
	\colhead{} 	&	
	\colhead{(\AA)} &
	\colhead{(km s$^{-1}$)} &
	\colhead{(km s$^{-1}$)}  
	}
			
\startdata
H$\alpha$ &       784.858 &        11.971 & -650.3 	&      -723.7 &      -18.1   \\
H$\alpha$ &       818.806 &        11.988 & -551.0 	&      -735.1 &      -26.6   \\
H$\alpha$ &       819.811 &        11.989 & 	-568.9	 &      -757.2 &       -24.4  \\
H$\alpha$ &       821.852 &        11.989 &  -584.0	&      -755.3 &       -23.4  \\
H$\alpha$ &       824.754 &        11.991 &  -588.7	&      -768.3 &        -27.1 \\
H$\alpha$ &       829.784 &        11.994 &  -575.6	&      -790.3 &        -27.5 \\
H$\alpha$ &       831.753 &        11.995 &  -559.1	&      -791.9 &        -28.3 \\
H$\alpha$ &       836.744 &        11.997 &  -542.4	&      -1028.7 &      -25.8   \\
\enddata
\end{deluxetable}

\clearpage

\end{document}